\documentclass{article}

\PassOptionsToPackage{numbers,compress}{natbib}

\usepackage[eandd, preprint]{neurips_2026}

\usepackage[utf8]{inputenc}
\usepackage[T1]{fontenc}
\PassOptionsToPackage{hyphens}{url}
\usepackage{hyperref}
\usepackage{url}
\usepackage{booktabs}
\usepackage{amsfonts}
\usepackage{amsmath}
\usepackage{amssymb}  
\usepackage{nicefrac}
\usepackage{microtype}
\usepackage{graphicx}
\usepackage{xcolor}
\usepackage{multirow}
\usepackage{enumitem}
\usepackage{subcaption}
\usepackage{colortbl}
\usepackage{listings}
\usepackage{threeparttable}  
\usepackage{fvextra}  
\usepackage{upquote}  
\usepackage{tcolorbox}
\tcbuselibrary{skins, breakable}
\usepackage{tikz}
\usetikzlibrary{positioning, fit, shapes.geometric}
\usepackage{wrapfig}  
\usepackage{tabularx}  
\usepackage{float}  
\usepackage{needspace}  

\definecolor{vsbg}{HTML}{1E1E1E}
\definecolor{vstext}{HTML}{F8F8F2}       
\definecolor{vsline}{HTML}{858585}
\definecolor{vsheading}{HTML}{569CD6}      
\definecolor{vsfunc}{HTML}{DCDCAA}         
\definecolor{vsbracket}{HTML}{FFD700}      

\graphicspath{{figures/}}

\renewcommand{\dagger}{\textsuperscript{\dag}}
\newcommand{\fmetric}[1]{\textbf{F#1}}
\newcommand{\cmetric}[1]{\textbf{C#1}}
\newcommand{\etype}[1]{\textbf{E#1}}

\newcommand{\bt}{\textasciigrave\textasciigrave\textasciigrave}  

\newlength{\promptboxwidth}
\newtcolorbox{promptbox}{%
  colback=vsbg, colframe=vsbg, boxrule=0pt, arc=2pt, outer arc=2pt,
  left=4pt, right=4pt, top=3pt, bottom=3pt,
  width=\dimexpr\promptboxwidth+12pt\relax,
  before skip=4pt, after skip=4pt
}

\title{LatentMD: Benchmarking Markdown Boundary\\Failures in LLM-Generated Text\thanks{Code and scoring tool: \url{https://anonymous.4open.science/r/LatentMD-code-B51D}.}}

\author{
  Sungjune Lee \\
  Seoul National University \\
  \texttt{casthood@snu.ac.kr} \\
  \And
  Myungjoo Kang\thanks{Corresponding author.} \\
  Seoul National University \\
  \texttt{mkang@snu.ac.kr} \\
}

\begin{document}

\maketitle

\begin{abstract}
Large language models (LLMs) increasingly generate Markdown that is consumed by renderers, agents, code extractors, and structured downstream pipelines.
Yet existing evaluations often conflate content quality with format adherence, leaving Markdown boundary failures under-measured.
We introduce LatentMD, a benchmark and evaluation protocol for diagnosing CommonMark-level fence-boundary failures in LLM-generated Markdown.
LatentMD separates content correctness from boundary correctness, enabling detection of outputs that are content-correct but boundary-broken.
The benchmark contains $4{,}179$ prompts and a CLI for scoring arbitrary model outputs.
Across $9$ LLMs and roughly $37{,}600$ generations, we find that Markdown boundary failures are widespread: $38.0\%$ of valid main-grid outputs are content-correct but boundary-broken, with substantial boundary breakage under unspecified prompts and in a small human-authored validation set.
Ablations show that failures are driven primarily by same-family symmetric-delimiter collisions rather than nesting alone, are only partially mitigated by prompt hints, and generalize to Python triple-quote docstrings while JSON remains robust as an asymmetric-delimiter control.
LatentMD provides a reproducible diagnostic target for parser-sensitive LLM evaluation.
\end{abstract}

\section{Introduction}
\label{sec:introduction}

LLM Markdown output is increasingly read by automated tools, not just human eyes~--- chat UIs render it on screen, agentic systems parse it to decide actions, and search-and-extraction systems pull code blocks out of it for reuse~\citep{Tan2024HtmlRAG}.
These tools rely on the implicit assumption that the output's structure is correct; when the structure breaks, the tools break with it.

This breaks regularly in the wild.
A user asks a chat model to write a Markdown tutorial about Python generators with two code examples; the model wraps its reply in a fenced block tagged \texttt{markdown} and embeds the two Python examples as inner fenced blocks.
In many chat UIs and developer tools, this output \textbf{visibly breaks}: the inner fences close the outer wrapper at the first length match, and the trailing content leaks outside the intended block as raw text or stripped fence characters.
In other cases the breakage is invisible at render time but still corrupts any tool that reads the document's spec-level structure.

This is not a contrived test case: the same fence-boundary failure has been filed across major LLM and developer tools, including OpenAI ChatKit, Google Gemini CLI, Microsoft VS Code, and open-source agentic projects~\citep{chatkitjs89,geminicli10515,vscode295126,goose8290,continue6059,openwebui5016}.
These incidents show that nested-fence breakage is a recurring deployment failure rather than a benchmark artefact.

Yet existing LLM-output benchmarks do not measure this failure.
They score content correctness and surface format adherence~\citep{Yang2025StructEval, Chen2025MDEval, Do2024FormatBiasEval, Zhou2023IFEval, Jiang2024FollowBench}, but none isolate whether the model can correctly track nested fence boundaries. A generation that is content-correct but boundary-broken would therefore be scored as \emph{passing} by these benchmarks. The same output would still corrupt any spec-strict downstream parser or renderer.
We close this gap with LatentMD, a $4{,}179$-prompt rule-based benchmark that adds an explicit \emph{Boundary} axis orthogonal to Content.
This isolates a content-correct but boundary-broken failure mode, the \emph{latent failure}. It supports paired causal and trend claims that aggregate-only benchmarks cannot.

\paragraph{Contributions.}
We make the following contributions.
\textbf{(a) Benchmark.}
We introduce LatentMD, the first systematic benchmark of LLM Markdown fence-boundary failures, comprising $4{,}179$ prompts that systematically vary outer wrapping, inner nesting, and delimiter properties.
\textbf{(b) Diagnostic methodology.}
We define an error-type taxonomy (premature closure, unclosed fence, latent delimiter collision) and a four-cell decomposition that separates content correctness from CommonMark-level boundary correctness, surfacing latent failures missed by content/format-only evaluations.
\textbf{(c) Empirical study.}
Across $9$ LLMs we show that Markdown boundary failures are widespread, driven primarily by same-family symmetric-delimiter collision, only partially mitigated by prompting, and echoed in Python but not JSON.
\textbf{(d) Reproducible artifact.}
We release prompts, generation scripts, and a stand-alone scoring tool for applying LatentMD to arbitrary model outputs.
\section{Background \& Related Work}
\label{sec:background}

\begin{wraptable}[15]{r}{0.58\linewidth}
  \vspace{-1.0em}
  \caption{Failure mode coverage. \checkmark: directly measured; \textit{partial}: indirectly detectable, no cause-level diagnosis; \textit{concept}: precursor without implementation; --: not covered. Per-cell: App.~\ref{app:coverage_justification}.}
  \label{tab:coverage_matrix}
  \centering
  \scriptsize
  \setlength{\tabcolsep}{3pt}
  \begin{tabular}{lcccc}
    \toprule
    \textbf{Failure Type} & \textbf{StructEval} & \textbf{FormatBias} & \textbf{IFEval} & \textbf{LatentMD} \\
    \midrule
    \etype{1} Premature Closure       & \textit{partial} & --               & --               & \checkmark \\
    \etype{2} Unclosed Fence          & \textit{partial} & --               & --               & \checkmark \\
    \etype{3} Latent Collision        & --              & --               & --               & \checkmark \\
    \midrule
    \fmetric{1} Instr. Compliance     & --              & --               & \textit{partial} & \checkmark \\
    \fmetric{2} Fence Balance         & \textit{partial} & --               & --               & \checkmark \\
    \fmetric{3} Safe Length           & --              & --               & --               & \checkmark \\
    \midrule
    4-cell Decomposition              & --              & \textit{concept} & --               & \checkmark \\
    Latent Failure Detection          & --              & --               & --               & \checkmark \\
    \bottomrule
  \end{tabular}
  \vspace{-1.0em}
\end{wraptable}

Existing structured-output benchmarks measure adjacent but non-overlapping properties.
StructEval~\citep{Yang2025StructEval} and StructTest~\citep{Chen2024StructTest} score AST-level structural matching across JSON, XML, and Markdown; MDEval~\citep{Chen2025MDEval} evaluates Markdown content quality; FormatBias~\citep{Do2024FormatBiasEval} argues for separating format from content; IFEval~\citep{Zhou2023IFEval} and FollowBench~\citep{Jiang2024FollowBench} measure format-instruction compliance; FMBench~\citep{Wang2026FMBench} benchmarks format-following; STED~\citep{Wang2025STED} addresses structured-output reliability.
None directly tests fence-balance or safe-length selection inside the same delimiter family~--- the precise mechanism that creates latent failures.
Formal-language work~\citep{Deletang2023Chomsky} has shown that autoregressive models struggle with balanced-bracket tracking on synthetic formal-language tasks. Our benchmark extends this observation to LLM Markdown generation, where nested fences impose the same balanced-bracket structure under realistic deployment conditions.
Table~\ref{tab:coverage_matrix} gives the per-failure-type coverage matrix.
\section{Benchmark Design}
\label{sec:benchmark_design}

\paragraph{Three Roles of Symmetric Delimiters.}
\label{sec:three_roles}
In Markdown fenced code blocks, the same character sequence (triple backticks \verb|```| or tildes \verb|~~~|) serves three distinct roles within a single generation: \textbf{(1) Wrapper} --- the outer fence enclosing the entire response; \textbf{(2) Content} --- inner fences that form the document's structure (e.g., code blocks within a tutorial); \textbf{(3) Literal} --- fence characters appearing as verbatim text (e.g., when explaining Markdown syntax). When an autoregressive model fails to disambiguate these three roles during generation, the output incurs a \emph{boundary-state tracking failure}, violating the delimiter nesting rules of CommonMark Section 4.5~\citep{CommonMark2024}. The benchmark axes below stratify generations by how each role is constrained.

\paragraph{Operational Basis.}
We adopt CommonMark Spec 0.31.2~\citep{CommonMark2024} as the operational reference for \emph{boundary failure}; this also applies to GFM~\citep{GFMSpec2019}, which inherits CommonMark's fenced-code rules unchanged. The spec serves as an externally specified reference, ensuring boundary-failure detection is reproducible rather than renderer-dependent.

\subsection{Experimental Axes}
\label{sec:axes}

The benchmark is organized around three orthogonal axes, each isolating a distinct source of boundary-state tracking failure. The \textbf{A-axis} manipulates whether the model must itself produce an outer fence, isolating wrapper-induced collision. The \textbf{B-axis} varies the inner nesting load, isolating literal-fence interference inside the response. The \textbf{D-axis} ablates delimiter properties (family, length, cross-family) to identify which property of symmetric delimiters drives the failure. A visual overview of all three axes is in Appendix~\ref{app:benchmark_axes_overview}.

\textbf{A-axis (Outer Constraint)} controls wrapper pressure. \textbf{A1 (Direct)} forbids an outer fence, giving a no-wrapper baseline for failures that do not require wrapping. \textbf{A2 (Wrapped)} requires a full-response fenced block, serving as the stress condition for wrapper-induced collision. \textbf{A3 (Unspecified)} gives no wrapper instruction and measures realistic base-rate self-wrapping. A1 is a negated instruction; LLM under-compliance~\citep{Jang2023NegatedPrompts} is isolated in Table~\ref{tab:a1_decomposition} (Step~3).

\textbf{B-axis (Inner Nesting)} controls same-character delimiter load. \textbf{B1 (Single)} has one code block and no nesting pressure. \textbf{B2 (Nested Single)} adds one literal Markdown rendering of that block, forcing content-vs-literal fence disambiguation; \textbf{B3 (Nested Multiple)} repeats this with two or more literal fence examples, increasing opportunities for premature closure. \textbf{B4 (Mixed Elements)} adds tables, blockquotes, or lists as non-fence distractors while retaining code-block structure.

\textbf{D-axis (Delimiter Ablations)} isolates which delimiter property drives failure. \textbf{D-Family} swaps the outer delimiter between backticks and tildes; \textbf{D-Outer-Length} fixes the outer length to 3, 4, or 5; \textbf{D-Inner-Run} fixes an inner same-family run of length 3, 4, or 5 to test a-priori \fmetric{3} safe-length prediction. \textbf{D-Cross-Family} uses different outer/inner delimiter families, removing same-family collision by design and distinguishing collision from generic nesting.

\paragraph{Full A$\times$B Crossing.}
Each prompt instantiates one A condition and one B condition. The $3$ A levels and $4$ B levels are fully crossed, giving $12$ A$\times$B cells per (TASK, LANG) pair. This full crossing supports within-task paired comparisons across treatments (\S\ref{sec:slot_generation}).

\subsection{Slot-Based Prompt Generation}
\label{sec:slot_generation}

Each prompt is constructed from a structural template whose two slots, \textbf{LANG} and \textbf{TASK}, are filled from external sources. Drawing slots from an external dataset (McEval~\citep{Chai2024McEval}) rather than model-generated content ensures the test distribution is independent of the evaluated model. \textbf{LANG} comprises $9$ programming languages (popularity consensus, Appendix~\ref{app:lang_selection}) and $3$ structured formats (Markdown, JSON, HTML), all restricted to languages with McEval task coverage. \textbf{TASK} provides $15$ task instances per LANG ($180$ in total), mechanically extracted from McEval instruction fields by stratified sampling that selects $5$ easy, $5$ middle, and $5$ hard tasks under a fixed random seed.

The benchmark totals 4{,}179 unique prompts. The breakdown is: 2{,}160 from the main A$\times$B grid, 1{,}440 from D-axis delimiter ablations, 432 from hint ablations, 27 from inline code-span probes, 90 from cross-format probes, and 30 human-authored natural-prompt validation prompts.

\begin{wrapfigure}[14]{r}{0.50\linewidth}
  \vspace{-1.0em}
  \centering
  \begin{tcolorbox}[
    colback=vsbg,
    colframe=vsbg,
    boxrule=0pt,
    arc=2pt,
    outer arc=2pt,
    left=4pt, right=4pt, top=3pt, bottom=3pt,
    width=\linewidth
  ]
\begin{Verbatim}[fontsize=\scriptsize, formatcom=\color{white}, baselinestretch=0.9]
Write a Markdown document that addresses
the following task. Include one <LANG>
code example. In the document, also show
the raw Markdown source for that code
example, including the opening fence,
the language tag, the code, and the
closing fence. Wrap your entire response
in a fenced markdown code block.

Task:
<TASK>
\end{Verbatim}
  \end{tcolorbox}
  \caption{Slot-based prompt template (A2$\times$B2). Per-condition examples: App.~\ref{app:prompt_templates}.}
  \label{fig:prompt_template_example}
  \vspace{-1.0em}
\end{wrapfigure}
\paragraph{Released Artifact.}
All prompts, templates, and the generation script are released as part of the artifact. We additionally release a stand-alone command-line tool that scores any model on the benchmark with the metrics defined in \S\ref{sec:evaluation_metrics}. The tool ingests a JSONL file of (\texttt{prompt\_id}, \texttt{response}) pairs and emits per-record labels and aggregate failure rates with Wilson 95\% confidence intervals. External users can therefore score arbitrary models without re-implementing the metrics.
Figure~\ref{fig:prompt_template_example} shows one instantiated prompt to anchor the slot-template structure; the slots \texttt{<LANG>} and \texttt{<TASK>} are filled per (LANG, TASK) pair.
\section{Evaluation Metrics}
\label{sec:evaluation_metrics}

We score each response along two orthogonal axes: \emph{boundary correctness} (fence balance and safe outer-fence length) and \emph{content correctness} (lightweight structural checks ensuring the response is a meaningful document).
Let $r \in \Sigma^{*}$ denote a response and $p \in \mathcal{P}_{\text{main}}$ a prompt — a combination of experimental axis settings and slot values, with full domain definitions in Appendix~\ref{app:format_metrics}.
Predicates that do not apply to a given prompt return \text{NA} (Not Available), which is ignored in the boolean conjunctions used below ($x \land \text{NA} = x$).

\subsection{Format Metrics (\fmetric{1}--\fmetric{3})}
\label{sec:format_metrics}

\textbf{\fmetric{1} (Instruction Compliance)} measures whether the model follows the A-axis wrapping directive: pass for A1 if the response does not contain an outer wrapper, pass for A2 if it does, \text{NA} for A3; it extends prior instruction-compliance metrics~\citep{Zhou2023IFEval} to the wrapper presence/absence case. \textbf{\fmetric{2} (Fence Balance)} checks CommonMark~\citep{CommonMark2024} fence balance with a stack discipline: opening fences push their family and length, and only bare same-family fences of at least the opening length can close the current block; the response passes iff the stack is empty at EOF, and the check is implemented on top of \texttt{markdown-it-py}~\citep{markdownitpy2024} with full pseudocode in Appendix~\ref{app:algorithms}. \textbf{\fmetric{3} (Safe Length)} checks whether the outer fence is long enough to survive any same-family run inside its content (\text{NA} when no outer fence exists), reported in two variants: \emph{a-priori} (used in the D-inner-run ablation, requiring an outer fence one character longer than the specified inner run) measures predictive compliance, and \emph{post-hoc} (used in the main A$\times$B analysis, with safe length recomputed from realized inner content) measures internal consistency.

\subsection{Content Metrics (\cmetric{1}--\cmetric{3})}
\label{sec:content_metrics}

Three content metrics ensure the response is a meaningful document rather than structurally empty output, in the spirit of FormatBias~\citep{Do2024FormatBiasEval} and prior structural checks~\citep{Yang2025StructEval,Chen2025MDEval}. \textbf{\cmetric{1} (Heading Structure)} passes when the response contains at least two headings. \textbf{\cmetric{2} (Code Language Match)} passes when at least one code block's info string matches the requested LANG (with alias map; fails if no code blocks are present). \textbf{\cmetric{3} (Block Compliance)} is a B-condition-aware check that the response contains the expected number of literal-syntax fenced blocks: zero for B1 and B4 (no nested literal fences), one for B2 (single nested fence), and two for B3 (multiple nested fences).

\subsection{Boundary Violation Error Types (\etype{0}--\etype{3})}
\label{sec:boundary_errors}

\begin{figure}[h]
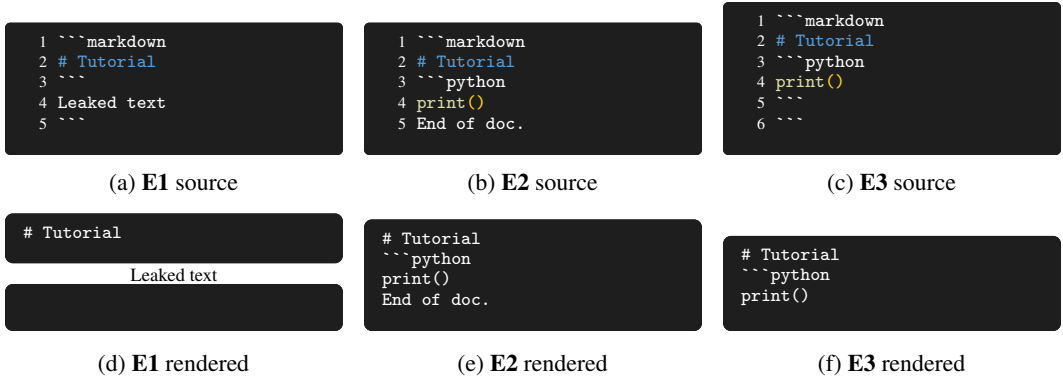

  \centering
  \begin{subfigure}[b]{0.32\linewidth}
    \centering
    \begin{mdsource}
\begin{Verbatim}[fontsize=\scriptsize, formatcom=\color{white}, commandchars=\\\{\}, numbers=left, numbersep=4pt, xleftmargin=1.5em]
```markdown
\textcolor{vsheading}{# Tutorial}
```
Leaked text
```
\end{Verbatim}
    \end{mdsource}
    \caption{\etype{1} source}
    \label{fig:e1_src}
  \end{subfigure}\hfill
  \begin{subfigure}[b]{0.32\linewidth}
    \centering
    \begin{mdsource}
\begin{Verbatim}[fontsize=\scriptsize, formatcom=\color{white}, commandchars=\\\{\}, numbers=left, numbersep=4pt, xleftmargin=1.5em]
```markdown
\textcolor{vsheading}{# Tutorial}
```python
\textcolor{vsfunc}{print}\textcolor{vsbracket}{()}
End of doc.
\end{Verbatim}
    \end{mdsource}
    \caption{\etype{2} source}
    \label{fig:e2_src}
  \end{subfigure}\hfill
  \begin{subfigure}[b]{0.32\linewidth}
    \centering
    \begin{mdsource}
\begin{Verbatim}[fontsize=\scriptsize, formatcom=\color{white}, commandchars=\\\{\}, numbers=left, numbersep=4pt, xleftmargin=1.5em]
```markdown
\textcolor{vsheading}{# Tutorial}
```python
\textcolor{vsfunc}{print}\textcolor{vsbracket}{()}
```
```
\end{Verbatim}
    \end{mdsource}
    \caption{\etype{3} source}
    \label{fig:e3_src}
  \end{subfigure}

  \vspace{0.5em}

  \begin{subfigure}[b]{0.32\linewidth}
    \centering
    \begin{mdblock}
\begin{Verbatim}[fontsize=\scriptsize, formatcom=\color{white}, commandchars=\\\{\}]
# Tutorial
\end{Verbatim}
    \end{mdblock}
    {\scriptsize\color{black}Leaked text}
    \begin{mdblock}
    \strut
    \end{mdblock}
    \caption{\etype{1} rendered}
    \label{fig:e1_rnd}
  \end{subfigure}\hfill
  \begin{subfigure}[b]{0.32\linewidth}
    \centering
    \begin{mdblock}
\begin{Verbatim}[fontsize=\scriptsize, formatcom=\color{white}, commandchars=\\\{\}]
# Tutorial
```python
print()
End of doc.
\end{Verbatim}
    \end{mdblock}
    \caption{\etype{2} rendered}
    \label{fig:e2_rnd}
  \end{subfigure}\hfill
  \begin{subfigure}[b]{0.32\linewidth}
    \centering
    \begin{mdblock}
\begin{Verbatim}[fontsize=\scriptsize, formatcom=\color{white}, commandchars=\\\{\}]
# Tutorial
```python
print()
\end{Verbatim}
    \end{mdblock}
    \caption{\etype{3} rendered}
    \label{fig:e3_rnd}
  \end{subfigure}
  \caption{Minimal source examples for \etype{1}/\etype{2}/\etype{3} under the LatentMD boundary classifier. Top row: source. Bottom row: how it renders to a reader. \textbf{(a, d) \etype{1}} (Premature Closure): the bare fence at line 3 matches the outer and prematurely closes it; line 5 opens a new bare-fence block that never closes, so $\fmetric{2}$ fails with an interior bare match. \textbf{(b, e) \etype{2}} (Unclosed Fence): the outer fence never closes, so all content stays trapped; $\fmetric{2}$ fails without an interior bare match. \textbf{(c, f) \etype{3}} (Latent Collision): inner and outer fences nest cleanly so $\fmetric{2}$ passes, yet the inner \texttt{\bt python} shares the outer's 3-backtick length, so $\fmetric{3}$ fails.}
  \label{fig:e_minimum_examples}
\end{figure}

Every generation receives one of four mutually exclusive and exhaustive error labels --- \etype{0} (Correct), \etype{1} (Premature Closure), \etype{2} (Unclosed Fence), or \etype{3} (Latent Collision) --- with conditions and interpretations given in Table~\ref{tab:e_types}, and a minimal worked example for each failure type shown in Figure~\ref{fig:e_minimum_examples}. \etype{1} fires when $\fmetric{2}$ fails and an interior bare same-family fence of length at least the outer-fence length is present; \etype{2} fires when $\fmetric{2}$ fails without such an interior bare match.

\begin{table}[h]
\caption{Boundary violation error types (\etype{0}--\etype{3}).}
\label{tab:e_types}
\centering
\small
\begin{tabular}{@{}lcl@{}}
\toprule
Label & Condition & Interpretation \\
\midrule
\etype{0} (Correct)            & $\fmetric{2}(r) \land \fmetric{3}(r)$       & boundary-correct \\
\etype{1} (Premature Closure)  & $\neg\fmetric{2}(r) \land \exists\,\text{bare match}$       & outer terminates early at a matching bare fence \\
\etype{2} (Unclosed Fence)     & $\neg\fmetric{2}(r) \land \nexists\,\text{bare match}$      & outer stays open at EOF (no matching bare fence) \\
\etype{3} (Latent Collision)   & $\fmetric{2}(r) \land \neg\fmetric{3}(r)$   & length collision masked by balanced match \\
\bottomrule
\end{tabular}
\end{table}

Per the CommonMark spec, a closing fence must contain only fence characters plus optional trailing whitespace, so a fence with an info string (e.g., \verb|```python|) can never close an outer fence; premature closure arises when the model writes bare triple backticks (\verb|```|) inside an inner block whose length matches the outer. When the response wraps an outer fence, the structural fix in all three failure cases is to satisfy $\fmetric{3}$, i.e., to make the outer fence strictly longer than the longest inner same-family run.

\newtcolorbox{mdsource}{%
  colback=vsbg,
  colframe=vsbg,
  boxrule=0pt,
  arc=2pt,
  outer arc=2pt,
  left=2pt, right=2pt, top=2pt, bottom=2pt,
  before skip=0pt, after skip=0pt
}

\newtcolorbox{mdblock}{%
  colback=vsbg,
  colframe=vsbg,
  boxrule=0pt,
  arc=3pt,
  outer arc=3pt,
  left=4pt, right=4pt, top=2pt, bottom=2pt,
  before skip=1pt, after skip=2pt,
  fontupper=\scriptsize\ttfamily\color{white}
}

\subsection{Four-Cell Decomposition and Latent Failure}
\label{sec:four_cell}
\label{sec:latent_failure}

Every generation is decomposed along two orthogonal axes. Formally:
\begin{align}
\mathrm{Boundary}(r) &:= \fmetric{2}(r) \land \fmetric{3}(r), \label{eq:boundary} \\
\mathrm{Content}(r, p) &:= \cmetric{1}(r) \land \cmetric{2}(r, p) \land \cmetric{3}(r, p). \label{eq:content}
\end{align}
\begin{wrapfigure}[8]{r}{0.50\linewidth}
  \vspace{-1.0em}
  \centering
  \begin{tikzpicture}[
    cell/.style={draw, minimum width=2.6cm, minimum height=0.7cm, align=center, font=\scriptsize},
    starred/.style={cell, fill=yellow!15, draw=red!60, very thick},
    axislabel/.style={font=\bfseries\scriptsize},
    rowlabel/.style={font=\bfseries\scriptsize, align=center}
  ]
    \node[axislabel] at (1.4, 1.00) {Boundary~$\checkmark$};
    \node[axislabel] at (4.1, 1.00) {Boundary~$\times$};
    \node[rowlabel] at (-0.65, 0.4)  {Content\\$\checkmark$};
    \node[rowlabel] at (-0.65, -0.5) {Content\\$\times$};
    \node[cell] (BC) at (1.4, 0.4)  {\textbf{Both correct}\\Spec-OK, on-topic};
    \node[cell] (FO) at (4.1, 0.4)  {\textbf{Format-only}\\Bdry OK, content fail};
    \node[starred] (LF) at (1.4, -0.5) {$\star$~\textbf{Latent failure}\\Content OK, bdry broken};
    \node[cell] (BW) at (4.1, -0.5) {\textbf{Both wrong}\\Bdry + content fail};
  \end{tikzpicture}
  \caption{Four-cell decomposition.}
  \label{fig:four_cell}
  \vspace{-1.0em}
\end{wrapfigure}
The four cells (Figure~\ref{fig:four_cell}) are mutually exclusive and exhaustive over valid generations.
The starred cell --- outputs that are content-correct but boundary-wrong --- is the \emph{latent failure}, defined formally in Eq.~(\ref{eq:lf}) below.
Prior content-only or format-only benchmarks judge such outputs as passing, even though they violate the spec at the boundary level.
Latent failures may render plausibly in lenient chat interfaces yet corrupt downstream consumers (agents, code extractors, structured parsers) that depend on spec-compliant boundary parsing; Appendix~\ref{app:latent_failure_example} shows a concrete instance, and Appendix~\ref{app:scope} details which Markdown constructs are in scope.

For a model $M$ and a prompt subset $\mathcal{P}^{\prime} \subseteq \mathcal{P}$, let $\mathcal{V} \subseteq \mathcal{P}^{\prime}$ denote the \emph{valid} subset --- prompts whose response was not externally truncated by hitting the token limit (see \S\ref{sec:truncation}).
We define latent failure and the latent failure rate (LFR) as
\begin{align}
\mathrm{LF}(r, p) &:= \neg \mathrm{Boundary}(r) \land \mathrm{Content}(r, p), \label{eq:lf} \\
\widehat{\mathrm{LFR}}_{M}(\mathcal{P}^{\prime}) &= \frac{1}{|\mathcal{V}|} \sum_{p \in \mathcal{V}} \mathbf{1}\{\mathrm{LF}(M(p), p)\}, \label{eq:lfr}
\end{align}
reported with the Wilson 95\% score interval~\citep{Wilson1927}.
This is the headline metric our benchmark surfaces, and that the research questions in \S\ref{sec:experiments} ultimately decompose.

\subsection{Validity Filtering}
\label{sec:truncation}

Each model in our evaluation reports a \texttt{finish\_reason} field on every generation indicating why decoding stopped.
We exclude generations whose \texttt{finish\_reason\,=\,length} (the model hit the \texttt{max\_tokens} limit), as these are token-truncation artefacts rather than complete free-form outputs.
No generation in our data triggered the safety \texttt{content\_filter} flag, so this case did not arise empirically.
Valid $N$ (after exclusion) is the denominator for all reported rates.
Per-model and per-LANG truncation statistics are in Appendix~\ref{app:truncation}.
\section{Experiments}
\label{sec:experiments}

We evaluate $9$ LLMs on LatentMD, following a mechanism-to-prevalence arc: trigger (wrapping $\times$ nesting), cause (same-family collision), mitigation limits (safe-length planning and hints), cross-format generalization, and latent-failure distribution.
We structure the results around $5$ RQs.
\textbf{(RQ1)} How does explicit outer-container wrapping interact with inner nesting complexity to trigger boundary-state tracking failures?
\textbf{(RQ2)} Are boundary failures driven by same-family delimiter collision rather than nesting complexity alone?
\textbf{(RQ3)} Can LLMs select a safe outer-fence length when inner-delimiter pressure is explicit, and do prompt-level mitigations close the gap?
\textbf{(RQ4)} Does the boundary-tracking failure generalize beyond Markdown fenced blocks to other structured-output formats?
\textbf{(RQ5)} How prevalent are latent failures, and how are they distributed across models, tiers, and languages?
Section~\ref{sec:a1_decomp} validates A1/A3/A2 stratification and human-authored prompts.

\subsection{Setup}
\label{sec:setup}

\begin{wraptable}[16]{r}{0.55\linewidth}
  \vspace{-1.0em}
  \caption{Models evaluated. ``Open'': open-weight (vLLM/HF-served); ``Closed'': closed-source API. Tiers reflect open-weight parameter count; closed-source counts are unavailable (Large/API group).}
  \label{tab:model_list}
  \centering
  \footnotesize
  \begin{tabular}{cllcc}
    \toprule
    \textbf{Tier} & \textbf{Model} & \textbf{Provider} & \textbf{Params} & \textbf{Type} \\
    \midrule
    \multirow{3}{*}{\rotatebox{90}{\scriptsize Small}}
    & Qwen3-7B          & Alibaba   & 7B  & Open   \\
    & Gemma-2-9B         & Google    & 9B  & Open   \\
    & Llama-3.1-8B       & Meta      & 8B  & Open   \\
    \midrule
    \multirow{2}{*}{\rotatebox{90}{\scriptsize Mid}}
    & Qwen3-32B          & Alibaba   & 32B & Open   \\
    & Gemma-3-27B        & Google    & 27B & Open   \\
    \midrule
    \multirow{4}{*}{\rotatebox{90}{\scriptsize Large}}
    & Llama-3.1-70B      & Meta      & 70B & Open   \\
    & GPT-4o             & OpenAI    & --  & Closed \\
    & Gemini-2.5-Flash   & Google    & --  & Closed \\
    & Claude-Sonnet-4    & Anthropic & --  & Closed \\
    \bottomrule
  \end{tabular}
  \vspace{-1.0em}
\end{wraptable}

We evaluate 9 models spanning 3 tiers (Small, Mid, Large/API) and both open-weight and closed-source API categories (Table~\ref{tab:model_list}).
Open-weight models are served locally via vLLM with HuggingFace Transformers as fallback for CUDA-constrained nodes; closed-source models~--- GPT-4o~\citep{OpenAI2024GPT4o}, Claude-Sonnet-4~\citep{Anthropic2024Claude}, Gemini-2.5-Flash~\citep{Google2025GeminiFlash}~--- are accessed through their APIs.
We use a common decoding configuration wherever supported: temperature$=0$ for greedy decoding, top\_p$=0.9$, and max\_tokens$=7168$ (the largest completion budget compatible with Gemma-2-9B's 8K context window after prompt overhead).
Open-weight runs fix seed$=42$; reasoning modes are disabled so outputs are pure Markdown; provider-specific top\_k and disable flags are in Appendix~\ref{app:hyperparameters}.
The full benchmark generates 4{,}179 unique prompts $\times$ 9 models $\approx$ 37{,}600 generations; we record \texttt{finish\_reason} for truncation filtering (\S\ref{sec:truncation}, Appendix~\ref{app:truncation}).
Main results report Wilson 95\% confidence intervals (CIs) over valid $N$ (Appendix~\ref{app:baseline_with_ci}) as descriptive binomial uncertainty over prompt instances.

\subsection{Wrapping \texorpdfstring{$\times$}{x} Nesting Triggers Boundary Failure (RQ1)}
\label{sec:main_results}

\begin{wrapfigure}[14]{r}{0.45\linewidth}
  \vspace{-1.0em}
  \centering
  \includegraphics[width=\linewidth]{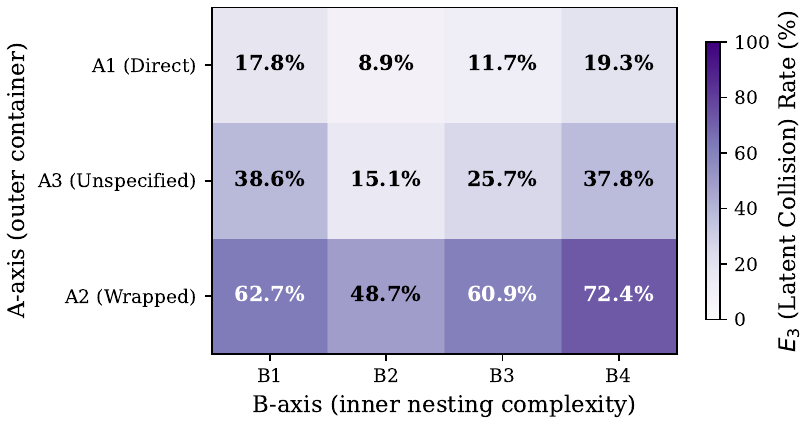}
  \caption{\etype{3} (latent collision) rate across A$\times$B, averaged over 9 models and LANG. Full per-model total boundary-failure rates ($\etype{1}\cup\etype{2}\cup\etype{3}$) in Appendix~\ref{app:full_results}.}
  \label{fig:main_heatmap}
  \vspace{-1.0em}
\end{wrapfigure}
Figure~\ref{fig:main_heatmap} shows the pooled \etype{3} pattern across the main A$\times$B grid.
We order A by experimental severity rather than condition ID: \textbf{A1} (no-wrap baseline), \textbf{A3} (unspecified, realistic anchor), and \textbf{A2} (forced-wrap stress test).
In the pooled \etype{3} view, all A1 cells remain below $20\%$.
\textbf{A3 (realistic anchor)} self-wraps $37\%$ of responses (Step~1 of Table~\ref{tab:a1_decomposition}) and pools to $40$--$55\%$ boundary failure across B conditions (Appendix~\ref{app:a_b_e_type}).
Total boundary failure under A2 reaches $100\%$ in many model$\times$B cells (Wilson 95\% CI $[97.9, 100]$, $n{=}180$ prompts per cell; Appendix~\ref{app:full_results}), confirming that A2 is a stress condition for testing whether wrapper-induced collision can drive failure to a ceiling, not a deployment prevalence estimate.
The \etype{3} mass concentrates on the A2 row ($49$--$72\%$) and peaks at A2$\times$B4, showing that wrapping itself triggers same-family delimiter collision even when inner content is structurally simple.
The lower \etype{3} rate at A2$\times$B2 should not be read as robustness: Appendix~\ref{app:a_b_e_type} shows substantial mass shifts into \etype{1}/\etype{2} while total boundary failure remains high.
Together, outer wrapping is the dominant trigger, while inner nesting modulates the \etype{1}/\etype{2}/\etype{3} failure mix; the full A$\times$B$\times$E breakdown is in Appendix~\ref{app:a_b_e_type}.

\subsection{Same-Family Collision, Not Nesting Alone, Drives Failure (RQ2)}
\label{sec:delimiter_ablation}

\begin{wraptable}[12]{r}{0.50\linewidth}
  \vspace{-1.0em}
  \caption{D-axis ablation: boundary-correct rate by delimiter condition. Boundary Correct = \fmetric{2}\,$\wedge$\,\fmetric{3}; higher is better. Per-group max in \textbf{bold}.}
  \label{tab:d_axis_breakdown}
  \centering
  \scriptsize
  \setlength{\tabcolsep}{3pt}
  \begin{tabular}{ll|ccc}
    \toprule
    \textbf{Condition} & \textbf{Value} & \textbf{Bdry OK (\%)} & \textbf{\fmetric{2} (\%)} & \textbf{\fmetric{3} (\%)} \\
    \midrule
    \multirow{2}{*}{Family} & Backtick      & 7.2 & 70.0 & 3.8 \\
                            & Tilde         & \textbf{37.0} & 83.2 & 36.7 \\
    \midrule
    \multirow{3}{*}{Outer length} & outer$=3$ & 10.3 & 69.8 & 7.0 \\
                            & outer$=4$       & \textbf{32.4} & 69.5 & 30.7 \\
                            & outer$=5$       & 28.4 & 62.5 & 25.9 \\
    \midrule
    \multirow{2}{*}{Cross-Family} & Bt outer & 22.2 & 73.3 & 20.3 \\
                                  & Ti outer & \textbf{35.3} & 74.0 & 35.3 \\
    \bottomrule
  \end{tabular}
  \vspace{-1.0em}
\end{wraptable}

\vspace{0.5em}

If boundary failure were caused primarily by generic nesting complexity, changing delimiter family while preserving nesting structure should not yield a selective improvement; if failures are driven by same-family symmetric-delimiter collision, manipulations that reduce or remove same-family collision should help most.
Switching the outer fence from backtick to tilde raises the boundary-correct rate from $7.2\%$ to $37.0\%$ (Table~\ref{tab:d_axis_breakdown} Family rows).
Explicit outer-length instructions raise the boundary-correct rate from $10.3\%$ at outer$=$3 to $32.4\%$ at outer$=$4, with diminishing returns at outer$=$5 ($28.4\%$).
The most direct diagnostic contrast comes from cross-family nesting: using different delimiter families for outer and inner fences structurally removes same-family collision while preserving nesting, raising the boundary-correct rate to $22.2\%$ for backtick-outer/tilde-inner and $35.3\%$ for tilde-outer/backtick-inner.
Residual failures remain high, indicating that delimiter collision is not the only failure source; nevertheless, the selective improvement supports same-family symmetric collision as a dominant driver rather than generic nesting alone.

\subsection{Safe-Length Selection Failure and Prompt Mitigations (RQ3)}
\label{sec:hint_ablation}

\begin{wrapfigure}{r}{0.46\linewidth}
  \vspace{-0.6em}
  \centering
  \includegraphics[width=\linewidth]{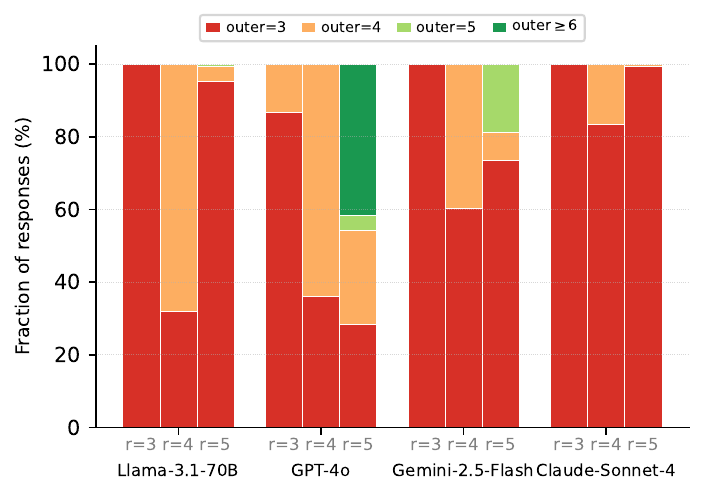}
  \caption{Outer-fence length distribution for the $4$ models with visible outer-length variation; full $9$-model table in Appendix~\ref{app:safe_length_distribution_full}.}
  \label{fig:safe_length_distribution}
  \vspace{-0.8em}
\end{wrapfigure}
The diagnosis of \emph{why} mitigation is necessary in the first place lies in the safe-length selection failure: even when the prompt explicitly states the inner-fence run length (D-inner-run ablation, $n{=}144$ per cell), $8$ of $9$ models achieve negligible \fmetric{3} a-priori pass rates, with most outputs still using an unsafe $3$-backtick outer fence; only GPT-4o adapts non-trivially (run$=$3: $13.2\%$, run$=$5: $41.7\%$).
Figure~\ref{fig:safe_length_distribution} shows the per-model distribution of outer-fence backtick lengths chosen.
Stacked colors are red (outer$=3$, fails all three inner-run conditions), orange (outer$=4$), light green (outer$=5$), dark green (outer~$\geq$~$6$); within each model, three sub-bars correspond to inner-run conditions r$=3,4,5$.
Among these, Llama-3.1-70B, GPT-4o, and Gemini-2.5-Flash partially respond by \emph{matching} the inner length (outer~$=$~run) but rarely exceed it, so these matched-length responses still fail the strict $>$ criterion; Claude-Sonnet-4 shows minor variation but little systematic safe-length planning.
The remaining $5$ models (omitted from the figure) use outer~$=3$ across nearly all runs, with \fmetric{3} pass rates $\leq 0.7\%$.

Three prompting interventions then test whether the failure can be mitigated without delimiter changes:
\textbf{H1} (boundary-awareness rule: ``outer fences must use more characters than inner fences''),
\textbf{H2} (one-shot example of correct nesting),
\textbf{H3} (length-specific instruction: ``use at least 4 backticks for the outermost fence'').
H3 is the strongest mitigation, dropping mean failure by $30.9$ percentage points ($96\%\!\rightarrow\!65\%$); H1 helps moderately ($-23$pp); H2 (one-shot) is the weakest ($-8$pp).
H3 helps because it supplies a concrete length cue, but it does not solve the general safe-length problem: ``at least 4'' remains unsafe whenever inner same-family runs are length $4$ or longer.
The per-model breakdown (Figure~\ref{fig:hint_ablation}, Appendix~\ref{app:hint_ablation_figure}) reveals strong model-level heterogeneity: the two Mid models barely respond to any hint, while the Large/API models split, with some dropping below $10\%$ under H3.
Thus, prompt hints reduce but do not close the safe-length gap.

\newpage
\subsection{Manipulation Check and Natural Prompt Validation}
\label{sec:a1_decomp}

\begin{wraptable}[10]{r}{0.43\linewidth}
  \vspace{-1.0em}
  \caption{A1 four-step decomposition (9-model average). Low Step~3 + high Step~4 isolates wrapper-induced boundary failure. Per-model: App.~\ref{app:a1_per_model}.}
  \label{tab:a1_decomposition}
  \centering
  \scriptsize
  \setlength{\tabcolsep}{1.5pt}
  \begin{tabularx}{\linewidth}{clXc}
    \toprule
    \textbf{Step} & \textbf{Metric} & \textbf{Description} & \textbf{Rate} \\
    \midrule
    1 & A3 wrap rate      & Base-rate wrap (no instr.)                          & 36.9\% \\
    2 & A1 non-comp.      & ``Do NOT wrap'' ignored (\fmetric{1})               & 19.0\% \\
    3 & A1 residual       & Obeyed no-wrap; bdry fail                           & 8.1\% \\
    4 & A2 bdry fail      & Boundary fail under wrap                            & 95.5\% \\
    \bottomrule
  \end{tabularx}
  \vspace{-1.0em}
\end{wraptable}

\paragraph{Manipulation check.}
The A-axis design assumes A1, A3, and A2 stratify wrapper pressure and boundary-failure risk (baseline $<$ realistic $<$ stress).
We verify this with a four-step decomposition (Table~\ref{tab:a1_decomposition}):
\textbf{Step~1} A3 wrap rate $36.9\%$ (models self-wrap roughly one third of the time without any instruction);
\textbf{Step~2} A1 non-compliance $19.0\%$ (responses that ignored ``do not wrap'');
\textbf{Step~3} compliance-conditional A1 residual $8.1\%$ (boundary failure among the responses that did obey ``do not wrap''~--- i.e.\ the failure that remains when no outer fence exists);
\textbf{Step~4} A2 boundary failure $95.5\%$.
The $8.1\% \rightarrow 95.5\%$ gap (Step~3 vs Step~4) confirms the manipulation is well-separated and the A2 ceiling is not a generic instruction-following artefact.
Two models (Gemma-3-27B, Llama-3.1-70B) exhibit non-trivial Step~3 residual ($34.5\%$, $40.7\%$), indicating a heterogeneous sub-population whose failures are not fully explained by outer wrapping alone.
Paired McNemar evidence (Appendix~\ref{app:mcnemar_a1_a2}) confirms the A2 increase is not due to task composition.
The $30$ human-authored prompts (Appendix~\ref{app:natural}) provide a second ecological check: high-elicitation prompts reach up to $56\%$ failure, broadly overlapping the A3 realistic-anchor range (\S\ref{sec:main_results}); the small set serves as a sanity check, not a population-level prevalence estimate.

\subsection{Failure Generalizes to Python but Not JSON (RQ4)}
\label{sec:cross_format}

To test whether symmetric-delimiter boundary failure is specific to Markdown or a more general property of LLM generation, we replicate the design in two contrasting target formats: Python (symmetric \texttt{\textquotedbl\textquotedbl\textquotedbl} triple-quote docstrings) and JSON (asymmetric \texttt{\{\}} brace nesting).
JSON achieves $100\%$ parse-validity across every model and every nesting depth (asymmetric-delimiter control), while Python triple-quote conditions vary widely: the three Python conditions escalate triple-quote pressure (K1: literal \texttt{\textquotedbl\textquotedbl\textquotedbl} inside docstring; K2: both \texttt{\textquotedbl\textquotedbl\textquotedbl}+\texttt{\textquotesingle\textquotesingle\textquotesingle} styles; K3: nested \texttt{\textquotedbl\textquotedbl\textquotedbl}), and JSON's K1/K2/K3 vary brace-nesting depth ($1$--$2$, $3$--$4$, or $\geq$$5$ levels).

\begin{wraptable}[8]{r}{0.55\linewidth}
  \vspace{-1.0em}
  \caption{Cross-format parse-validity (\%) per tier (Small/Mid/Large/API $=$ 3/2/4 models). Per-model: App.~\ref{app:crossformat}.}
  \label{tab:crossformat_per_tier}
  \centering
  \scriptsize
  \setlength{\tabcolsep}{3pt}
  \begin{tabular}{l|ccc|ccc}
    \toprule
    & \multicolumn{3}{c|}{\textbf{Python}} & \multicolumn{3}{c}{\textbf{JSON}} \\
    \textbf{Tier} & \textbf{K1} & \textbf{K2} & \textbf{K3} & \textbf{K1} & \textbf{K2} & \textbf{K3} \\
    \midrule
    Small    & 97.8 & 93.3 & 91.1 & 100.0 & 100.0 & 100.0 \\
    Mid      & 56.7 & 73.3 & 66.7 & 100.0 & 100.0 & 100.0 \\
    Large/API & 31.7 & 68.3 & 55.0 & 100.0 & 100.0 & 100.0 \\
    \bottomrule
  \end{tabular}
  \vspace{-1.0em}
\end{wraptable}

The tier-aggregated view (Table~\ref{tab:crossformat_per_tier}) inverts the usual ``larger is safer'' expectation: under Python K1 the Small tier averages $97.8\%$ pass rate, the Mid tier $56.7\%$, and the Large/API tier only $31.7\%$, with the K1 drop driven primarily by GPT-4o and Gemini-2.5-Flash, both hitting $0\%$ as literal delimiter placement terminates the docstring prematurely.
An analogous symmetric-delimiter boundary failure observed in Markdown thus reappears under Python triple-quote, indicating that scale or frontier-model status alone does not eliminate boundary-tracking failures; the per-model breakdown (Appendix~\ref{app:crossformat}) shows that exact failure modes and model ordering are format- and condition-dependent.
Inline code spans provide a related but noisier delimiter-handling probe: pooled run$=$1/2 failures are high, but Appendix~\ref{app:inline} shows the aggregate mixes interpretation defaults with genuine boundary-collision errors, so we treat it only as complementary evidence.

\subsection{38\% of Generations Are Latent Failures (RQ5)}
\label{sec:latent_analysis}

\begin{table}[t]
  \caption{Per-model content correctness and latent failure rates with Wilson 95\% CIs (valid $N$ on main A$\times$B grid, post-truncation). Content Correct = \cmetric{1}--\cmetric{3} all pass; Latent Failure = content-correct $\wedge\ \neg(\fmetric{2}\wedge\fmetric{3})$ (either fence balance or safe length fails). Each cell: \emph{rate} \texttt{[lower, upper]}.}
  \label{tab:uniquely_captured}
  \centering
  \footnotesize
  \setlength{\tabcolsep}{3pt}
  \begin{tabular}{cl|c|cc}
    \toprule
    & \textbf{Model} & \textbf{Valid $N$} & \textbf{Content Correct (\%)} & \textbf{Latent Failure (\%)} \\
    \midrule
    \multirow{3}{*}{\rotatebox{90}{\scriptsize Small}}
    & Qwen3-7B             & 2{,}141 & 69.3 [67.3, 71.2] & 48.1 [46.0, 50.2] \\
    & Gemma-2-9B           & 2{,}157 & 15.5 [14.0, 17.1] & \phantom{0}1.4 [\phantom{0}1.0, \phantom{0}2.0] \\
    & Llama-3.1-8B         & 2{,}146 & 48.2 [46.1, 50.3] & 14.5 [13.1, 16.1] \\
    \midrule
    \multirow{2}{*}{\rotatebox{90}{\scriptsize Mid}}
    & Qwen3-32B            & 2{,}152 & 71.6 [69.7, 73.5] & 46.1 [44.0, 48.2] \\
    & Gemma-3-27B          & 2{,}158 & 76.0 [74.2, 77.8] & 73.8 [71.9, 75.6] \\
    \midrule
    \multirow{4}{*}{\rotatebox{90}{\scriptsize Large/API}}
    & Llama-3.1-70B        & 2{,}159 & 69.1 [67.1, 71.0] & 35.7 [33.7, 37.8] \\
    & GPT-4o               & 2{,}160 & 80.0 [78.3, 81.7] & 43.3 [41.3, 45.4] \\
    & Gemini-2.5-Flash     & 1{,}895 & 76.1 [74.1, 78.0] & 49.9 [47.6, 52.1] \\
    & Claude-Sonnet-4      & 2{,}160 & 93.0 [91.9, 94.0] & 30.6 [28.7, 32.6] \\
    \midrule
    & \textbf{Overall} & 19{,}128 & 66.4 [65.7, 67.1] & 38.0 [37.3, 38.7] \\
    \bottomrule
  \end{tabular}
\end{table}

The four-cell decomposition reveals that $38.0\%$ of main-grid generations are latent failures~--- content-correct outputs with spec-level boundary violations invisible to prior format-adherence benchmarks (Table~\ref{tab:uniquely_captured})~--- and among content-correct generations $57.2\%$ fail boundary metrics ($38.0/66.4$), so any scoring that ignores fence-level boundary state misses this entire mass.
Per-model variation is large: Gemma-3-27B pairs the highest latent failure ($73.8\%$) with high content correctness ($76.0\%$), while Gemma-2-9B has only $1.4\%$ latent failure but the lowest content correctness ($15.5\%$), so its low latent rate reflects content failure rather than boundary success.
The per-language stratification (Appendix~\ref{app:latent_per_lang}) exposes a counterintuitive ordering~--- code-heavy languages (Shell $46.2\%$, C\# $45.1\%$, C++ $42.9\%$) carry the highest latent failure rate while Markdown carries the lowest ($9.2\%$), an artefact of Markdown's lowest content correctness ($21.6\%$) pre-empting latent classification rather than a genuine boundary success.
The Cochran--Armitage trend test shows a significant population-level monotonic trend in \etype{3} (latent collision) rate with inner nesting complexity (pooled $Z = +6.17$, $p < 10^{-3}$; Appendix~\ref{app:cochran_armitage}).
Together, these results establish latent failure as a prevalent ($38.0\%$), heterogeneously distributed (extremes $1.4\%$ to $73.8\%$), language-pervasive, and dose-responsive failure mode that cannot be detected by content-only or format-only scoring.

\paragraph{Robustness to content-metric choice.}
The headline Latent Failure rate is $38.0\%$ under the primary Content definition ($\cmetric{1}$--$\cmetric{3}$) versus $33.1\%$ under Full-C (adding $\cmetric{4}$ keyword-based Task Completion), with every model showing a lower but directionally consistent rate under Full-C; the finding does not invert under stricter content gating (Appendix~\ref{app:full_c_sensitivity}).

\section{Limitations and Future Work}
\label{sec:limitations}

LatentMD is a diagnostic benchmark for Markdown fence-boundary failures, not a universal ranking of LLM reliability.
Its boundary judgments are defined against CommonMark 0.31.2, but deployment impact depends on the downstream parser or renderer: some failures may be harmless in lenient chat interfaces while corrupting spec-strict tools such as agents, code extractors, or parsers.

The benchmark focuses primarily on Markdown fenced code blocks, with smaller cross-format probes for Python triple-quote docstrings and JSON; our natural-prompt validation set is small and serves as an ecological sanity check rather than a field prevalence estimate.
Future work should extend this boundary-evaluation framework to other structured-text formats (\LaTeX\ verbatim environments, YAML block scalars, HTML \texttt{<pre>} regions), expand the natural-prompt component with larger and more diverse user-derived prompts, and use LatentMD to track model progress and evaluate delimiter-aware generation or decoding-time mitigations.
Several of these directions are taken up in Appendix~\ref{app:additional}, which decomposes the headline rate by elicitation condition, measures downstream harm under real renderers and extractors, evaluates seven newer models, expands the natural-prompt set to $248$ organic requests, and quantifies the benchmark's effective diagnostic diversity.
\section{Conclusion}
\label{sec:conclusion}

We presented LatentMD, a benchmark and evaluation protocol for diagnosing Markdown boundary failures in LLM-generated outputs.
By separating content correctness from CommonMark-level boundary correctness, LatentMD reveals failures that can be missed by content-only or surface format evaluations.
Across $9$ LLMs, we find that boundary failures are widespread, driven primarily by same-family symmetric-delimiter collisions, and only partially mitigated by prompting; $38.0\%$ of main-grid generations are content-correct but boundary-broken.
An analogous failure appears in Python triple-quote docstrings while JSON remains robust as an asymmetric-delimiter control; scale or frontier-model status alone does not eliminate symmetric-delimiter boundary-tracking failures.

These results suggest that structured-text evaluation should measure parser-level boundary correctness, not only visible rendering, instruction compliance, or content quality.
LatentMD provides a reproducible protocol, prompt suite, diagnostic taxonomy, and scoring tool for evaluating whether LLM-generated Markdown is safe for downstream tools that depend on spec-compliant structure.

\bibliographystyle{plainnat}
\bibliography{references}

\begin{thebibliography}{27}
\providecommand{\natexlab}[1]{#1}
\providecommand{\url}[1]{\texttt{#1}}
\expandafter\ifx\csname urlstyle\endcsname\relax
  \providecommand{\doi}[1]{doi: #1}\else
  \providecommand{\doi}{doi: \begingroup \urlstyle{rm}\Url}\fi

\bibitem[{Anthropic}(2025)]{Anthropic2024Claude}
{Anthropic}.
\newblock Claude {Sonnet} 4.
\newblock \url{https://www.anthropic.com/news/claude-4}, 2025.
\newblock Accessed: 2026-04-01.

\bibitem[{Block}(2026)]{goose8290}
{Block}.
\newblock Desktop markdown rendering bug: fenced code blocks can break message
  layout.
\newblock \url{https://github.com/block/goose/issues/8290}, 2026.
\newblock Goose AI agent issue.

\bibitem[Chai et~al.(2025)Chai, Liu, Yang, Yin, Jin, Liu, Sun, Zhang, Ren, Guo,
  Wang, Wang, Wu, Wang, Li, Yang, Duan, Zhang, and Li]{Chai2024McEval}
Linzheng Chai, Shukai Liu, Jian Yang, Yuwei Yin, Ke~Jin, Jiaheng Liu, Tao Sun,
  Ge~Zhang, Changyu Ren, Hongcheng Guo, Noah Wang, Boyang Wang, Xianjie Wu,
  Bing Wang, Tongliang Li, Liqun Yang, Sufeng Duan, Zhaoxiang Zhang, and
  Zhoujun Li.
\newblock {McEval}: Massively multilingual code evaluation.
\newblock In \emph{The Thirteenth International Conference on Learning
  Representations (ICLR)}, 2025.
\newblock URL \url{https://openreview.net/forum?id=UunCPtPOlZ}.

\bibitem[Chen et~al.(2024)Chen, Jiao, Ravaut, Farruque, Nguyen, Qin, Dey, Ding,
  Xiong, Joty, and Zhou]{Chen2024StructTest}
Hailin Chen, Fangkai Jiao, Mathieu Ravaut, Nawshad Farruque, Xuan~Phi Nguyen,
  Chengwei Qin, Manan Dey, Bosheng Ding, Caiming Xiong, Shafiq Joty, and Yingbo
  Zhou.
\newblock {StructTest}: Benchmarking {LLM}s' reasoning through compositional
  structured outputs.
\newblock \emph{arXiv preprint arXiv:2412.18011}, 2024.

\bibitem[Chen et~al.(2025)Chen, Liu, Shi, Wang, Chen, Zhao, and
  Ren]{Chen2025MDEval}
Zhongpu Chen, Yinfeng Liu, Long Shi, Zhi-Jie Wang, Xingyan Chen, Yu~Zhao, and
  Fuji Ren.
\newblock {MDEval}: Evaluating and enhancing markdown awareness in large
  language models.
\newblock In \emph{Proceedings of the ACM on Web Conference 2025 (WWW '25)},
  pages 2981--2991. Association for Computing Machinery, 2025.
\newblock \doi{10.1145/3696410.3714674}.

\bibitem[{Continue}(2025)]{continue6059}
{Continue}.
\newblock When {LLM} generate markdown with code start with three backquote,
  usually it will be broken.
\newblock \url{https://github.com/continuedev/continue/issues/6059}, 2025.
\newblock Continue IDE extension issue.

\bibitem[D{\'e}l{\'e}tang et~al.(2023)D{\'e}l{\'e}tang, Ruoss, Grau-Moya,
  Genewein, Buesing, Catt, Catt, Mattern, Hutter, Legg, and
  Ortega]{Deletang2023Chomsky}
Gr{\'e}goire D{\'e}l{\'e}tang, Anian Ruoss, Jordi Grau-Moya, Tim Genewein, Lars
  Buesing, Elliot Catt, Marcus Catt, Tim Mattern, Marcus Hutter, Shane Legg,
  and Pedro~A. Ortega.
\newblock Neural networks and the {Chomsky} hierarchy.
\newblock In \emph{International Conference on Learning Representations
  (ICLR)}, 2023.

\bibitem[{Do Xuan Long} et~al.(2025){Do Xuan Long}, {Ngoc-Hai Nguyen},
  {Tiviatis Sim}, {Hieu Dao}, {Shafiq Joty}, {Kenji Kawaguchi}, {Nancy F.
  Chen}, and {Min-Yen Kan}]{Do2024FormatBiasEval}
{Do Xuan Long}, {Ngoc-Hai Nguyen}, {Tiviatis Sim}, {Hieu Dao}, {Shafiq Joty},
  {Kenji Kawaguchi}, {Nancy F. Chen}, and {Min-Yen Kan}.
\newblock {LLMs} are biased towards output formats! systematically evaluating
  and mitigating output format bias of {LLMs}.
\newblock In \emph{Proceedings of the 2025 Conference of the Nations of the
  Americas Chapter of the Association for Computational Linguistics: Human
  Language Technologies (NAACL-HLT)}, pages 299--330. Association for
  Computational Linguistics, 2025.
\newblock \doi{10.18653/v1/2025.naacl-long.15}.

\bibitem[{ExecutableBookProject}(2024)]{markdownitpy2024}
{ExecutableBookProject}.
\newblock \texttt{markdown-it-py}: {Markdown} parser in {Python}.
\newblock \url{https://github.com/executablebooks/markdown-it-py}, 2024.
\newblock Version 4.0.0 (released 2024-08-11); CommonMark 0.31.2 compliant.

\bibitem[{GitHub}(2019)]{GFMSpec2019}
{GitHub}.
\newblock {GitHub Flavored Markdown} spec.
\newblock \url{https://github.github.com/gfm/}, 2019.
\newblock Accessed: 2026-04-01.

\bibitem[{Google}(2025)]{geminicli10515}
{Google}.
\newblock {CLI} {Markdown} rendering is broken for triple-backtick code blocks.
\newblock \url{https://github.com/google-gemini/gemini-cli/issues/10515}, 2025.
\newblock Google Gemini CLI issue.

\bibitem[{Google DeepMind}(2025)]{Google2025GeminiFlash}
{Google DeepMind}.
\newblock {Gemini} 2.5 flash.
\newblock \url{https://deepmind.google/technologies/gemini/flash/}, 2025.
\newblock Accessed: 2026-04-01.

\bibitem[Jang et~al.(2023)Jang, Ye, and Seo]{Jang2023NegatedPrompts}
Joel Jang, Seonghyeon Ye, and Minjoon Seo.
\newblock Can large language models truly understand prompts? {A} case study
  with negated prompts.
\newblock In \emph{Proceedings of the 1st Transfer Learning for Natural
  Language Processing Workshop, PMLR 203}, 2023.

\bibitem[Jiang et~al.(2024)Jiang, Wang, Zeng, Zhong, Li, Mi, Shang, Jiang, Liu,
  and Wang]{Jiang2024FollowBench}
Yuxin Jiang, Yufei Wang, Xingshan Zeng, Wanjun Zhong, Liangyou Li, Fei Mi,
  Lifeng Shang, Xin Jiang, Qun Liu, and Wei Wang.
\newblock {FollowBench}: A multi-level fine-grained constraints following
  benchmark for large language models.
\newblock In \emph{Proceedings of the 62nd Annual Meeting of the Association
  for Computational Linguistics (ACL)}, pages 4667--4688, 2024.

\bibitem[MacFarlane et~al.(2024)]{CommonMark2024}
John MacFarlane et~al.
\newblock {CommonMark} spec version 0.31.2.
\newblock \url{https://spec.commonmark.org/0.31.2/}, 2024.
\newblock Accessed: 2026-04-01.

\bibitem[{Microsoft}(2026)]{vscode295126}
{Microsoft}.
\newblock create\_file tool wraps content in extra backtick fences when file
  contains markdown code blocks.
\newblock \url{https://github.com/microsoft/vscode/issues/295126}, 2026.
\newblock VS Code Copilot Chat issue.

\bibitem[{Open WebUI}(2024)]{openwebui5016}
{Open WebUI}.
\newblock Incorrect rendering of nested code blocks in {Markdown}.
\newblock \url{https://github.com/open-webui/open-webui/issues/5016}, 2024.
\newblock Open WebUI issue.

\bibitem[{OpenAI}(2024)]{OpenAI2024GPT4o}
{OpenAI}.
\newblock Hello {GPT-4o}.
\newblock \url{https://openai.com/index/hello-gpt-4o/}, 2024.
\newblock Accessed: 2026-04-01.

\bibitem[{OpenAI}(2025)]{chatkitjs89}
{OpenAI}.
\newblock Nested triple-backtick code blocks break when generating {Markdown}
  files in responses.
\newblock \url{https://github.com/openai/chatkit-js/issues/89}, 2025.
\newblock OpenAI official rendering library issue.

\bibitem[Tan et~al.(2025)Tan, Dou, Wang, Wang, Chen, and Wen]{Tan2024HtmlRAG}
Jiejun Tan, Zhicheng Dou, Wen Wang, Mang Wang, Weipeng Chen, and Ji-Rong Wen.
\newblock {HtmlRAG}: {HTML} is better than plain text for modeling retrieved
  knowledge in {RAG} systems.
\newblock In \emph{Proceedings of the ACM on Web Conference 2025 (WWW '25)},
  pages 1733--1746. Association for Computing Machinery, 2025.
\newblock \doi{10.1145/3696410.3714546}.

\bibitem[Wang et~al.(2025)Wang, Yu, Zhang, Jiang, Song, He, Liu, and
  Deb]{Wang2025STED}
Guanghui Wang, Jinze Yu, Xing Zhang, Dayuan Jiang, Yin Song, Peiyang He,
  Xuefeng Liu, and Tomal Deb.
\newblock {STED} and consistency scoring: A framework for evaluating {LLM}
  structured output reliability.
\newblock In \emph{NeurIPS 2025 Workshop on Structured Probabilistic Inference
  \& Generative Modeling (SPIGM)}, 2025.
\newblock URL \url{https://openreview.net/forum?id=rSCV1hTZvF}.

\bibitem[Wang et~al.(2026)Wang, Zhou, and Ding]{Wang2026FMBench}
Yaoting Wang, Yun Zhou, and Henghui Ding.
\newblock {FMBench}: Adaptive large language model output formatting.
\newblock \emph{arXiv preprint arXiv:2602.06384}, 2026.

\bibitem[Wilson(1927)]{Wilson1927}
Edwin~B. Wilson.
\newblock Probable inference, the law of succession, and statistical inference.
\newblock \emph{Journal of the American Statistical Association}, 22\penalty0
  (158):\penalty0 209--212, 1927.

\bibitem[Yang et~al.(2026)Yang, Jiang, He, Siu, Zhang, Liao, Li, Zeng, Jia,
  Wang, Schneider, Ruan, Ma, Lyu, Wang, Lu, Do, Jiang, Nie, and
  Chen]{Yang2025StructEval}
Jialin Yang, Dongfu Jiang, Tony He, Sherman Siu, Yuxuan Zhang, Disen Liao,
  Zhuofeng Li, Huaye Zeng, Yiming Jia, Haozhe Wang, Benjamin Schneider, Chi
  Ruan, Wentao Ma, Zhiheng Lyu, Yifei Wang, Yi~Lu, Quy~Duc Do, Ziyan Jiang,
  Ping Nie, and Wenhu Chen.
\newblock {StructEval}: Benchmarking {LLM}s' capabilities to generate
  structural outputs.
\newblock \emph{Transactions on Machine Learning Research}, 2026.
\newblock URL \url{https://openreview.net/forum?id=buDwV7LUA7}.

\bibitem[Zhao et~al.(2024)Zhao, Ren, Hessel, Cardie, Choi, and Deng]{Zhao2024}
Wenting Zhao, Xiang Ren, Jack Hessel, Claire Cardie, Yejin Choi, and Yuntian
  Deng.
\newblock {WildChat}: 1m {ChatGPT} interaction logs in the wild.
\newblock In \emph{International Conference on Learning Representations
  (ICLR)}, 2024.

\bibitem[Zheng et~al.(2024)Zheng, Chiang, Sheng, Li, Zhuang, Wu, Zhuang, Li,
  Lin, Xing, Gonzalez, Stoica, and Zhang]{Zheng2024}
Lianmin Zheng, Wei-Lin Chiang, Ying Sheng, Tianle Li, Siyuan Zhuang, Zhanghao
  Wu, Yonghao Zhuang, Zhuohan Li, Zi~Lin, Eric~P. Xing, Joseph~E. Gonzalez, Ion
  Stoica, and Hao Zhang.
\newblock {LMSYS-Chat-1M}: A large-scale real-world {LLM} conversation dataset.
\newblock In \emph{International Conference on Learning Representations
  (ICLR)}, 2024.

\bibitem[Zhou et~al.(2023)Zhou, Lu, Mishra, Brahma, Basu, Luan, Zhou, and
  Hou]{Zhou2023IFEval}
Jeffrey Zhou, Tianjian Lu, Swaroop Mishra, Siddhartha Brahma, Sujoy Basu,
  Yi~Luan, Denny Zhou, and Le~Hou.
\newblock Instruction-following evaluation for large language models.
\newblock \emph{arXiv preprint arXiv:2311.07911}, 2023.

\end{thebibliography}

\newpage
\appendix

\section{Metric and Algorithm Details}
\label{app:metric_details}

\subsection{Format Metrics: Full Definitions}
\label{app:format_metrics}

\paragraph{Notation, recap.}
$\Sigma^{*}$ is the response (Markdown text) space; $\mathcal{P}_{\text{main}} = \mathcal{A} \times \mathcal{B} \times \mathcal{L} \times \mathcal{T}$ is the main prompt space (\S\ref{sec:evaluation_metrics}), and a prompt $p = (a, b, \ell, t) \in \mathcal{P}_{\text{main}}$ is a tuple of an A-axis condition $a$, a B-axis condition $b$, a LANG $\ell$, and a TASK $t$. The four slot domains are: $\mathcal{A} = \{\text{A1}, \text{A2}, \text{A3}\}$ (A-axis levels, \S\ref{sec:axes}); $\mathcal{B} = \{\text{B1}, \text{B2}, \text{B3}, \text{B4}\}$ (B-axis levels, \S\ref{sec:axes}); $\mathcal{L}$ contains the 12 LANG values (nine programming languages plus three structured formats, \S\ref{sec:slot_generation}); and $\mathcal{T}$ contains the 15 TASK instances per LANG (180 in total, \S\ref{sec:slot_generation}). A model $M : \mathcal{P} \to \Sigma^{*}$ under greedy decoding maps each prompt to a single response.
We use the indicator $\mathbf{1}\{\cdot\} \in \{0, 1\}$, the boolean conjunction $\wedge$ extended to $\{0, 1, \text{NA}\}$ via $x \wedge \text{NA} := x$ (so $\text{NA}$ is the identity), and the negation $\neg$ defined on $\{0, 1\}$ only.

\paragraph{Format Metric Definitions.}
For $r \in \Sigma^{*}$, $p = (a, b, \ell, t) \in \mathcal{P}_{\text{main}}$:
\begin{flalign}
\fmetric{1}(r, p) &= \begin{cases}
  \mathbf{1}\{\neg \mathrm{HasOuterFence}(r)\} & \text{if } a = \text{A1}, \\
  \mathbf{1}\{\mathrm{HasOuterFence}(r)\} & \text{if } a = \text{A2}, \\
  \text{NA} & \text{if } a = \text{A3}.
\end{cases} \\
\fmetric{2}(r) &= \mathbf{1}\{\,\textsc{StackBalance}(r) = \emptyset\,\} \quad \text{(CommonMark Spec, Section 4.5)} \\
\fmetric{3}^{\text{post}}(r) &= \begin{cases}
  \mathbf{1}\{\,\mathrm{outerLen}(r) > \max\nolimits_{\text{inner}} \mathrm{run}(r)\,\} & \text{if } \mathrm{HasOuterFence}(r), \\
  \text{NA} & \text{otherwise}.
\end{cases} \\
\fmetric{3}^{\text{apr}}(r, p) &= \mathbf{1}\{\,\mathrm{outerLen}(r) > \mathrm{innerRun}(p)\,\} \quad \text{(D-inner-run only)}
\end{flalign}
Here $\textsc{StackBalance}$ is the parser pseudocoded in Appendix~\ref{app:algorithms}, returning the residual stack contents at end-of-text; $\mathrm{outerLen}(r)$ is the length of the outermost opening fence; $\mathrm{run}(r)$ scans inner content for the longest same-family backtick/tilde run.

\paragraph{Content Metric Definitions.}
\begin{flalign}
\cmetric{1}(r) &= \mathbf{1}\{\,|\{h \in r : h \text{ is an ATX or Setext heading}\}| \geq 2\,\} \\
\cmetric{2}(r, p) &= \mathbf{1}\{\,\exists \text{ code block } b \in r : \mathrm{infoStr}(b) \in \mathrm{alias}(\ell)\,\} \\
\cmetric{3}(r, p) &= \mathbf{1}\{\,n_{\mathrm{literal}}(r, p) \geq k(b\text{-condition of } p)\,\}
\end{flalign}
where $n_{\mathrm{literal}}(r, p)$ counts fenced blocks that serve as literal Markdown-source examples under the B-condition templates; $k(\cdot)$ encodes the prompt-specified literal-block count --- $k(\text{B1}) = k(\text{B4}) = 0$ (no inner block), $k(\text{B2}) = 1$ (single inner block), $k(\text{B3}) = 2$ (multiple inner blocks) --- and $\mathrm{alias}(\ell)$ is the LANG-alias map released with the artefact.
$\cmetric{2}$ fails when the response contains no code blocks (matching the released scoring CLI).

\paragraph{Cells, Indicator Functions, and Aggregate Estimators.}
The four-cell decomposition (\S\ref{sec:four_cell}) defines four indicators on $(r, p)$:
\begin{flalign}
\mathrm{both\_correct}(r, p) &:= \mathrm{Boundary}(r) \wedge \mathrm{Content}(r, p) \\
\mathrm{LF}(r, p) &:= \neg \mathrm{Boundary}(r) \wedge \mathrm{Content}(r, p) \quad \text{(latent failure)} \\
\mathrm{format\_only}(r, p) &:= \mathrm{Boundary}(r) \wedge \neg \mathrm{Content}(r, p) \\
\mathrm{both\_wrong}(r, p) &:= \neg \mathrm{Boundary}(r) \wedge \neg \mathrm{Content}(r, p)
\end{flalign}
For $\mathrm{Boundary} := \fmetric{2} \wedge \fmetric{3}^{\text{post}}$ and $\mathrm{Content} := \cmetric{1} \wedge \cmetric{2} \wedge \cmetric{3}$ with the $\text{NA}$-as-identity convention applied to $\fmetric{3}^{\text{post}}$ when no outer fence is present (the only NA case in the framework), the four indicators sum to $1$ on every valid response (mutual exclusion and exhaustiveness).
Aggregate rates are $\widehat{\mathrm{LFR}}_{M}(\mathcal{P}^{\prime}) = |\mathcal{V}|^{-1} \sum_{p \in \mathcal{V}} \mathrm{LF}(M(p), p)$, with $\mathcal{V} \subseteq \mathcal{P}^{\prime}$ the valid (non-truncated) subset; analogous estimators apply to the other three cells.

\paragraph{Deleted Metrics.}
The initial metric design included four additional format metrics that were removed during the final framework consolidation:
\emph{F3-old} (CommonMark Parse Success) was subsumed by the pipeline pre-filter;
\emph{F4-old} (AST Structural Match) was replaced by descriptive block-count statistics;
\emph{F5-old} (Embedded Block Preservation) was decomposed into \cmetric{2} (Code Language Match) and \cmetric{3} (Block Compliance);
\emph{F6-old} (Render Equivalence) was superseded by the latent failure definition (\S\ref{sec:latent_failure}), which uses the four-cell decomposition instead of a rendering-based proxy.

\paragraph{Two Compliance Layers: Prompt vs Specification.}
The F-metrics span two distinct compliance layers that are easy to conflate. \fmetric{1} (Instruction Compliance) is \emph{prompt-level}: it asks ``did the model follow the explicit A-axis directive (wrap / do-not-wrap)?'' and is judged against the prompt itself. \fmetric{2} (Fence Balance) and \fmetric{3} (Safe Length, post-hoc variant) are \emph{specification-level}: they ask ``does the produced text conform to CommonMark closing-fence rules?'' and are judged against the spec without reference to the prompt. \fmetric{3} additionally carries an \emph{a-priori} variant used in the D-inner-run ablation that is again prompt-level: it tests whether the model picked a safe outer length given an explicit inner-run constraint. Keeping these layers explicit clarifies what each metric isolates: \fmetric{1} measures intent-following, \fmetric{2}/\fmetric{3} measure spec-conformant production, and the post-hoc/a-priori variants of \fmetric{3} separate reactive from predictive safety.

\subsection{Auto-Detection Algorithms}
\label{app:algorithms}


\paragraph{Stack-Based Fence Balance Checker (\fmetric{2}).}
The pseudocode tracks opening fences via a LIFO stack and pops on each matching closing fence:

\begin{tcolorbox}[colback=vsbg, colframe=vsbg, boxsep=2pt, left=4pt, right=4pt, top=2pt, bottom=2pt, sharp corners]
\begin{Verbatim}[fontsize=\small, formatcom=\color{vstext}]
function check_fence_balance(text):
    stack = []
    for each line in text:
        if line matches OPENING_FENCE_PATTERN:
            extract (family, length, info_str)
            push (family, length) onto stack
        elif line matches CLOSING_FENCE_PATTERN:
            extract (family, length)
            # closing fence must carry no info string
            if stack is not empty:
                (ofam, olen) = stack.top()
                if family == ofam and length >= olen:
                    pop stack
    return stack is empty  # True = balanced
\end{Verbatim}
\end{tcolorbox}

\paragraph{Safe Length Calculator (\fmetric{3}).}
The pseudocode scans interior content for the longest run of the outer fence character and verifies the strict-inequality safety condition:

\begin{tcolorbox}[colback=vsbg, colframe=vsbg, boxsep=2pt, left=4pt, right=4pt, top=2pt, bottom=2pt, sharp corners]
\begin{Verbatim}[fontsize=\small, formatcom=\color{vstext}]
function check_safe_length(text, ofam, olen):
    max_run = 0
    for each line in text (excluding outer fences):
        for each run of ofam characters in line:
            max_run = max(max_run, run_length)
    return olen > max_run  # strict inequality
\end{Verbatim}
\end{tcolorbox}

Additional pseudo-code for \etype{0}--\etype{3} classification and the 4-cell decomposition algorithm is provided in the code release.

\subsection{Scope Justification}
\label{app:scope}

Our analysis focuses on \emph{paired delimiters that simultaneously serve as opening and closing boundaries}: fenced code blocks (backtick, tilde) and inline code spans (backtick).
These constructs exhibit the symmetric-delimiter problem: the same character sequence must be interpreted as wrapper, content, or literal depending on generation state.
Single-marker constructs (blockquotes \verb|>|, lists \verb|-|, headings \verb|#|) do not form paired boundaries and thus do not create delimiter-collision opportunities.
Inline emphasis (\verb|*|, \verb|**|) operates within a single line and does not expose the multi-line boundary-state tracking challenge we investigate.
YAML front matter is excluded because \verb|---| markers are not defined as paired delimiters in CommonMark 0.31.2 and their semantics vary across Markdown extensions.

\subsection{Coverage Matrix: Cell-Level Justification}
\label{app:coverage_justification}

\paragraph{StructEval ``partial'' for \etype{1}/\etype{2}.}
StructEval includes syntactic-validity and structural-element checks for structured outputs, so boundary failures that visibly alter parseability or required structural elements can be indirectly reflected in its scores.
However, StructEval does not inspect CommonMark fence-closing semantics directly, distinguish \etype{1} from \etype{2}, or compare outer and inner fence lengths.
Hence ``partial'': some consequential symptoms may be detectable, but cause-level fence-boundary diagnosis is not.

\paragraph{StructEval ``--'' for \etype{3}.}
StructEval does not perform fence length comparisons and does not include the opening/closing fence length relationship in its metric definitions.
\etype{3} (Latent Collision) is a fence-length-specific check and is fundamentally undetectable by StructEval.

\paragraph{StructEval ``partial'' for \fmetric{2}.}
StructEval's syntax-validity checks can indirectly penalize malformed structured outputs when fence imbalance affects parseability or required elements.
However, it does not implement a stack-based CommonMark fence-balance checker or diagnose the specific cause of imbalance.

\paragraph{IFEval ``partial'' for \fmetric{1}.}
IFEval evaluates general instruction-following with verifiable constraints, which conceptually overlaps with our \fmetric{1} wrapping-instruction compliance.
However, it does not target the specific wrapper-presence/absence manipulation used here, nor does it condition downstream CommonMark boundary metrics on compliance.

\paragraph{FormatBias ``concept'' for 4-cell.}
FormatBias argues that format and content should be evaluated separately, which conceptually aligns with our 4-cell decomposition.
However, FormatBias does not implement a boundary-correct/wrong axis and does not perform an actual 2$\times$2 decomposition.
It is a conceptual precursor, hence ``concept.''

\paragraph{Remaining ``--'' cells.}
For each remaining ``--'' cell, we reviewed the respective benchmark's metric definitions and confirmed that no metric is capable of detecting the corresponding failure type.

\paragraph{Benchmarks omitted from Table~\ref{tab:coverage_matrix}.}
MDEval~\citep{Chen2025MDEval} and FMBench~\citep{Wang2026FMBench} are excluded from the matrix columns because their metric definitions do not directly cover our fence-boundary failure types.
MDEval evaluates Markdown awareness, readability, and content-structure quality without inspecting CommonMark fence-closing semantics or outer/inner fence-length safety.
FMBench evaluates adaptive Markdown output formatting under diverse structural and layout constraints, including mixed content and code-block formatting, but does not perform per-fence stack-balance checks, fence-length comparisons, or same-family nesting diagnosis.
Including all-\texttt{--} columns would inflate the matrix without informational content, so we restrict it to benchmarks with at least one partial or positive coverage cell.

\subsection{Latent Failure Downstream Example}
\label{app:latent_failure_example}

\begin{table}[h]
  \caption{Concrete latent-failure example. Model \textbf{GPT-4o} on prompt \texttt{main\_A2\_B1\_python\_Python\_35} (A-cond=A2, B-cond=B1, task=Python/35). The response passes all content checks (\cmetric{1}$\wedge$\cmetric{2}$\wedge$\cmetric{3}) but the inner code-fence run length collides with the outer wrapper, so \fmetric{3} fails (\etype{3} latent collision). A naive regex extractor or visual rendering would flag this as ``correct''; only spec-level CommonMark parsing surfaces the boundary violation.}
  \label{tab:latent_failure_example}
  \centering
  \scriptsize
  \begin{tabular}{p{0.92\linewidth}}
    \toprule
    \textbf{Model output (excerpt)} \\
    \midrule
    \begin{minipage}{\linewidth}
    \vspace{2pt}
    \begin{tcolorbox}[colback=vsbg, colframe=vsbg, boxrule=0pt, arc=2pt, outer arc=2pt, left=2pt, right=2pt, top=2pt, bottom=2pt, width=\linewidth]
\begin{Verbatim}[fontsize=\scriptsize, formatcom=\color{white}, commandchars=\\\{\}, numbers=left, numbersep=4pt, xleftmargin=1.5em, frame=none, baselinestretch=0.85, breaklines=true, breakanywhere=true, breakindent=0pt, breaksymbolleft=, breaksymbolright=]
```markdown
\textcolor{vsheading}{# Calculating the Number of Apples Tao Tao Can Reach}

In this document, we will discuss how to calculate the number of apples that Tao Tao can reach. Tao Tao can reach a certain height with her hand stretched upwards, and she can also use a stool to extend her reach by an additional 30 centimeters. We will write a Python function to determine how many apples are within her reach.

\textcolor{vsheading}{## Problem Statement}

Given a list of apple heights and Tao Tao's maximum reach, we need to calculate how many apples she can reach. The function should consider both her natural reach and the additional height provided by a stool.

\textcolor{vsheading}{### Function Signature}

```python
\textcolor{vsheading}{def} \textcolor{vsfunc}{count_apples_reachable}\textcolor{vsbracket}{(}apple_heights: List\textcolor{vsbracket}{[}int\textcolor{vsbracket}{]}, tao_reach: int\textcolor{vsbracket}{)} -> int:
```

\textcolor{vsheading}{### Parameters}

[... abbreviated ...]
\end{Verbatim}
    \end{tcolorbox}
    \vspace{2pt}
    \end{minipage} \\
    \midrule
    \textbf{Diagnosis:} \cmetric{1}/\cmetric{2}/\cmetric{3} all pass (heading present, language-tag matches, code block of expected type present), so the response is judged \emph{Content-correct}. However, an inner code block uses a backtick run that matches the outer wrapper, breaking \fmetric{3} (safe-length) — the CommonMark spec terminates the outer fence at the first matching closing run, so the document parses with mis-nested blocks (\etype{3}). \\
    \bottomrule
  \end{tabular}
\end{table}

This example illustrates how a latent failure---a generation that is content-correct but boundary-wrong (\fmetric{2} or \fmetric{3} fail)---produces different outcomes depending on the downstream consumer.
A naive regex-based extractor may either truncate at the first apparent closing fence or, in balanced-but-unsafe cases, miss the latent length collision entirely.
LatentMD's $\fmetric{3}$ check identifies unsafe same-family fence-length collisions (\etype{3}), which can cause spec-strict downstream parsers or extractors to mis-nest or truncate the document.
This demonstrates why spec-level boundary checking is necessary even when visual rendering appears correct.

\paragraph{Anonymized real-world incident.}
Beyond engineered prompts, the same failure mode has been observed organically by end users on production chat platforms.
In one incident witnessed by the authors during a routine usage of a major chat assistant, the model produced a Markdown tutorial about Python that visually rendered as expected inside the chat UI; when the user copied the output into a CommonMark-strict Markdown editor for archival, all code blocks beyond the first silently merged with the surrounding prose, corrupting the document's structure.
The phenomenology aligns precisely with the latent failure cell quantified in our slot-based experiments (Table~\ref{tab:uniquely_captured}): visually normal output, structurally broken at the spec level, surfaced only when downstream tooling diverges from the lenient renderer used for the initial display.
We do not identify the chat platform here; the public bug-tracker references in \S\ref{sec:introduction} provide independently verifiable instances of the same pattern across multiple platforms.

\section{Benchmark Design}
\label{app:benchmark_design}

\subsection{Visual Overview of Axes}
\label{app:benchmark_axes_overview}

\begin{figure}[H]
  \centering
  \begin{tikzpicture}[
    font=\small,
    abox/.style={draw=blue!45!cyan, fill=blue!8, rounded corners=3pt,
                 minimum width=3.0cm, minimum height=0.95cm,
                 align=center, font=\small, inner sep=2pt, line width=0.4pt},
    bbox/.style={draw=orange!70, fill=orange!12, rounded corners=3pt,
                 minimum width=2.3cm, minimum height=0.95cm,
                 align=center, font=\small, inner sep=2pt, line width=0.4pt},
    dbox/.style={draw=teal!55!green, fill=green!13, rounded corners=3pt,
                 minimum width=2.3cm, minimum height=0.95cm,
                 align=center, font=\small, inner sep=2pt, line width=0.4pt},
    summary/.style={draw=brown!50, fill=yellow!12, rounded corners=4pt,
                    inner sep=7pt, align=center, font=\small, line width=0.5pt}
  ]

  \node[align=right, anchor=east] at (-0.05, 0)
    {\textbf{A-axis}\\[-1pt]{\scriptsize\itshape\color{black!55}Outer constraint}};
  \node[abox] (a1) at (2.10, 0) {\textbf{A1}\\Direct};
  \node[abox] (a2) at (5.40, 0) {\textbf{A2}\\Wrapped};
  \node[abox] (a3) at (8.70, 0) {\textbf{A3}\\Unspecified};

  \node[align=right, anchor=east] at (-0.05, -1.45)
    {\textbf{B-axis}\\[-1pt]{\scriptsize\itshape\color{black!55}Inner structure}};
  \node[bbox] (b1) at (1.55, -1.45) {\textbf{B1}\\Single};
  \node[bbox] (b2) at (4.10, -1.45) {\textbf{B2}\\Nested single};
  \node[bbox] (b3) at (6.65, -1.45) {\textbf{B3}\\Nested multi};
  \node[bbox] (b4) at (9.20, -1.45) {\textbf{B4}\\Mixed};

  \node[align=right, anchor=east] at (-0.05, -2.90)
    {\textbf{D-axis}\\[-1pt]{\scriptsize\itshape\color{black!55}Delimiter ablations}};
  \node[dbox] (d1) at (1.55, -2.90) {\textbf{Family}\\\scriptsize backtick / tilde};
  \node[dbox] (d2) at (4.10, -2.90) {\textbf{Length}\\\scriptsize outer 3 / 4 / 5};
  \node[dbox] (d3) at (6.65, -2.90) {\textbf{Inner run}\\\scriptsize 3 / 4 / 5};
  \node[dbox] (d4) at (9.20, -2.90) {\textbf{Cross-fam}\\\scriptsize mixed pair};

  \node[anchor=center, color=black!70] at (5.4, -4.10)
    {Slot-based generation: LANG (12) $\times$ TASK (15 per LANG, from McEval) $=$ 180 base pairs};
  \node[anchor=center, color=black!70] at (5.4, -4.65)
    {Main: A (3) $\times$ B (4) $\times$ 180 base pairs $=$ 2{,}160 prompts};

  \node[summary] (sum) at (5.4, -5.65) {%
    Main 2{,}160 $+$ D-axis 1{,}440 $+$ Hint 432 $+$ Inline 27 $+$ Cross-format 90 $+$ Natural 30 \\[2pt]
    \textbf{\color{brown!75!black}$=$ 4{,}179 unique prompts $\times$ 9 models $\approx$ 37{,}600 generations} ($T = 0$)
  };
  \end{tikzpicture}
  \caption{Visual overview of the three benchmark axes (cross-reference for \S\ref{sec:axes}). A controls outer constraint (3 levels), B controls inner nesting (4 levels), and D enumerates delimiter ablations. Total: 4,179 prompts (breakdown in \S\ref{sec:slot_generation}).}
  \label{fig:benchmark_axes}
\end{figure}

\subsection{Evaluation Pipeline}
\label{app:eval_pipeline}

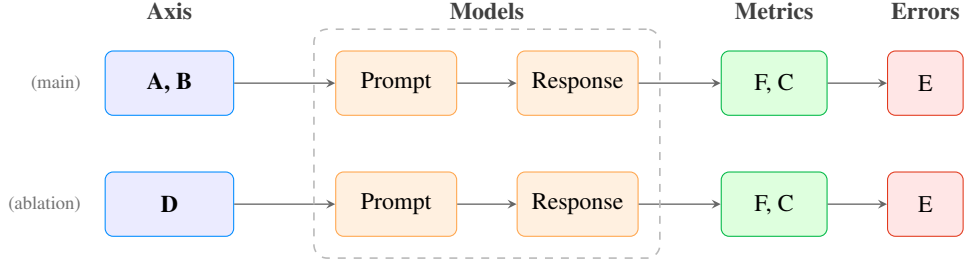
\begin{figure}[H]
  \centering
  \begin{tikzpicture}[
    font=\small,
    axisbox/.style={draw=blue!45!cyan, fill=blue!8, rounded corners=3pt,
                    minimum width=1.7cm, minimum height=0.85cm,
                    align=center, line width=0.4pt},
    prbox/.style={draw=orange!70, fill=orange!12, rounded corners=3pt,
                  minimum width=1.6cm, minimum height=0.85cm,
                  align=center, line width=0.4pt},
    metricbox/.style={draw=teal!55!green, fill=green!13, rounded corners=3pt,
                      minimum width=1.4cm, minimum height=0.85cm,
                      align=center, line width=0.4pt},
    errorbox/.style={draw=red!55!brown, fill=red!10, rounded corners=3pt,
                     minimum width=1.0cm, minimum height=0.85cm,
                     align=center, line width=0.4pt},
    models/.style={draw=gray!55, dashed, rounded corners=5pt,
                   line width=0.6pt, inner sep=8pt},
    arr/.style={->, >=stealth, line width=0.5pt, color=black!60},
  ]
    \node[font=\small\bfseries\color{black!75}] at (0, 1.95) {Axis};
    \node[font=\small\bfseries\color{black!75}] at (8, 1.95) {Metrics};
    \node[font=\small\bfseries\color{black!75}] at (10, 1.95) {Errors};

    \node[axisbox] (ab) at (0, 1.0) {\textbf{A, B}};
    \node[prbox] (p1) at (3, 1.0) {Prompt};
    \node[prbox] (r1) at (5.4, 1.0) {Response};
    \node[metricbox] (m1) at (8, 1.0) {F, C};
    \node[errorbox] (e1) at (10, 1.0) {E};
    \draw[arr] (ab) -- (p1);
    \draw[arr] (p1) -- (r1);
    \draw[arr] (r1) -- (m1);
    \draw[arr] (m1) -- (e1);

    \node[axisbox] (d) at (0, -0.6) {\textbf{D}};
    \node[prbox] (p2) at (3, -0.6) {Prompt};
    \node[prbox] (r2) at (5.4, -0.6) {Response};
    \node[metricbox] (m2) at (8, -0.6) {F, C};
    \node[errorbox] (e2) at (10, -0.6) {E};
    \draw[arr] (d) -- (p2);
    \draw[arr] (p2) -- (r2);
    \draw[arr] (r2) -- (m2);
    \draw[arr] (m2) -- (e2);

    \node[models, fit=(p1)(r1)(p2)(r2)] (modbox) {};
    \node[font=\small\bfseries\color{black!75}, anchor=south] at (modbox.north) {Models};

    \node[font=\scriptsize\color{black!55}, anchor=east] at (-1.05, 1.0) {(main)};
    \node[font=\scriptsize\color{black!55}, anchor=east] at (-1.05, -0.6) {(ablation)};
  \end{tikzpicture}
  \caption{Evaluation pipeline overview. Each row represents one experimental track. The \textbf{main} track conditions on A and B values, while the \textbf{ablation} track conditions on D values. In both tracks, the axis values are rendered into a Prompt; one of nine Models produces a Response; the Response is scored by Format ($\fmetric{1}$--$\fmetric{3}$) and Content ($\cmetric{1}$--$\cmetric{3}$) metrics; the metric outcomes determine one of four boundary-violation labels ($\etype{0}$--$\etype{3}$). The dashed wrapper marks the model-interaction stage, the only stochastic component when temperature $> 0$ (we use $T = 0$ throughout).}
  \label{fig:eval_pipeline}
\end{figure}

\subsection{LANG Selection: Programming Stratum Consensus}
\label{app:lang_selection}

The nine programming languages in the LANG pool are selected by a popularity consensus rule applied to three independent rankings published in 2024-2025: the TIOBE Programming Community Index (December 2025 snapshot), the GitHub Octoverse 2024 most-used languages list, and the Stack Overflow 2025 Developer Survey ``most popular technologies'' table.

A language is included in the programming stratum if and only if it appears in at least two of the three rankings within the top thirty entries. This two-of-three consensus reduces the influence of any single ranking's idiosyncrasies (e.g., TIOBE's search-engine bias, GitHub's repository-creation bias, or Stack Overflow's questionnaire selection bias) while still admitting any language that two independent sources agree is widely used.

After applying the consensus rule, we further restrict to languages that have at least one instruction-tuned task in the McEval~\citep{Chai2024McEval} benchmark, which is required by our slot-based template (see \S\ref{sec:slot_generation}). The final nine languages are: Python, JavaScript, TypeScript, Java, C++, C, C\#, Go, and Shell.

The structured-format stratum (three additional values --- Markdown, JSON, and HTML) is then drawn from the McEval non-programming task set, since these formats are well-defined by their respective specifications and do not need a popularity ranking.

\subsection{Prompt Templates: Per-Condition Examples}
\label{app:prompt_templates}
\label{app:per_condition_examples}

The full per-condition prompt set (9 YAML files, 4{,}179 prompts) is released as part of the LatentMD dataset; the examples below show one representative prompt per axis condition, with the two filled-in slots written as \texttt{<LANG>} and \texttt{<TASK>} (cf.\ Figure~\ref{fig:prompt_template_example}).

\paragraph{A2 $\times$ B2 Example.}
\mbox{}

\begin{promptbox}
\begin{Verbatim}[fontsize=\scriptsize, formatcom=\color{white}, baselinestretch=0.9]
Write a Markdown document that addresses the following task.
Include one <LANG> code example. In the document, also show
the raw Markdown source for that code example, including the
opening fence, the language tag, the code, and the closing fence.
Wrap your entire response in a fenced markdown code block.

Task:
<TASK>
\end{Verbatim}
\end{promptbox}

\needspace{8\baselineskip}
\paragraph{A3 $\times$ B2 Example (no wrap directive --- realistic anchor).}
\mbox{}

\begin{promptbox}
\begin{Verbatim}[fontsize=\scriptsize, formatcom=\color{white}, baselinestretch=0.9]
Write a Markdown document that addresses the following task.
Include one <LANG> code example. In the document, also show
the raw Markdown source for that code example, including the
opening fence, the language tag, the code, and the closing fence.

Task:
<TASK>
\end{Verbatim}
\end{promptbox}

\paragraph{A2 $\times$ B1 Example (simple, no nested fence).}
\mbox{}

\begin{promptbox}
\begin{Verbatim}[fontsize=\scriptsize, formatcom=\color{white}, baselinestretch=0.9]
Write a Markdown document that addresses the following task.
Include exactly one <LANG> code example.
Wrap your entire response in a fenced markdown code block.

Task:
<TASK>
\end{Verbatim}
\end{promptbox}

\paragraph{A2 $\times$ B3 Example (multiple nested fences).}
\mbox{}

\begin{promptbox}
\begin{Verbatim}[fontsize=\scriptsize, formatcom=\color{white}, baselinestretch=0.9]
Write a Markdown document that addresses the following task.
Include <LANG> code examples. In at least two places in the
document, show the raw Markdown source for the code examples,
including the opening fence, the language tag, the code, and
the closing fence.
Wrap your entire response in a fenced markdown code block.

Task:
<TASK>
\end{Verbatim}
\end{promptbox}

\paragraph{A2 $\times$ B4 Example (mixed elements).}
\mbox{}

\begin{promptbox}
\begin{Verbatim}[fontsize=\scriptsize, formatcom=\color{white}, baselinestretch=0.9]
Write a Markdown document that addresses the following task.
Include <LANG> code examples, a comparison table, a blockquote
with a citation, and a numbered list of steps.
Wrap your entire response in a fenced markdown code block.

Task:
<TASK>
\end{Verbatim}
\end{promptbox}

\paragraph{D-Family (Tilde) Example.}
\mbox{}

\begin{promptbox}
\begin{Verbatim}[fontsize=\scriptsize, formatcom=\color{white}, baselinestretch=0.9]
[A2 x B2 template as above] +
Use tilde fences (~~~) instead of backtick fences for all code
blocks in your response.
\end{Verbatim}
\end{promptbox}

\paragraph{D-Inner-Run Example.}
\mbox{}

\begin{promptbox}
\begin{Verbatim}[fontsize=\scriptsize, formatcom=\color{white}, baselinestretch=0.9]
[A2 x B4 template as above] +
When writing inner code blocks, use exactly 4 backticks for
the opening fence.
\end{Verbatim}
\end{promptbox}

\paragraph{D-Length Example.}
\mbox{}

\begin{promptbox}
\begin{Verbatim}[fontsize=\scriptsize, formatcom=\color{white}, baselinestretch=0.9]
[A2 x B2 template as above] +
When writing fenced code blocks, use exactly 5 backticks for
the outermost fence.
\end{Verbatim}
\end{promptbox}

\paragraph{D-Cross-Family Example.}
\mbox{}

\begin{promptbox}
\begin{Verbatim}[fontsize=\scriptsize, formatcom=\color{white}, baselinestretch=0.9]
[A2 x B2 template as above] +
Use backtick fences for your outer wrapper, and tilde fences
for any inner code block examples.
\end{Verbatim}
\end{promptbox}

\needspace{8\baselineskip}
\paragraph{Hint H1 (Boundary-Awareness Rule) Example.}
\mbox{}

\begin{promptbox}
\begin{Verbatim}[fontsize=\scriptsize, formatcom=\color{white}, baselinestretch=0.9]
[A2 x B2 template as above] +
Hint: Maintain correct nesting of all code fences. Outer fences
must use MORE fence characters than any same-type fence appearing
inside the content.
\end{Verbatim}
\end{promptbox}

\paragraph{Hint H2 (One-Shot) Example.}
\mbox{}

\begin{promptbox}
\begin{Verbatim}[fontsize=\scriptsize, formatcom=\color{white}, baselinestretch=0.9]
[A2 x B2 template as above] +
Hint: Here is an example of correct nesting:
````markdown
# Title
```python
print('hello')
```
````
Now write yours following this pattern.
\end{Verbatim}
\end{promptbox}

\paragraph{Hint H3 (Length-Specific) Example.}
\mbox{}

\begin{promptbox}
\begin{Verbatim}[fontsize=\scriptsize, formatcom=\color{white}, baselinestretch=0.9]
[A2 x B2 template as above] +
Hint: Use at least 4 backticks for your outermost fence.
\end{Verbatim}
\end{promptbox}

\paragraph{Cross-Format Python K1 Example.}
\mbox{}

\begin{promptbox}
\begin{Verbatim}[fontsize=\scriptsize, formatcom=\color{white}, baselinestretch=0.9]
Write a Python function that addresses the following task. The
function must have a docstring that mentions the triple-quote
delimiter """ as part of the documentation text.

Task:
<TASK>
\end{Verbatim}
\end{promptbox}

\paragraph{Cross-Format JSON K1 Example (asymmetric-delimiter control).}
\mbox{}

\begin{promptbox}
\begin{Verbatim}[fontsize=\scriptsize, formatcom=\color{white}, baselinestretch=0.9]
Write a JSON document that addresses the following task. Use 1
to 2 levels of nested objects or arrays.

Task:
<TASK>
\end{Verbatim}
\end{promptbox}

\paragraph{Inline I1, run$=$1 Example.}
\mbox{}

\begin{promptbox}
\begin{Verbatim}[fontsize=\scriptsize, formatcom=\color{white}, baselinestretch=0.9]
Write one sentence in Markdown that contains a single inline
code span whose displayed content is the text: `foo`.
\end{Verbatim}
\end{promptbox}

\section{Experimental Setup}
\label{app:experimental_setup}

\subsection{Hyperparameter Table}
\label{app:hyperparameters}

\begin{table}[h]
  \caption{Full hyperparameter configuration for all experiments.}
  \label{tab:hyperparameters}
  \centering
  \small
  \begin{tabular}{lll}
    \toprule
    \textbf{Parameter} & \textbf{Value} & \textbf{Notes} \\
    \midrule
    Temperature     & 0 & Greedy; sweep in App.~\ref{app:temperature} \\
    top\_p          & 0.9 & Cross-API compatible \\
    top\_k          & 40  & Gemini/Anthropic; ignored by OpenAI \\
    max\_tokens     & 7168 & Fits Gemma-2 8K context \\
    seed            & 42 & Open-weight only \\
    Serving (open)  & vLLM / HF Transformers & See Table~\ref{tab:model_list} \\
    Serving (closed) & Official APIs & OpenAI, Google, Anthropic \\
    \bottomrule
  \end{tabular}
\end{table}

Open-weight models were served on NVIDIA A100 40GB, RTX 6000 Ada 48GB, H100 SXM 80GB, and RTX 3090 24GB GPUs, via vLLM (HuggingFace Transformers for A100 due to CUDA version constraints).
Closed-source models were queried from a desktop workstation via official APIs.
Under temperature-0 greedy decoding, open-weight outputs are deterministic up to floating-point precision and therefore independent of the specific GPU used.

\paragraph{Approximate runtime.}
Representative wall-clock measurements include $\sim$8 hours for Llama-3.1-8B on a single A100 and $\sim$52 hours for Llama-3.1-70B on 5 A100 GPUs (full main grid via HuggingFace Transformers).
Open-weight ablation generations ran on a 2$\times$H100 SXM cloud node within $\sim$32 wall-clock hours using vLLM.
Together with the remaining open-weight runs, aggregate open-weight compute was on the order of several hundred GPU-hours, dominated by Llama-3.1-70B main-grid generation.
Closed-source API generations were queried in parallel from a single workstation, with wall-clock time governed by provider latency and rate limits; aggregate API spend is itemized in Appendix~\ref{app:compute_cost}.

\paragraph{Reasoning-mode disable flags.}
Three models in our suite ship with optional ``thinking'' / extended-reasoning modes: Qwen3 (both 7B and 32B), Gemini-2.5-Flash, and Claude-Sonnet-4.
We disable all of them so that every response is a single Markdown answer rather than a reasoning trace followed by an answer.
Per-model flags: Qwen3 via \texttt{enable\_thinking=False}, Gemini-2.5-Flash via \texttt{thinkingBudget=0}, and Claude-Sonnet-4 via its default-off extended-thinking setting (no \texttt{thinking} field in the request).
The remaining six models (Gemma-2-9B, Gemma-3-27B, Llama-3.1-8B/70B, GPT-4o) do not expose a reasoning-mode toggle and were used as-is.

\paragraph{Closed-source group classification.}
GPT-4o, Claude-Sonnet-4, and Gemini-2.5-Flash do not publish parameter counts, so they cannot be placed in a parameter-defined size tier in the strict sense.
We group them with the largest open-weight model (Llama-3.1-70B) under a unified ``Large/API'' label in Table~\ref{tab:model_list}, treating ``Tier'' as an operational rather than a parameter-strict grouping.
The Small and Mid tiers are defined by parameter count among open-weight models only.
This grouping affects descriptive aggregations (e.g., the tier averages in Table~\ref{tab:crossformat_per_tier}) but does not enter any model-level statistical claim, all of which are reported per individual model.

\subsection{Closed-source API Usage and Cost}
\label{app:compute_cost}

Closed-source API calls were issued from a workstation, with approximate usage costs of $\$67$ (Anthropic Claude-Sonnet-4), $\$25$ (OpenAI GPT-4o), and $\$30$ (Google Gemini-2.5-Flash plus a small Flash-Lite probe), totalling $\approx\$122$ across the three closed APIs for the full benchmark ($4{,}179$ prompts $\times$ $3$ closed models, covering both main grid and ablations).
These figures are based on provider billing records at the time of the experiments and are intended only as a reproducibility reference; actual costs may vary with provider pricing, rate limits, batching, and response length.

\begin{table}[h]
  \caption{Closed-API inference cost for the full benchmark (approximate, in USD).}
  \label{tab:compute_cost}
  \centering
  \small
  \begin{tabular}{llp{6.0cm}}
    \toprule
    \textbf{API} & \textbf{Cost} & \textbf{Coverage} \\
    \midrule
    Anthropic Claude-Sonnet-4      & $\approx \$67$  & Main + Ablation \\
    OpenAI GPT-4o                  & $\approx \$25$  & Main + Ablation \\
    Google Gemini-2.5-Flash/Lite   & $\approx \$30$  & Main + Ablation (Flash + Lite probe) \\
    \midrule
    \textbf{Total}                 & $\approx \mathbf{\$122}$ & 3 closed APIs, full benchmark \\
    \bottomrule
  \end{tabular}
\end{table}

\section{Truncation Filtering and Natural-Prompt Validation}
\label{app:filter_validation}

\subsection{Truncation Statistics}
\label{app:truncation}

\begin{table}[h]
  \caption{Truncation statistics per model (main + ablation experiments combined). Generations with \texttt{finish\_reason = ``length''} are excluded from all metric computations. Valid $N$ is used as the denominator for Wilson confidence intervals.}
  \label{tab:truncation_report}
  \centering
  \scriptsize
  \begin{tabular}{lccc}
    \toprule
    \textbf{Model} & \textbf{Total $N$} & \textbf{Truncated} & \textbf{Valid $N$} \\
    \midrule
    Qwen3-7B           & 4179 & 30 & 4149 \\
    Gemma-2-9B         & 4179 & 6 & 4173 \\
    Llama-3.1-8B       & 4179 & 34 & 4145 \\
    Qwen3-32B          & 4179 & 22 & 4157 \\
    Gemma-3-27B        & 4179 & 8 & 4171 \\
    Llama-3.1-70B      & 4179 & 4 & 4175 \\
    GPT-4o             & 4179 & 0 & 4179 \\
    Gemini-2.5-Flash   & 4179 & 503 & 3676 \\
    Claude-Sonnet-4    & 4179 & 1 & 4178 \\
    \midrule
    \textbf{Total}   & 37611 & 608 & 37003 \\
    \bottomrule
  \end{tabular}
\end{table}

Responses with \texttt{finish\_reason=``length''} are excluded from all boundary metrics and reduce the effective sample size (valid~$N$) for the affected model.
Under identical conditions (\texttt{max\_tokens=7168}, temperature$=0$), Gemini-2.5-Flash exhibits a truncation rate of ${\sim}$12--13\%, while GPT-4o and Claude-Sonnet-4 show 0\%.
This discrepancy arises because Gemini generates 2--3$\times$ longer outputs for the same prompts and occasionally enters degenerate repetition loops (e.g., repeating a single character until the token limit).
This pattern is consistent with the observed repetition loops in our generations.
No post-processing (e.g., trailing whitespace removal) is applied to any model's output; all responses are evaluated as-is.

\paragraph{Excluded model (Gemini-2.5-Flash-Lite).}
We additionally ran Gemini-2.5-Flash-Lite on the full $4{,}179$-prompt suite as a within-Google-family probe. Its responses are even more verbose than Flash's (avg ${\sim}18$K characters per response vs Flash's ${\sim}14$K), yielding $803/4{,}179 = 19.2\%$ truncation --- markedly above Flash's $12.0\%$, producing an effective sample size ($3{,}376$) that diverges from the 9 paper models ($\geq 3{,}676$) widely enough to substantially weaken per-cell comparability across stratified analyses. We therefore exclude Flash-Lite from the main analysis cohort; per-record results are released alongside the 9-model data for downstream verification.

\subsection{Per-Language Truncation Distribution}
\label{app:truncation_per_lang}

The aggregate truncation report above pools across languages.
Stratified by language (Table~\ref{tab:truncation_per_lang}), Gemini-2.5-Flash's $12.3\%$ overall truncation rate concentrates in Go ($26.1\%$), JavaScript ($22.2\%$), TypeScript ($22.2\%$), and C++ ($17.2\%$) and is much lower in Markdown ($3.9\%$) and Shell ($2.2\%$).
Other models show $<1\%$ truncation across all languages.
The pattern is consistent with the verbose-output hypothesis (Gemini generates 2--3$\times$ longer responses for verbose-syntax languages and exhausts the $7168$-token budget).

\begin{table}[h]
  \caption{Truncation rate (\%) per model and language. Truncation = \texttt{finish\_reason} == \texttt{length} (max\_tokens budget exhausted). Each cell is computed on $N{\approx}45$ prompts (3 A $\times$ 4 B $\times$ $\sim$3.75 TASK per LANG, before truncation filtering). Bold marks model rows with $>$5\% overall truncation.}
  \label{tab:truncation_per_lang}
  \centering
  \scriptsize
  \setlength{\tabcolsep}{2pt}
  \begin{tabular}{cl|cccccccccccc|c}
    \toprule
    & \textbf{Model} & \rotatebox{60}{\scriptsize Python} & \rotatebox{60}{\scriptsize JavaScript} & \rotatebox{60}{\scriptsize Java} & \rotatebox{60}{\scriptsize C++} & \rotatebox{60}{\scriptsize C\#} & \rotatebox{60}{\scriptsize TypeScript} & \rotatebox{60}{\scriptsize Shell} & \rotatebox{60}{\scriptsize C} & \rotatebox{60}{\scriptsize Go} & \rotatebox{60}{\scriptsize HTML} & \rotatebox{60}{\scriptsize JSON} & \rotatebox{60}{\scriptsize Markdown} & \textbf{Total} \\
    \midrule
    \multirow{3}{*}{\rotatebox{90}{\scriptsize Small}}
    & Qwen3-7B                 & 2.8 & 0.0 & 0.0 & 2.2 & 0.6 & 0.6 & 0.6 & 1.7 & 1.7 & 0.6 & 0.0 & 0.0 & 0.9 \\
    & Gemma-2-9B               & 0.0 & 0.0 & 0.0 & 0.0 & 0.0 & 0.0 & 1.1 & 0.0 & 0.0 & 0.0 & 0.6 & 0.0 & 0.1 \\
    & Llama-3.1-8B             & 3.9 & 0.0 & 0.0 & 0.0 & 0.0 & 0.0 & 0.0 & 0.0 & 0.0 & 1.1 & 0.0 & 2.8 & 0.6 \\
    \midrule
    \multirow{2}{*}{\rotatebox{90}{\scriptsize Mid}}
    & Qwen3-32B                & 1.7 & 0.0 & 0.0 & 1.1 & 0.0 & 0.0 & 1.1 & 0.0 & 0.6 & 0.0 & 0.0 & 0.0 & 0.4 \\
    & Gemma-3-27B              & 0.0 & 0.0 & 0.0 & 0.0 & 0.0 & 0.0 & 0.0 & 0.0 & 0.6 & 0.0 & 0.0 & 0.6 & 0.1 \\
    \midrule
    \multirow{4}{*}{\rotatebox{90}{\scriptsize Large}}
    & Llama-3.1-70B            & 0.0 & 0.0 & 0.0 & 0.0 & 0.0 & 0.0 & 0.0 & 0.0 & 0.0 & 0.6 & 0.0 & 0.0 & 0.0 \\
    & GPT-4o                   & 0.0 & 0.0 & 0.0 & 0.0 & 0.0 & 0.0 & 0.0 & 0.0 & 0.0 & 0.0 & 0.0 & 0.0 & 0.0 \\
    & \textbf{Gemini-2.5-Flash} & 7.8 & 22.2 & 5.0 & 17.2 & 6.1 & 22.2 & 2.2 & 13.9 & 26.1 & 7.2 & 13.3 & 3.9 & \textbf{12.3} \\
    & Claude-Sonnet-4          & 0.0 & 0.0 & 0.0 & 0.0 & 0.0 & 0.0 & 0.0 & 0.0 & 0.0 & 0.0 & 0.0 & 0.0 & 0.0 \\
    \midrule
    & \textbf{Overall} & 1.8 & 2.5 & 0.6 & 2.3 & 0.7 & 2.5 & 0.6 & 1.7 & 3.2 & 1.0 & 1.5 & 0.8 & 1.6 \\
    \bottomrule
  \end{tabular}
\end{table}

\subsection{Natural Prompt Validation}
\label{app:natural}

\begin{table}[h]
  \caption{Natural prompt validation results. 30 human-authored prompts $\times$ 9 models across three elicitation levels. Fail Rate = boundary-incorrect responses / total responses observed for that prompt (truncated responses excluded).}
  \label{tab:natural_prompts}
  \centering
  \scriptsize
  \begin{tabular}{clcc}
    \toprule
    \textbf{\#} & \textbf{Prompt (abbreviated)} & \textbf{Level} & \textbf{Fail Rate} \\
    \midrule
    \multicolumn{4}{l}{\emph{High elicitation}} \\
    1  & Tutorial on fenced code blocks                 & High   & 11.1\% \\
    2  & Display raw Markdown source in a document      & High   & 55.6\% \\
    3  & Markdown syntax cheat sheet (raw + rendered)   & High   & 28.6\% \\
    4  & Tutorial on nesting code blocks                & High   & 22.2\% \\
    5  & Static site generator docs: fenced examples    & High   & 22.2\% \\
    6  & Reference card: literal triple backticks       & High   & 33.3\% \\
    7  & Blog post on Markdown pitfalls                 & High   & 44.4\% \\
    8  & Markdown linter docs: correct/incorrect        & High   & 37.5\% \\
    9  & GitHub README: raw Markdown source             & High   & 22.2\% \\
    10 & CommonMark spec: fence lengths and languages   & High   & 55.6\% \\
    \multicolumn{4}{l}{\emph{Medium elicitation}} \\
    11 & Python README for data pipeline                & Medium & 0.0\% \\
    12 & JavaScript async/await guide                   & Medium & 0.0\% \\
    13 & REST API docs with curl examples               & Medium & 11.1\% \\
    14 & README.md for Rust CLI tool                    & Medium & 0.0\% \\
    15 & Overview of sorting algorithms (C++)           & Medium & 0.0\% \\
    16 & API reference for Go package                   & Medium & 11.1\% \\
    17 & TypeScript React getting-started guide         & Medium & 0.0\% \\
    18 & Python list comprehensions vs for loops        & Medium & 33.3\% \\
    19 & SQL joins quick reference                      & Medium & 0.0\% \\
    20 & JS to TypeScript migration guide               & Medium & 0.0\% \\
    \multicolumn{4}{l}{\emph{Low elicitation}} \\
    21 & Python virtual environment setup               & Low    & 0.0\% \\
    22 & Debugging memory leaks in Node.js              & Low    & 0.0\% \\
    23 & JWT authentication implementation              & Low    & 0.0\% \\
    24 & Git rebase vs merge                            & Low    & 0.0\% \\
    25 & PostgreSQL database setup                      & Low    & 0.0\% \\
    26 & Docker environment variables                   & Low    & 0.0\% \\
    27 & Optimizing slow SQL queries                    & Low    & 0.0\% \\
    28 & Webhooks: how they work and setup              & Low    & 0.0\% \\
    29 & Nginx reverse proxy configuration              & Low    & 0.0\% \\
    30 & Automated testing in CI/CD pipeline            & Low    & 0.0\% \\
    \bottomrule
  \end{tabular}
\end{table}

Thirty human-authored prompts---phrased as a developer might naturally write them---serve as an external validity check.
Prompts span three elicitation levels: 10 \emph{high} (explicitly request nested code blocks or raw Markdown syntax), 10 \emph{medium} (request code examples without mentioning nesting), and 10 \emph{low} (technical topics where code blocks are likely but not explicitly requested).
With $N=30$, we report descriptive failure-rate ranges and treat this set as an ecological sanity check rather than a population-level prevalence estimate.


\section{Main Results: Full Tables}
\label{app:main_results}

\subsection{Full A\texorpdfstring{$\times$}{x}B\texorpdfstring{$\times$}{x}Model Results}
\label{app:full_results}

\begin{table}[H]
  \caption{Boundary failure rate (\%) per model across A$\times$B conditions (= \etype{1}$\cup$\etype{2}$\cup$\etype{3}, i.e.\ \fmetric{2} or \fmetric{3} fail). Columns are ordered by experimental severity rather than condition ID: A1 (no-wrap baseline) $\rightarrow$ A3 (unspecified, realistic anchor) $\rightarrow$ A2 (forced-wrap stress test). Per-cell mechanism breakdown (\etype{1}/\etype{2}/\etype{3}) is in Appendix~\ref{app:a_b_e_per_model}.}
  \label{tab:main_results}
  \centering
  \scriptsize
  \setlength{\tabcolsep}{1.5pt}
  \begin{tabular}{cl|cccc|cccc|cccc}
    \toprule
    & & \multicolumn{4}{c|}{\textbf{A1 (Direct)}} & \multicolumn{4}{c|}{\textbf{A3 (Unspecified)}} & \multicolumn{4}{c}{\textbf{A2 (Wrapped)}} \\
    & \textbf{Model} & B1 & B2 & B3 & B4 & B1 & B2 & B3 & B4 & B1 & B2 & B3 & B4 \\
    \midrule
    \multirow{3}{*}{\rotatebox{90}{\scriptsize Small}}
    & Qwen3-7B             & 5.6 & 23.5 & 44.1 & 0.0 & 91.0 & 98.3 & 92.7 & 77.6 & 90.5 & 97.8 & 98.3 & 92.2 \\
    & Gemma-2-9B           & 4.4 & 7.8 & 5.6 & 0.0 & 1.7 & 31.7 & 18.3 & 0.0 & 60.6 & 97.8 & 93.3 & 87.7 \\
    & Llama-3.1-8B         & 0.0 & 7.2 & 2.8 & 0.6 & 0.6 & 52.2 & 26.8 & 1.7 & 93.3 & 91.1 & 96.1 & 86.4 \\
    \midrule
    \multirow{2}{*}{\rotatebox{90}{\scriptsize Mid}}
    & Qwen3-32B            & 52.2 & 18.9 & 18.9 & 75.8 & 71.7 & 41.1 & 37.4 & 46.9 & 96.1 & 97.8 & 99.4 & 98.9 \\
    & Gemma-3-27B          & 88.8 & 75.0 & 84.4 & 93.3 & 95.0 & 93.9 & 96.1 & 92.8 & 95.0 & 100.0 & 100.0 & 96.1 \\
    \midrule
    \multirow{4}{*}{\rotatebox{90}{\scriptsize Large}}
    & Llama-3.1-70B        & 0.0 & 83.9 & 78.9 & 0.0 & 0.0 & 80.6 & 76.7 & 0.0 & 93.3 & 100.0 & 100.0 & 95.0 \\
    & GPT-4o               & 16.7 & 1.7 & 1.1 & 11.1 & 83.3 & 2.2 & 10.6 & 93.3 & 93.3 & 100.0 & 98.9 & 97.8 \\
    & Gemini-2.5-Flash     & 15.3 & 22.5 & 20.0 & 2.9 & 75.5 & 98.8 & 96.4 & 51.5 & 93.2 & 100.0 & 100.0 & 99.3 \\
    & Claude-Sonnet-4      & 0.0 & 0.0 & 0.6 & 0.0 & 0.0 & 0.0 & 0.0 & 0.0 & 99.4 & 100.0 & 100.0 & 100.0 \\
    \bottomrule
  \end{tabular}

  \vspace{0.15em}
  {\footnotesize B1=single, B2=nested single, B3=nested multiple, B4=mixed. A3 (Unspecified) reflects realistic prompts: models self-wrap $\sim$37\% of responses (Step~1 of Table~\ref{tab:a1_decomposition}); their boundary fail rate sits between A1 (low) and A2 (stress-test ceiling).}
\end{table}

Table~\ref{tab:main_results} reports the per-model boundary failure rate (\etype{1}$\cup$\etype{2}$\cup$\etype{3}) for every A$\times$B cell; the body \S\ref{sec:main_results} cites the pooled view (Figure~\ref{fig:main_heatmap}) and the per-condition severity ordering. Complete per-metric breakdowns (\fmetric{1}--\fmetric{3}, \cmetric{1}--\cmetric{4}, \etype{0}--\etype{3}) for every cell and model are also provided in the supplementary materials as structured JSONL files.

\subsection{Main Results with Wilson 95\% Confidence Intervals}
\label{app:baseline_with_ci}

\begin{table}[h]
  \caption{Boundary failure rate (\%) per model under \textbf{A1 (Direct)} prompts, with Wilson 95\% confidence intervals over valid $N$ (after truncation filtering). Each cell shows \emph{rate} \texttt{[CI lower, CI upper]}. Wilson's small-sample correction yields tight intervals near the [0, 100] extremes.}
  \label{tab:baseline_with_ci}
  \centering
  \scriptsize
  \setlength{\tabcolsep}{3pt}
  \begin{tabular}{cl|cccc}
    \toprule
    & \textbf{Model} & B1 & B2 & B3 & B4 \\
    \midrule
    \multirow{3}{*}{\rotatebox{90}{\scriptsize Small}}
    & Qwen3-7B           & 5.6 [3.1,10.0] & 23.5 [17.8,30.2] & 44.1 [37.0,51.4] & 0.0 [0.0,2.1] \\
    & Gemma-2-9B         & 4.4 [2.3,8.5] & 7.8 [4.7,12.6] & 5.6 [3.0,9.9] & 0.0 [0.0,2.1] \\
    & Llama-3.1-8B       & 0.0 [0.0,2.1] & 7.2 [4.3,12.0] & 2.8 [1.2,6.4] & 0.6 [0.1,3.1] \\
    \midrule
    \multirow{2}{*}{\rotatebox{90}{\scriptsize Mid}}
    & Qwen3-32B          & 52.2 [45.0,59.4] & 18.9 [13.8,25.2] & 18.9 [13.8,25.2] & 75.8 [69.0,81.5] \\
    & Gemma-3-27B        & 88.8 [83.4,92.7] & 75.0 [68.2,80.8] & 84.4 [78.4,89.0] & 93.3 [88.7,96.2] \\
    \midrule
    \multirow{4}{*}{\rotatebox{90}{\scriptsize Large}}
    & Llama-3.1-70B      & 0.0 [0.0,2.1] & 83.9 [77.8,88.5] & 78.9 [72.4,84.2] & 0.0 [0.0,2.1] \\
    & GPT-4o             & 16.7 [11.9,22.8] & 1.7 [0.6,4.8] & 1.1 [0.3,4.0] & 11.1 [7.3,16.5] \\
    & Gemini-2.5-Flash   & 15.3 [10.6,21.7] & 22.5 [16.8,29.3] & 20.0 [14.6,26.7] & 2.9 [1.1,7.3] \\
    & Claude-Sonnet-4    & 0.0 [0.0,2.1] & 0.0 [0.0,2.1] & 0.6 [0.1,3.1] & 0.0 [0.0,2.1] \\
    \bottomrule
  \end{tabular}
\end{table}

\begin{table}[h]
  \caption{Boundary failure rate (\%) per model under \textbf{A2 (Wrapped)} prompts, with Wilson 95\% confidence intervals over valid $N$ (after truncation filtering). Each cell shows \emph{rate} \texttt{[CI lower, CI upper]}.}
  \label{tab:baseline_with_ci_a2}
  \centering
  \scriptsize
  \setlength{\tabcolsep}{3pt}
  \begin{tabular}{cl|cccc}
    \toprule
    & \textbf{Model} & B1 & B2 & B3 & B4 \\
    \midrule
    \multirow{3}{*}{\rotatebox{90}{\scriptsize Small}}
    & Qwen3-7B           & 90.5 [85.3,94.0] & 97.8 [94.4,99.1] & 98.3 [95.1,99.4] & 92.2 [87.3,95.3] \\
    & Gemma-2-9B         & 60.6 [53.3,67.4] & 97.8 [94.4,99.1] & 93.3 [88.7,96.2] & 87.7 [82.1,91.7] \\
    & Llama-3.1-8B       & 93.3 [88.6,96.1] & 91.1 [86.1,94.5] & 96.1 [92.1,98.1] & 86.4 [80.6,90.7] \\
    \midrule
    \multirow{2}{*}{\rotatebox{90}{\scriptsize Mid}}
    & Qwen3-32B          & 96.1 [92.2,98.1] & 97.8 [94.4,99.1] & 99.4 [96.9,99.9] & 98.9 [96.0,99.7] \\
    & Gemma-3-27B        & 95.0 [90.8,97.4] & 100.0 [97.9,100.0] & 100.0 [97.9,100.0] & 96.1 [92.2,98.1] \\
    \midrule
    \multirow{4}{*}{\rotatebox{90}{\scriptsize Large}}
    & Llama-3.1-70B      & 93.3 [88.7,96.2] & 100.0 [97.9,100.0] & 100.0 [97.9,100.0] & 95.0 [90.8,97.4] \\
    & GPT-4o             & 93.3 [88.7,96.2] & 100.0 [97.9,100.0] & 98.9 [96.0,99.7] & 97.8 [94.4,99.1] \\
    & Gemini-2.5-Flash   & 93.2 [88.2,96.2] & 100.0 [97.7,100.0] & 100.0 [97.6,100.0] & 99.3 [96.3,99.9] \\
    & Claude-Sonnet-4    & 99.4 [96.9,99.9] & 100.0 [97.9,100.0] & 100.0 [97.9,100.0] & 100.0 [97.9,100.0] \\
    \bottomrule
  \end{tabular}
\end{table}

This table is the per-cell counterpart of Table~\ref{tab:main_results} (body), augmented with Wilson 95\% score intervals for the boundary failure rate.

\paragraph{Why a CI under $\boldsymbol{T=0}$?}
Greedy decoding is intended to suppress the \emph{generation-level} variance of $M(p)$ to (approximately) zero, so repeated sampling is not used to estimate within-prompt stochasticity.
The reported intervals instead quantify a different source of uncertainty: the \emph{prompt-sampling} variance induced by drawing 15 TASKs per LANG (5\,Easy\,$+$\,5\,Mid\,$+$\,5\,Hard at $\text{seed}{=}42$) from a larger McEval-restricted pool.
A different seed would yield a different 180-prompt subset, and hence a slightly different observed pass rate; the Wilson interval provides an approximate $95\%$ binomial uncertainty interval over the sampled prompt instances.
Within-prompt comparisons (A1 vs A2 on the same TASK $\times$ LANG) are handled by paired McNemar tests (Appendix~\ref{app:mcnemar_a1_a2}), which are insensitive to this between-prompt sampling variance and therefore complement, rather than duplicate, the Wilson CI.

\paragraph{Why Wilson rather than Wald.}
The Wilson interval is preferred over the normal approximation because failure rates near $0\%$ and $100\%$ are common in our data, where the normal (Wald) interval would yield negative or super-unit bounds, or collapse to a degenerate point at the extremes (e.g., $\widehat{p} = 1.0$, $n = 180$ gives Wald $= [1.0, 1.0]$, asserting certainty).
Wilson's small-sample correction keeps both endpoints in $[0, 100]$ while preserving nominal coverage; for the same case it gives the asymmetric $[97.9\%, 100\%]$, properly reflecting upper-bound uncertainty.
Inspection by row shows that the dominant per-condition failure rates ($\geq 90\%$ for A2 cells across most models) have intervals of width $\leq 8$pp, supporting the qualitative claim that A2 failures are not borderline.

\subsection{Outer-Fence Length Distribution (All Nine Models)}
\label{app:safe_length_distribution_full}

This subsection provides the full per-model breakdown of outer-fence length choices on D-inner-run prompts, complementing the four-model body figure in \S\ref{sec:hint_ablation}.
For each (model, inner-run condition r), Table~\ref{tab:safe_length} reports the percentage of responses choosing each outer-fence backtick length (out of $n{=}144$ records per cell) and the resulting \fmetric{3} a-priori pass rate (sum of cells with outer~$>$~r).

\begin{table}[h]
  \centering
  \scriptsize
\begin{tabular}{ll|cccc|c}
  \toprule
  \textbf{Model} & \textbf{r} & \textbf{outer$=$3} & \textbf{outer$=$4} & \textbf{outer$=$5} & \textbf{outer$\geq$6} & \textbf{\fmetric{3} pass} \\
  \midrule
  \multirow{3}{*}{Qwen3-7B}
                    & 3 & 100.0 &  0.0 &  0.0 &  0.0 &  0.0 \\
                    & 4 & 100.0 &  0.0 &  0.0 &  0.0 &  0.0 \\
                    & 5 & 100.0 &  0.0 &  0.0 &  0.0 &  0.0 \\
  \midrule
  \multirow{3}{*}{Gemma-2-9B}
                    & 3 & 100.0 &  0.0 &  0.0 &  0.0 &  0.0 \\
                    & 4 & 100.0 &  0.0 &  0.0 &  0.0 &  0.0 \\
                    & 5 & 100.0 &  0.0 &  0.0 &  0.0 &  0.0 \\
  \midrule
  \multirow{3}{*}{Llama-3.1-8B}
                    & 3 & 100.0 &  0.0 &  0.0 &  0.0 &  0.0 \\
                    & 4 & 100.0 &  0.0 &  0.0 &  0.0 &  0.0 \\
                    & 5 & 100.0 &  0.0 &  0.0 &  0.0 &  0.0 \\
  \midrule
  \multirow{3}{*}{Qwen3-32B}
                    & 3 & 100.0 &  0.0 &  0.0 &  0.0 &  0.0 \\
                    & 4 &  99.3 &  0.7 &  0.0 &  0.0 &  0.7 \\
                    & 5 &  99.3 &  0.0 &  0.0 &  0.7 &  0.7 \\
  \midrule
  \multirow{3}{*}{Gemma-3-27B}
                    & 3 & 100.0 &  0.0 &  0.0 &  0.0 &  0.0 \\
                    & 4 & 100.0 &  0.0 &  0.0 &  0.0 &  0.0 \\
                    & 5 & 100.0 &  0.0 &  0.0 &  0.0 &  0.0 \\
  \midrule
  \multirow{3}{*}{Llama-3.1-70B}
                    & 3 & 100.0 &  0.0 &  0.0 &  0.0 &  0.0 \\
                    & 4 &  31.9 & 68.1 &  0.0 &  0.0 &  0.0 \\
                    & 5 &  95.1 &  4.2 &  0.7 &  0.0 &  0.0 \\
  \midrule
  \multirow{3}{*}{GPT-4o}
                    & 3 &  86.8 & 13.2 &  0.0 &  0.0 & \textbf{13.2} \\
                    & 4 &  36.1 & 63.9 &  0.0 &  0.0 &  0.0 \\
                    & 5 &  28.5 & 25.7 &  4.2 & 41.7 & \textbf{41.7} \\
  \midrule
  \multirow{3}{*}{Gemini-2.5-Flash}
                    & 3 & 100.0 &  0.0 &  0.0 &  0.0 &  0.0 \\
                    & 4 &  60.4 & 39.6 &  0.0 &  0.0 &  0.0 \\
                    & 5 &  73.6 &  7.6 & 18.8 &  0.0 &  0.0 \\
  \midrule
  \multirow{3}{*}{Claude-Sonnet-4}
                    & 3 & 100.0 &  0.0 &  0.0 &  0.0 &  0.0 \\
                    & 4 &  83.3 & 16.7 &  0.0 &  0.0 &  0.0 \\
                    & 5 &  99.3 &  0.7 &  0.0 &  0.0 &  0.0 \\
  \bottomrule
\end{tabular}

  \caption{Outer-fence length distribution and \fmetric{3} a-priori pass rate per (model, inner-run condition r) on D-inner-run prompts. Each row sums to $100\%$ across the four outer-length columns ($n{=}144$). The \fmetric{3} pass column equals the sum of cells with outer~$>$~r. Most models default to outer~$=3$ regardless of r; a few (Llama-3.1-70B, GPT-4o, Gemini-2.5-Flash, Claude-Sonnet-4) partially adapt by matching the inner length (outer~$=$~r), but only GPT-4o non-trivially exceeds it.}
  \label{tab:safe_length}
\end{table}

\subsection{Hint Ablation: Per-Model Heatmap}
\label{app:hint_ablation_figure}

\begin{figure}[H]
  \centering
  \includegraphics[width=0.95\linewidth]{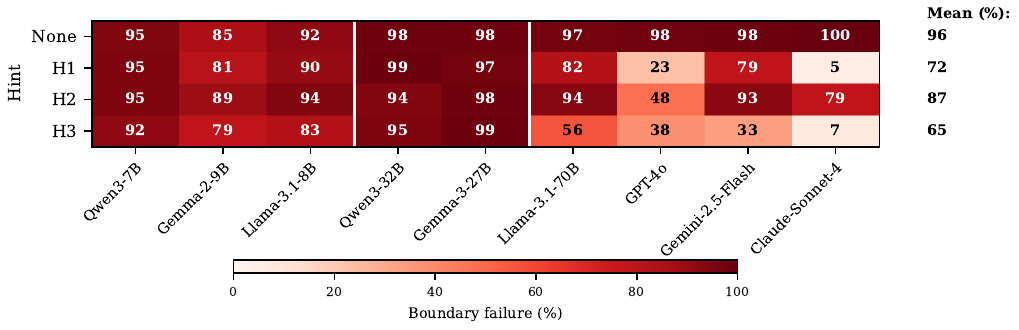}
  \caption{Per-model boundary failure rate (\%) across hint conditions (rows: None / H1 / H2 / H3; columns: 9 models in tier order). Cell color: deeper red indicates higher failure rate. Right column: row means across models. Body summary: \S\ref{sec:hint_ablation}.}
  \label{fig:hint_ablation}
\end{figure}

\section{Manipulation Checks and Robustness Diagnostics}
\label{app:robustness_diagnostics}

\subsection{A1 Per-Model Decomposition}
\label{app:a1_per_model}

Per-model breakdown of the four-step A1 decomposition (Table~\ref{tab:a1_decomposition} in \S\ref{sec:a1_decomp} shows the 9-model averages; Table~\ref{tab:a1_per_model} below shows the dispersion behind each step).

\begin{table}[h]
  \caption{Per-model A1 decomposition. Step 1: A3 wrap rate; Step 2: A1 non-compliance (\fmetric{1}); Step 3: compliance-conditional A1 residual boundary failure; Step 4: A2 boundary failure rate.}
  \label{tab:a1_per_model}
  \centering
  \small
  \begin{tabular}{lcccc}
    \toprule
    \textbf{Model} & \textbf{Step 1} & \textbf{Step 2} & \textbf{Step 3} & \textbf{Step 4} \\
    \midrule
    Qwen3-7B         & 77.6\% & 18.2\% & 1.5\% & 94.7\% \\
    Gemma-2-9B        & 0.1\% & 5.8\% & 3.1\% & 85.7\% \\
    Llama-3.1-8B      & 0.1\% & 0.4\% & 2.4\% & 91.7\% \\
    Qwen3-32B         & 46.2\% & 43.3\% & 1.5\% & 98.0\% \\
    Gemma-3-27B       & 83.8\% & 79.4\% & 34.5\% & 97.9\% \\
    Llama-3.1-70B     & 0.0\% & 0.0\% & 40.7\% & 97.1\% \\
    GPT-4o            & 47.9\% & 7.2\% & 0.7\% & 98.1\% \\
    Gemini-2.5-Flash  & 82.5\% & 16.0\% & 3.2\% & 98.4\% \\
    Claude-Sonnet-4   & 0.0\% & 0.0\% & 0.1\% & 99.9\% \\
    \bottomrule
  \end{tabular}
\end{table}

\subsection{Full-C Sensitivity Analysis}
\label{app:full_c_sensitivity}

Our pre-specified escalation rule (\S\ref{sec:content_metrics}) demotes \cmetric{4} (keyword-based Task Completion) when the Kendall's $\tau$ between Full-C (\cmetric{1}$\wedge$\cmetric{2}$\wedge$(\cmetric{4}$\geq$0.7)) and Structural-C (\cmetric{1}$\wedge$\cmetric{2}) model rankings on the ``content-correct, boundary-wrong'' cell falls below 0.8.
We use \cmetric{1}$\wedge$\cmetric{2} for the ranking-stability trigger because \cmetric{3} is a template-structure check; the final primary Content definition retains \cmetric{3} as specified in \S\ref{sec:content_metrics}.
On our data, $\tau=0.667<0.8$, so \cmetric{4} is demoted and the primary Content judgment is \cmetric{1}$\wedge$\cmetric{2}$\wedge$\cmetric{3} throughout the main text.
For transparency we retain the Full-C (with \cmetric{4}) four-cell decomposition here; the headline Latent Failure rate under Full-C is $33.1\%$ (overall), versus $38.0\%$ under the primary definition without \cmetric{4}.
The direction of model rankings is preserved (all 9 models exhibit higher Latent Failure under the primary definition than under Full-C), and the most extreme cases---Gemma-3-27B (73.8\% vs 57.0\%) and Llama-3.1-8B (14.5\% vs 10.8\%)---are shifted by the content-filter relaxation, consistent with the view that Full-C's keyword gate was removing structurally correct responses from the Content-Correct side; the primary $\cmetric{1}$--$\cmetric{3}$ definition is therefore the more conservative choice for boundary-failure attribution.
Full per-model numbers under both definitions are provided as \texttt{LatentMD/results/tables/four\_cell.json} in the supplementary materials (\texttt{primary\_struct\_only\_c} and \texttt{sensitivity\_full\_c} entries).

\section{Stratified Error-Type Analyses}
\label{app:e_type_stratified}

This section breaks down the error-type distribution (\etype{0} correct / \etype{1} premature / \etype{2} unclosed / \etype{3} latent collision) along four orthogonal stratifications: language (12 LANG values), model size tier (Small/Mid/Large), inner-nesting condition (B1--B4), and McEval task difficulty (easy/middle/hard).
All numbers are pooled across the 9 main-experiment models and exclude truncated responses.

\subsection{Per-Language Distribution}
\label{app:e_type_per_lang}
\begin{table}[h]
  \caption{Error type (\etype{0}--\etype{3}) distribution per LANG, pooled across all 9 models. $N$ = number of valid records per language.}
  \label{tab:e_type_per_lang}
  \centering
  \scriptsize
  \begin{tabular}{lc|cccc}
    \toprule
    \textbf{LANG} & \textbf{$N$} & \textbf{\etype{0} (\%)} & \textbf{\etype{1} (\%)} & \textbf{\etype{2} (\%)} & \textbf{\etype{3} (\%)} \\
    \midrule
    Python         & 1591 & 43.6 & 11.0 & 12.3 & 33.1 \\
    JavaScript     & 1580 & 45.0 & 8.9 & 12.0 & 34.1 \\
    Java           & 1611 & 43.3 & 10.2 & 16.1 & 30.4 \\
    C++            & 1583 & 43.6 & 8.6 & 11.6 & 36.2 \\
    C\#            & 1608 & 41.0 & 13.1 & 12.8 & 33.1 \\
    TypeScript     & 1579 & 42.6 & 8.4 & 13.7 & 35.3 \\
    Shell          & 1611 & 37.6 & 9.9 & 12.2 & 40.2 \\
    C              & 1592 & 41.9 & 8.0 & 13.9 & 36.2 \\
    Go             & 1568 & 43.7 & 9.2 & 14.5 & 32.5 \\
    HTML           & 1603 & 44.4 & 6.7 & 10.0 & 39.0 \\
    JSON           & 1595 & 38.6 & 10.6 & 10.7 & 40.2 \\
    Markdown       & 1607 & 64.8 & 4.2 & 3.3 & 27.7 \\
    \bottomrule
  \end{tabular}
\end{table}

Markdown stands out as the easiest LANG (\etype{0}~$=$~64.8\%) while JSON, Shell, and HTML show the highest \etype{3} rates ($\geq$39\%); the spread is consistent with the view that \etype{3} (latent collision) is driven by the syntactic similarity between code-fence delimiters and inner content tokens, not by language semantics.

\subsection{Per-Model-Size-Tier Distribution}
\label{app:e_type_per_tier}
\begin{table}[h]
  \caption{Error type distribution per model size tier (Small=3, Mid=2, Large=4 models).}
  \label{tab:e_type_per_tier}
  \centering
  \scriptsize
  \begin{tabular}{lc|cccc}
    \toprule
    \textbf{Tier} & \textbf{$N$} & \textbf{\etype{0} (\%)} & \textbf{\etype{1} (\%)} & \textbf{\etype{2} (\%)} & \textbf{\etype{3} (\%)} \\
    \midrule
    Small          & 6444 & 53.4 & 10.8 & 17.5 & 18.3 \\
    Mid            & 4310 & 22.3 & 11.9 & 7.0 & 58.8 \\
    Large          & 8374 & 48.3 & 6.3 & 10.2 & 35.3 \\
    \bottomrule
  \end{tabular}
\end{table}

The ordering is non-monotonic in tier --- Mid models (Qwen3-32B and Gemma-3-27B) have the lowest \etype{0} rate (22.3\%) and highest \etype{3} (58.8\%), driven by Gemma-3-27B's outlier behavior; Small and Large tiers are closer to each other (53.4\% vs 48.3\% on \etype{0}).

\subsection{Per-B-Condition Distribution}
\label{app:e_type_per_b_condition}
\begin{table}[h]
  \caption{Error type distribution per B-condition (inner nesting complexity), pooled across all 9 models. \etype{3} (latent collision) climbs sharply with B complexity.}
  \label{tab:e_type_per_b_condition}
  \centering
  \scriptsize
  \begin{tabular}{lc|cccc}
    \toprule
    \textbf{B-cond} & \textbf{$N$} & \textbf{\etype{0} (\%)} & \textbf{\etype{1} (\%)} & \textbf{\etype{2} (\%)} & \textbf{\etype{3} (\%)} \\
    \midrule
    B1             & 4800 & 47.7 & 3.0 & 9.7 & 39.6 \\
    B2             & 4812 & 40.1 & 19.4 & 16.4 & 24.1 \\
    B3             & 4793 & 41.1 & 10.7 & 15.7 & 32.5 \\
    B4             & 4723 & 48.0 & 3.0 & 5.7 & 43.3 \\
    \bottomrule
  \end{tabular}
\end{table}

\etype{1} (premature closure) peaks at B2 (19.4\%) where the response must include a code example \emph{plus} a literal Markdown explanation --- the premature-closure failure mode is most strongly activated by exactly one layer of literal-fence content.
\etype{3} (latent collision) is high at B1 (39.6\%) and B4 (43.3\%) where the inner code block competes most directly with the outer wrapper run length.

\subsection{Per-Difficulty Distribution}
\label{app:e_type_per_difficulty}
\begin{table}[h]
  \caption{Error type distribution per task difficulty, pooled across all 9 models. Difficulty levels (easy/middle/hard) are taken from McEval task metadata (\url{LatentMD/src/prompt_gen/_mceval_tasks.py} in the released code).}
  \label{tab:e_type_per_difficulty}
  \centering
  \scriptsize
  \begin{tabular}{lc|cccc}
    \toprule
    \textbf{Difficulty} & \textbf{$N$} & \textbf{\etype{0} (\%)} & \textbf{\etype{1} (\%)} & \textbf{\etype{2} (\%)} & \textbf{\etype{3} (\%)} \\
    \midrule
    Easy           & 6437 & 43.6 & 8.7 & 11.4 & 36.3 \\
    Middle         & 6374 & 45.2 & 9.0 & 11.8 & 33.9 \\
    Hard           & 6317 & 43.7 & 9.5 & 12.6 & 34.2 \\
    \bottomrule
  \end{tabular}
\end{table}

The distribution is essentially flat across McEval task difficulty levels (\etype{0} 43.6/45.2/43.7\%), suggesting that boundary-state tracking failures are not primarily explained by McEval semantic task difficulty.

\subsection{A\texorpdfstring{$\times$}{x}B Joint Distribution (Pooled)}
\label{app:a_b_e_type}

The body Table~\ref{tab:main_results} reports the boundary failure rate (\etype{1}$\cup$\etype{2}$\cup$\etype{3}) per cell.
Table~\ref{tab:a_b_e_type} pools the full \etype{0}/\etype{1}/\etype{2}/\etype{3} distribution across all 9 models to surface which boundary mechanism each (A, B) cell preferentially triggers (A3 included).

\begin{table}[h]
  \caption{Error-type composition per A$\times$B cell, averaged across all 9 models. Each row sums to 100\% (rounding aside); \etype{0} = correct (\fmetric{2}$\wedge$\fmetric{3} pass), \etype{1} = premature outer-fence closure, \etype{2} = unclosed outer fence, \etype{3} = latent collision (\fmetric{2} pass but \fmetric{3} fail). \textbf{Bold} marks the dominant error type per row (excluding \etype{0}), exposing which boundary mechanism each (A, B) cell preferentially triggers.}
  \label{tab:a_b_e_type}
  \centering
  \scriptsize
  \begin{tabular}{ll|cccc}
    \toprule
    \textbf{A} & \textbf{B} & \etype{0} (\%) & \etype{1} (\%) & \etype{2} (\%) & \etype{3} (\%) \\
    \midrule
    A1 & B1    & 79.7    & 1.2    & 1.4    & \textbf{17.8} \\
       & B2    & 73.3    & 5.1    & \textbf{12.8}    & 8.9 \\
       & B3    & 71.5    & 5.2    & 11.6    & \textbf{11.7} \\
       & B4    & 79.6    & 0.3    & 0.8    & \textbf{19.3} \\
    \midrule
    A2 & B1    & 9.5    & 5.7    & 22.1    & \textbf{62.7} \\
       & B2    & 1.7    & 37.1    & 12.4    & \textbf{48.7} \\
       & B3    & 1.6    & 20.9    & 16.6    & \textbf{60.9} \\
       & B4    & 5.2    & 8.1    & 14.3    & \textbf{72.4} \\
    \midrule
    A3 & B1    & 53.5    & 2.2    & 5.7    & \textbf{38.6} \\
       & B2    & 44.6    & 16.7    & \textbf{23.5}    & 15.1 \\
       & B3    & 49.5    & 6.2    & 18.6    & \textbf{25.7} \\
       & B4    & 59.6    & 0.9    & 1.8    & \textbf{37.8} \\
    \bottomrule
  \end{tabular}
\end{table}

Two patterns merit attention.
\textbf{(i)} The A2 row is \etype{3}-dominant across all B conditions (49--72\% latent collision); explicit wrapping makes \etype{3} the dominant failure mode, while inner nesting modulates the \etype{1}/\etype{2}/\etype{3} mix (e.g., A2$\times$B2 still carries substantial \etype{1} mass).
\textbf{(ii)} A1$\times$B2 and A1$\times$B3 are the only A1 cells where \etype{2} (unclosed fence) becomes the dominant failure mode (12.8\% and 11.6\%); these are the conditions where the model must produce literal-syntax inner blocks without an outer wrapper, and a non-trivial fraction of responses leave an inner fence open. A3 cells fall between A1 and A2 with \etype{3} as the dominant failure type (15--39\%), reflecting the base-rate wrapping behavior measured in Step~1 of the A1 decomposition (Table~\ref{tab:a1_decomposition}).

\subsection{A\texorpdfstring{$\times$}{x}B Joint Distribution (Per-Model)}
\label{app:a_b_e_per_model}

Table~\ref{tab:a_b_e_per_model} disaggregates the same A$\times$B$\times$E breakdown to each of the 9 models.
The per-model view reveals mechanism fingerprints that the pooled summary obscures: some Large/API models, especially Claude-Sonnet-4 and GPT-4o, are \etype{3}-dominant under A2 (latent collision $\geq$92\% of the cell mass), while small models such as Gemma-2-9B and Llama-3.1-8B exhibit substantial \etype{2} (unclosed fence) mass under A2, indicating that smaller models additionally leave fences unclosed rather than failing only by length-safety.

\begin{table}[h]
  \caption{Per-model error mechanism breakdown across A$\times$B conditions. Each cell shows \etype{1}\,/\,\etype{2}\,/\,\etype{3} percentages (rounded to integers); \etype{0} (correct) is implied as $100-\mathrm{sum}$, and the body Table~\ref{tab:main_results} boundary failure rate equals the sum. \textbf{Bold} marks the dominant mechanism per cell. Pooled (model-averaged) version: Table~\ref{tab:a_b_e_type}.}
  \label{tab:a_b_e_per_model}
  \centering
  \scriptsize
  \setlength{\tabcolsep}{1.5pt}
  \begin{tabular}{cl|cccc|cccc}
    \toprule
    & & \multicolumn{4}{c|}{\textbf{A1 (Direct)}} & \multicolumn{4}{c}{\textbf{A2 (Wrapped)}} \\
    & \textbf{Model} & B1 & B2 & B3 & B4 & B1 & B2 & B3 & B4 \\
    \midrule
    \multirow{3}{*}{\rotatebox{90}{\scriptsize Small}}
    & Qwen3-7B             & \textbf{2}/1/\textbf{2} & \textbf{19}/1/3 & 14/4/\textbf{26} & 0/0/0 & 11/5/\textbf{75} & \textbf{74}/6/18 & 22/5/\textbf{72} & 9/0/\textbf{83} \\
    & Gemma-2-9B           & 2/\textbf{3}/0 & 2/\textbf{6}/0 & 2/\textbf{4}/0 & 0/0/0 & 9/\textbf{48}/3 & \textbf{66}/21/12 & 33/\textbf{45}/15 & 2/\textbf{79}/7 \\
    & Llama-3.1-8B         & 0/0/0 & 1/\textbf{6}/0 & 0/\textbf{3}/0 & 0/\textbf{1}/0 & 6/\textbf{64}/23 & 19/\textbf{38}/33 & 13/\textbf{65}/18 & 6/31/\textbf{49} \\
    \midrule
    \multirow{2}{*}{\rotatebox{90}{\scriptsize Mid}}
    & Qwen3-32B            & 0/1/\textbf{51} & 2/2/\textbf{15} & 2/1/\textbf{16} & 0/1/\textbf{75} & 6/5/\textbf{86} & \textbf{46}/7/44 & 38/2/\textbf{60} & 31/2/\textbf{66} \\
    & Gemma-3-27B          & 6/7/\textbf{77} & 16/12/\textbf{47} & 27/4/\textbf{53} & 3/6/\textbf{85} & 2/22/\textbf{72} & 22/13/\textbf{64} & 31/11/\textbf{58} & 2/2/\textbf{92} \\
    \midrule
    \multirow{4}{*}{\rotatebox{90}{\scriptsize Large}}
    & Llama-3.1-70B        & 0/0/0 & 0/\textbf{84}/0 & 0/\textbf{79}/0 & 0/0/0 & 8/\textbf{53}/32 & 18/27/\textbf{56} & 28/23/\textbf{49} & 8/14/\textbf{72} \\
    & GPT-4o               & 0/0/\textbf{17} & 0/\textbf{2}/0 & 0/\textbf{1}/0 & 0/0/\textbf{11} & 2/0/\textbf{92} & 2/0/\textbf{98} & 2/0/\textbf{97} & 0/0/\textbf{98} \\
    & Gemini-2.5-Flash     & 1/1/\textbf{13} & 7/2/\textbf{14} & 2/8/\textbf{10} & 0/1/\textbf{2} & 9/1/\textbf{83} & 47/0/\textbf{53} & 19/0/\textbf{81} & 15/0/\textbf{84} \\
    & Claude-Sonnet-4      & 0/0/0 & 0/0/0 & 0/\textbf{1}/0 & 0/0/0 & 0/0/\textbf{99} & 41/0/\textbf{59} & 1/0/\textbf{99} & 0/0/\textbf{100} \\
    \bottomrule
  \end{tabular}
\end{table}

\subsection{Per-Language Latent Failure Distribution}
\label{app:latent_per_lang}

\begin{table}[h]
  \caption{Latent failure rate per language (pooled across all 9 models) with Wilson 95\% confidence intervals (computed on valid $N$ after truncation filtering). Latent Failure = content-correct $\wedge$ boundary-wrong (\fmetric{2}/\fmetric{3} fail). Languages are sorted by latent failure rate (descending). Each cell shows \emph{rate} \texttt{[CI lower, CI upper]}. Code-heavy languages (Shell, C\#, C++, TypeScript) have the highest latent failure rate (43--46\%): models produce structurally correct content but break the outer fence boundary. Markdown is the lowest (9\%) only because its content correctness is also lowest (22\%)~--- latent failure requires content-correct $=$ True, so when the model fails to produce a valid Markdown document it can never enter the latent-failure cell.}
  \label{tab:latent_per_lang}
  \centering
  \scriptsize
  \setlength{\tabcolsep}{4pt}
  \begin{tabular}{l|c|cc}
    \toprule
    \textbf{LANG} & \textbf{Valid $N$} & \textbf{Content Correct (\%)} & \textbf{Latent Failure (\%)} \\
    \midrule
    Shell          & 1{,}611 & 74.7 [72.5, 76.7] & 46.2 [43.8, 48.6] \\
    C\#            & 1{,}608 & 75.9 [73.7, 77.9] & 45.1 [42.7, 47.5] \\
    C++            & 1{,}583 & 74.0 [71.8, 76.1] & 42.9 [40.5, 45.3] \\
    TypeScript     & 1{,}579 & 74.9 [72.7, 76.9] & 42.8 [40.4, 45.3] \\
    JavaScript     & 1{,}580 & 75.0 [72.8, 77.1] & 42.2 [39.7, 44.6] \\
    Python         & 1{,}591 & 73.7 [71.5, 75.8] & 41.5 [39.1, 44.0] \\
    C              & 1{,}592 & 70.5 [68.3, 72.7] & 41.5 [39.1, 44.0] \\
    Go             & 1{,}568 & 72.4 [70.1, 74.5] & 40.1 [37.7, 42.6] \\
    JSON           & 1{,}595 & 63.3 [60.9, 65.7] & 39.0 [36.6, 41.4] \\
    Java           & 1{,}611 & 68.5 [66.2, 70.7] & 36.8 [34.5, 39.2] \\
    HTML           & 1{,}603 & 53.2 [50.8, 55.6] & 28.9 [26.8, 31.2] \\
    Markdown       & 1{,}607 & 21.6 [19.7, 23.7] & \phantom{0}9.2 [\phantom{0}7.9, 10.7] \\
    \midrule
    \textbf{Overall}  & 19{,}128 & \textbf{66.4 [65.7, 67.1]} & \textbf{38.0 [37.3, 38.7]} \\
    \bottomrule
  \end{tabular}
\end{table}

Among code-heavy languages the latent failure rate is uniformly high (Shell $46.2\%$, C\# $45.1\%$, C++ $42.9\%$, TypeScript $42.8\%$, JavaScript $42.2\%$, Python $41.5\%$, C $41.5\%$, Go $40.1\%$).
Markup languages have lower latent failure (HTML $28.9\%$, JSON $39.0\%$, Java $36.8\%$); Markdown is the lowest ($9.2\%$) but only because its content correctness is also lowest ($21.6\%$)~--- latent failure requires content-correct $=$ True, so when the model fails to produce a valid Markdown document it cannot enter the latent-failure cell at all.
The substantive finding is that wherever the model can produce a meaningful document, the document is roughly equally likely to be boundary-broken across all programming languages.

\section{Joint and Cascade Failure Analyses}
\label{app:joint_cascade}

\subsection{F1\texorpdfstring{$\times$}{x}F2 Joint Failure Analysis}
\label{app:f1_f2_cross}

\begin{table}[h]
  \caption{F1$\times$F2 joint distribution per model (\% of valid main experiment responses, $N{\approx}2160$ each). F1 = Instruction Compliance (A1+A2 directive followed), F2 = Fence Balance (closing fence present and length-matched). Cells sum to 100\% per row (rounding aside).}
  \label{tab:f1_f2_cross}
  \centering
  \scriptsize
  \begin{tabular}{cl|cccc}
    \toprule
    & \textbf{Model} & \textbf{F1+F2+} & \textbf{F1+F2$-$} & \textbf{F1$-$F2+} & \textbf{F1$-$F2$-$} \\
    \midrule
    \multirow{3}{*}{\rotatebox{90}{\scriptsize Small}}
    & Qwen3-7B           & 49.2 & 10.4 & 24.8 & 15.6 \\
    & Gemma-2-9B         & 35.4 & 10.9 & 33.6 & 20.1 \\
    & Llama-3.1-8B       & 43.2 & 4.9 & 28.9 & 23.0 \\
    \midrule
    \multirow{2}{*}{\rotatebox{90}{\scriptsize Mid}}
    & Qwen3-32B          & 40.5 & 10.7 & 45.4 & 3.3 \\
    & Gemma-3-27B        & 29.1 & 7.0 & 47.0 & 16.8 \\
    \midrule
    \multirow{4}{*}{\rotatebox{90}{\scriptsize Large}}
    & Llama-3.1-70B      & 38.1 & 18.9 & 20.2 & 22.7 \\
    & GPT-4o             & 63.6 & 0.6 & 35.5 & 0.3 \\
    & Gemini-2.5-Flash   & 52.9 & 8.6 & 27.1 & 11.5 \\
    & Claude-Sonnet-4    & 63.1 & 3.5 & 33.3 & 0.0 \\
    \bottomrule
  \end{tabular}
\end{table}

The four cells decompose responses by whether the model followed the A1/A2 instruction (\fmetric{1}) and whether the closing fence is present and length-matched (\fmetric{2}).
The \emph{F1$+$F2$-$} cell isolates instruction-compliant but boundary-broken outputs---the locus of the wrapper-induced collisions discussed in \S\ref{sec:latent_analysis}.
The \emph{F1$-$F2$+$} cell, in contrast, captures responses that ignored the wrap directive yet still produced a balanced inner block; \fmetric{1} and \fmetric{2} are not independent in either direction.

\subsection{Block Compliance by Boundary Condition}
\label{app:c3_per_b_condition}

\begin{table}[h]
  \caption{\cmetric{3} (Block Compliance) pass rate (\%) per model, stratified by B-axis (inner nesting complexity). \cmetric{3} requires the response to contain a code block matching the requested structural type. B1=single code block, B2=nested single, B3=nested multiple, B4=mixed elements. Computed on main experiment responses ($N{\approx}540$ per (model, B) cell).}
  \label{tab:c3_per_b_condition}
  \centering
  \scriptsize
  \begin{tabular}{cl|cccc}
    \toprule
    & \textbf{Model} & \textbf{B1} & \textbf{B2} & \textbf{B3} & \textbf{B4} \\
    \midrule
    \multirow{3}{*}{\rotatebox{90}{\scriptsize Small}}
    & Qwen3-7B           & 100.0 & 90.5 & 41.9 & 100.0 \\
    & Gemma-2-9B         & 100.0 & 40.6 & 5.9 & 100.0 \\
    & Llama-3.1-8B       & 100.0 & 58.7 & 5.6 & 100.0 \\
    \midrule
    \multirow{2}{*}{\rotatebox{90}{\scriptsize Mid}}
    & Qwen3-32B          & 100.0 & 88.1 & 84.8 & 100.0 \\
    & Gemma-3-27B        & 100.0 & 94.2 & 85.2 & 100.0 \\
    \midrule
    \multirow{4}{*}{\rotatebox{90}{\scriptsize Large}}
    & Llama-3.1-70B      & 100.0 & 92.0 & 36.7 & 100.0 \\
    & GPT-4o             & 100.0 & 92.8 & 49.6 & 100.0 \\
    & Gemini-2.5-Flash   & 100.0 & 85.1 & 64.6 & 100.0 \\
    & Claude-Sonnet-4    & 100.0 & 95.9 & 85.0 & 100.0 \\
    \bottomrule
  \end{tabular}
\end{table}

\cmetric{3} (Block Compliance) measures whether the response contains the number of literal Markdown-source fenced blocks required by the B-condition, independent of fence balance (\fmetric{2}) or length safety (\fmetric{3}).
Because \cmetric{3} requires zero literal-source blocks for B1 and B4, these cells are near-saturated by construction; the informative contrast is B2 (one literal-source block required) and especially B3 (two required), where small models drop to single digits while large models retain $\geq$50\%.
This stratification motivates separating Content (\cmetric{1}$\wedge$\cmetric{2}$\wedge$\cmetric{3}) from Boundary (\fmetric{2}$\wedge$\fmetric{3}): a B3 response can pass Content yet still fail at the fence boundary.

\subsection{E2 Unclosed-Fence Sub-Patterns}
\label{app:e2_subpattern}

\begin{table}[h]
  \caption{Decomposition of \etype{2} (Unclosed Fence) responses by re-parsing the raw response text and inspecting the \emph{last} fence-line that follows the outer opening. Sub-patterns: \emph{No close} (no fence-line at all $\Rightarrow$ model ended in prose); \emph{Short close} (last fence is same family but shorter than the outer, so CommonMark \S 4.5 ignored it as a close); \emph{Wrong family} (last fence is the opposite family, e.g.\ tilde when outer is backtick); \emph{Imbalanced mid} (last fence matches outer length+family but the stack stayed non-empty, indicating internally inconsistent nesting earlier); \emph{No outer} (no opening fence detected, mostly A1 responses without wrapper). Percentages are of \emph{total} responses per model; the five sub-pattern rates per row sum to the \etype{2} total.}
  \label{tab:e2_subpattern}
  \centering
  \scriptsize
  \setlength{\tabcolsep}{3pt}
  \begin{tabular}{lc|ccccc}
    \toprule
    \textbf{Model} & \textbf{\etype{2} (\%)} & \textbf{No close} & \textbf{Short close} & \textbf{Wrong family} & \textbf{Imbal.\ mid} & \textbf{No outer} \\
    \midrule
    Qwen3-7B           & 6.9 & 0.0 & 0.0 & 0.0 & 6.9 & 0.0 \\
    Gemma-2-9B         & 21.4 & 0.0 & 0.0 & 0.0 & 21.4 & 0.0 \\
    Llama-3.1-8B       & 24.0 & 0.0 & 0.0 & 0.0 & 24.0 & 0.0 \\
    Qwen3-32B          & 3.5 & 0.0 & 0.0 & 0.0 & 3.5 & 0.0 \\
    Gemma-3-27B        & 10.5 & 0.0 & 0.0 & 0.0 & 10.5 & 0.0 \\
    Llama-3.1-70B      & 36.5 & 0.0 & 0.0 & 0.0 & 36.5 & 0.0 \\
    GPT-4o             & 0.5 & 0.0 & 0.0 & 0.0 & 0.5 & 0.0 \\
    Gemini-2.5-Flash   & 2.8 & 0.1 & 0.0 & 0.0 & 2.7 & 0.0 \\
    Claude-Sonnet-4    & 0.0 & 0.0 & 0.0 & 0.0 & 0.0 & 0.0 \\
    \midrule
    \textbf{Overall}      & 11.9      & 0.0      & 0.0      & 0.0      & 11.9      & 0.0 \\
    \bottomrule
  \end{tabular}
\end{table}

We re-parse the raw response text of every \etype{2} record with a stack-based CommonMark fence walker to characterize \emph{how} the outer fence ended up unclosed.
Across all 9 models, nearly every \etype{2} response falls into the \emph{Imbalanced mid} sub-pattern: the model successfully wrote at least one matching inner-block close but the cumulative stack ended non-empty.
Pure \emph{No close} (model wrote no fence-line at all after the outer opening) accounts for $<0.2\%$ of total responses, even on the worst model.
\emph{Short close} attempts (a closing fence shorter than the outer) are also negligible because CommonMark requires fence runs of length $\geq 3$, so any plausible close attempt would already match a 3-backtick outer.
The takeaway: \etype{2} is not a ``model gave up writing'' failure mode but rather a \emph{nesting-bookkeeping} failure---models produce structurally complex output where the cumulative open/close balance silently leaves the outer wrapper unclosed.

\subsection{Error Cascade Analysis}
\label{app:error_cascade}

\begin{table}[h]
  \caption{Error cascade: probability the response is content-correct conditional on whether \etype{1} (premature closure) fired. A large gap between the two columns indicates that early-closure failures tend to truncate or omit content (cascade effect); a small gap means boundary failures and content failures are decoupled.}
  \label{tab:error_cascade}
  \centering
  \scriptsize
  \begin{tabular}{lcc|cc}
    \toprule
    \textbf{Model} & $N_{\etype{1}=\text{true}}$ & P(C $\mid$ \etype{1}) (\%) & $N_{\etype{1}=\text{false}}$ & P(C $\mid$ $\neg$\etype{1}) (\%) \\
    \midrule
    Qwen3-7B           & 407 & 91.2 & 1734 & 64.1 \\
    Gemma-2-9B         & 207 & 5.8 & 1950 & 16.5 \\
    Llama-3.1-8B       & 83 & 45.8 & 2063 & 48.3 \\
    Qwen3-32B          & 226 & 54.0 & 1926 & 73.7 \\
    Gemma-3-27B        & 288 & 87.2 & 1870 & 74.3 \\
    Llama-3.1-70B      & 113 & 77.0 & 2046 & 68.6 \\
    GPT-4o             & 10 & 50.0 & 2150 & 80.2 \\
    Gemini-2.5-Flash   & 326 & 77.6 & 1569 & 75.8 \\
    Claude-Sonnet-4    & 75 & 100.0 & 2085 & 92.8 \\
    \midrule
    \textbf{Overall}     & 1735 & 70.0 & 17393 & 66.1 \\
    \bottomrule
  \end{tabular}
\end{table}

The conditional gap between $P(\text{content-correct} \mid \etype{1})$ and $P(\text{content-correct} \mid \neg\etype{1})$ is small in aggregate (70.0\% vs 66.1\%), suggesting that premature-closure failures do not systematically truncate or omit content; content correctness and fence-tracking failure are not tightly coupled in this diagnostic.
Per-model rows show high variance (e.g., GPT-4o: 50.0/80.2\%, Claude-Sonnet-4: 100.0/92.8\%); small $N_{\etype{1}}$ in some rows (GPT-4o $n{=}10$, Claude $n{=}75$) limits the per-model interpretation.

\section{Statistical Tests}
\label{app:statistical_tests}

\begin{table}[h]
  \caption{Statistical test plan. All families of multiple comparisons use Holm--Bonferroni correction where applicable.}
  \label{tab:stat_tests}
  \centering
  \small
  \begin{tabular}{lll}
    \toprule
    \textbf{Comparison} & \textbf{Test} & \textbf{Effect Size} \\
    \midrule
    Between models (failure rate)     & Chi-squared / Fisher's exact & Cram{\'e}r's $V$ \\
    Within model, between conditions  & McNemar's test (paired)       & Odds ratio \\
    Outer length $\times$ failure     & Cochran-Armitage trend        & -- \\
    C3 sensitivity (model ordering)   & Kendall's $\tau$              & -- \\
    \bottomrule
  \end{tabular}
  \vspace{0.3em}

  {\small Confidence intervals: 95\% Wilson score intervals (binomial proportion, valid $N$ denominator). Stability: CV and ICC(1,1) from 30 prompts $\times$ 5 runs.}
\end{table}

\subsection{Inter-Model Agreement (Fleiss' Kappa)}
\label{app:fleiss_kappa}

\begin{table}[h]
  \caption{Fleiss' kappa for inter-model agreement on the binary \emph{boundary-correct} judgment (\fmetric{2}$\wedge$\fmetric{3}). Computed per tier (Small/Mid/Large) and overall across the 9 models. Higher kappa indicates that models tend to fail (or pass) on the same prompts; values near 0 indicate near-independent failure modes.}
  \label{tab:fleiss_kappa}
  \centering
  \scriptsize
  \begin{tabular}{lccc}
    \toprule
    \textbf{Group} & \textbf{Models} & \textbf{Prompts $N$} & \textbf{Fleiss' $\kappa$} \\
    \midrule
    Small              & 3 & 2125 & 0.433 \\
    Mid                & 2 & 2150 & 0.008 \\
    Large              & 4 & 1895 & 0.392 \\
    Overall (all 9)    & 9 & 1868 & 0.349 \\
    \bottomrule
  \end{tabular}
\end{table}

Aggregate Fleiss' $\kappa = 0.349$ across the 9 models indicates moderate but far-from-perfect agreement on which prompts cause boundary failures.
The Mid tier's near-zero $\kappa$ ($+0.008$) reflects Gemma-3-27B's anomalously high failure rate combined with Qwen3-32B's much milder profile --- the two models show little agreement on which prompts fail, indicating that boundary-failure susceptibility is not a single ``hard prompt'' axis but at least partially model-specific.

\subsection{Cochran--Armitage Trend Test}
\label{app:cochran_armitage}

\begin{table}[h]
  \caption{Cochran--Armitage trend test for \etype{3} (latent collision) rate as a function of B-condition complexity (B1$\rightarrow$B2$\rightarrow$B3$\rightarrow$B4 treated as equally-spaced dose levels). A positive $Z$ indicates that \etype{3} rate increases monotonically with nesting complexity. All p-values two-sided.}
  \label{tab:cochran_armitage}
  \centering
  \scriptsize
  \begin{tabular}{lcccc|cc}
    \toprule
    \textbf{Model} & \textbf{B1 (\%)} & \textbf{B2 (\%)} & \textbf{B3 (\%)} & \textbf{B4 (\%)} & $Z$ & $p$ \\
    \midrule
    Qwen3-7B           & 53.7 & 13.2 & 47.0 & 52.6 & +3.20 & 0.00137 \\
    Gemma-2-9B         & 1.1 & 3.9 & 5.0 & 2.2 & +1.36 & 0.174 \\
    Llama-3.1-8B       & 7.6 & 11.1 & 6.0 & 16.3 & +3.56 & $<10^{-3}$ \\
    Qwen3-32B          & 67.0 & 30.1 & 36.1 & 62.1 & -0.95 & 0.343 \\
    Gemma-3-27B        & 73.7 & 56.4 & 58.5 & 86.1 & +4.43 & $<10^{-3}$ \\
    Llama-3.1-70B      & 10.6 & 18.5 & 16.3 & 24.1 & +5.27 & $<10^{-3}$ \\
    GPT-4o             & 63.5 & 33.5 & 35.6 & 67.0 & +1.31 & 0.191 \\
    Gemini-2.5-Flash   & 46.7 & 31.3 & 57.5 & 45.5 & +2.32 & 0.0205 \\
    Claude-Sonnet-4    & 33.1 & 19.8 & 33.0 & 33.3 & +1.56 & 0.12 \\
    \midrule
    \textbf{Overall (pooled)} & 39.6 & 24.1 & 32.5 & 43.3 & +6.17 & $<10^{-3}$ \\
    \bottomrule
  \end{tabular}
\end{table}

The pooled Cochran--Armitage Z-statistic is $+6.17$ ($p<10^{-3}$), providing evidence for a positive ordered trend in \etype{3} (latent collision) rate with B-condition complexity, despite non-monotone per-cell profiles in some models.
At the per-model level, 5/9 models show statistically significant positive trends; the four exceptions (Gemma-2-9B, Qwen3-32B, GPT-4o, Claude-Sonnet-4) either have very low \etype{3} rates throughout (Gemma-2-9B) or show U-shaped rather than monotone profiles (Qwen3-32B, GPT-4o have lower \etype{3} at B2/B3 than at B1/B4).

\subsection{Chi-Squared Pairwise Comparisons}
\label{app:chi_squared_pairwise}

\begin{table}[h]
  \caption{Chi-squared pairwise model comparisons of boundary-correct judgments on the main experiment ($N=2{,}160$ prompts per model). Cells show Cram\'er's $V$ effect size; significance markers reflect Holm-Bonferroni-corrected $p$-values: $^{*}p<0.05$, $^{**}p<0.01$, $^{***}p<0.001$. Lower triangle only; diagonal blank.}
  \label{tab:chi_squared_pairwise}
  \centering
  \scriptsize
  \setlength{\tabcolsep}{2.5pt}
  \begin{tabular}{l|cccccccc}
    \toprule
    & Q7B & G2-9B & L3.1-8B & Q32B & G3-27B & L3.1-70B & GPT-4o & Gemini-2.5-Flash \\
    \midrule
    G2-9B & $0.33^{***}$ &  &  &  &  &  &  &  \\
    L3.1-8B & $0.29^{***}$ & $0.04^{**}$ &  &  &  &  &  &  \\
    Q32B & $0.05^{**}$ & $0.29^{***}$ & $0.25^{***}$ &  &  &  &  &  \\
    G3-27B & $0.31^{***}$ & $0.61^{***}$ & $0.57^{***}$ & $0.36^{***}$ &  &  &  &  \\
    L3.1-70B & $0.09^{***}$ & $0.25^{***}$ & $0.21^{***}$ & $0.04^{*}$ & $0.39^{***}$ &  &  &  \\
    GPT-4o & $0.17^{***}$ & $0.17^{***}$ & $0.13^{***}$ & $0.12^{***}$ & $0.46^{***}$ & $0.08^{***}$ &  &  \\
    Gemini-2.5-Flash & $0.02$ & $0.31^{***}$ & $0.27^{***}$ & $0.02$ & $0.34^{***}$ & $0.06^{***}$ & $0.14^{***}$ &  \\
    Claude-S-4 & $0.34^{***}$ & $0.01$ & $0.05^{**}$ & $0.30^{***}$ & $0.61^{***}$ & $0.26^{***}$ & $0.18^{***}$ & $0.32^{***}$ \\
    \bottomrule
  \end{tabular}
\end{table}

Pairwise comparison of boundary-correct judgments between every pair of the 9 models.
Cram\'er's $V$ in the lower triangle ranges from $0.01$ (Claude vs Gemma-2-9B, negligible effect size) to $0.61$ (Claude vs Gemma-3-27B, very large effect).
Effect sizes correlate with the per-model \emph{Latent Failure} rates reported in Table~\ref{tab:uniquely_captured}: model pairs with similar latent-failure profiles cluster together with low $V$, while pairs spanning the catastrophic and proficient extremes show $V \geq 0.5$.
The four highest-$V$ pairs all involve Gemma-3-27B (the dominant outlier) or Claude-Sonnet-4 (the dominant proficient).

\subsection{McNemar Paired Test (A1 vs A2)}
\label{app:mcnemar_a1_a2}

\begin{table}[h]
  \caption{McNemar paired test for boundary-correct judgments under A1 (Direct) versus A2 (Wrapped), with each prompt observed under both conditions. $b$ = discordant pairs A1$=$pass / A2$=$fail; $c$ = discordant pairs A1$=$fail / A2$=$pass. The odds ratio $b/c$ quantifies the asymmetry: A2 is dramatically worse than A1 in nearly every discordant case.}
  \label{tab:mcnemar_a1_a2}
  \centering
  \scriptsize
  \begin{tabular}{lc}
    \toprule
    \textbf{Quantity} & \textbf{Value} \\
    \midrule
    Discordant pairs $b$ (A1$=$pass, A2$=$fail) & 4,509 \\
    Discordant pairs $c$ (A1$=$fail, A2$=$pass) & 4 \\
    Odds ratio $b/c$                              & 1127.25 \\
    McNemar $\chi^2$ statistic                  & 4495.02 \\
    $p$-value (two-sided)                          & $<10^{-3}$ \\
    \bottomrule
  \end{tabular}
\end{table}

The McNemar paired test asks: of prompts where A1 and A2 outcomes differ, how asymmetric is the difference?
The answer is decisively asymmetric ($b/c \approx 1127$): in $4{,}509$ paired prompts the model passes under A1 (no wrapper) but fails under A2 (wrapper required), while only $4$ prompts show the reverse pattern.
This is the within-prompt analogue of the marginal $\sim$95\% A2 failure observed in Table~\ref{tab:main_results}: the wrapper directive does not just shift the marginal failure rate---it flips the per-prompt outcome in nearly every discordant case.

\section{Cross-Format and Auxiliary Probes}
\label{app:auxiliary_probes}

\subsection{Cross-Format Analysis}
\label{app:crossformat}

\begin{table}[h]
  \caption{Cross-format parse-validity rate (\%) per model. \textbf{Python} conditions test escalating triple-quote (\texttt{\textquotedbl\textquotedbl\textquotedbl} / \texttt{\textquotesingle\textquotesingle\textquotesingle}) collision pressure inside docstrings (K1: literal mention; K2: both delimiter styles; K3: nested triple-quote in docstring code example). \textbf{JSON} conditions test escalating brace-nesting depth (K1: 1--2 levels; K2: 3--4 levels; K3: $\geq$5 levels). Pass = the extracted code block parses successfully via Python \texttt{ast.parse} or \texttt{json.loads}; failure = syntax/parse error. JSON serves as a partial negative control: its asymmetric braces do not exhibit the same length-matching collision as Markdown fences, so we expect higher pass rates regardless of nesting depth.}
  \label{tab:crossformat_results}
  \centering
  \scriptsize
  \begin{tabular}{cl|ccc|ccc}
    \toprule
    & & \multicolumn{3}{c|}{\textbf{Python (triple-quote collision)}} & \multicolumn{3}{c}{\textbf{JSON (nesting depth)}} \\
    & \textbf{Model} & \textbf{K1} & \textbf{K2} & \textbf{K3} & \textbf{K1} & \textbf{K2} & \textbf{K3} \\
    \midrule
    \multirow{3}{*}{\rotatebox{90}{\scriptsize Small}}
    & Qwen3-7B           & 93.3 & 93.3 & 80.0 & 100.0 & 100.0 & 100.0 \\
    & Gemma-2-9B         & 100.0 & 93.3 & 100.0 & 100.0 & 100.0 & 100.0 \\
    & Llama-3.1-8B       & 100.0 & 93.3 & 93.3 & 100.0 & 100.0 & 100.0 \\
    \midrule
    \multirow{2}{*}{\rotatebox{90}{\scriptsize Mid}}
    & Qwen3-32B          & 40.0 & 60.0 & 86.7 & 100.0 & 100.0 & 100.0 \\
    & Gemma-3-27B        & 73.3 & 86.7 & 46.7 & 100.0 & 100.0 & 100.0 \\
    \midrule
    \multirow{4}{*}{\rotatebox{90}{\scriptsize Large}}
    & Llama-3.1-70B      & 46.7 & 86.7 & 26.7 & 100.0 & 100.0 & 100.0 \\
    & GPT-4o             & 0.0 & 93.3 & 80.0 & 100.0 & 100.0 & 100.0 \\
    & Gemini-2.5-Flash   & 0.0 & 73.3 & 20.0 & 100.0 & 100.0 & 100.0 \\
    & Claude-Sonnet-4    & 80.0 & 20.0 & 93.3 & 100.0 & 100.0 & 100.0 \\
    \bottomrule
  \end{tabular}
\end{table}

We test whether the symmetric-delimiter collision phenomenon extends beyond Markdown by replicating the experiment in two contrasting target formats: \textbf{Python} (symmetric triple-quote docstrings) and \textbf{JSON} (asymmetric braces).
Each prompt asks the model to produce a syntactically valid Python function or JSON document under escalating delimiter-collision pressure (K1$\rightarrow$K3); we evaluate by attempting to parse the extracted code block with Python \texttt{ast.parse} or \texttt{json.loads}.

The pattern in Table~\ref{tab:crossformat_results} mirrors the Markdown finding: \emph{JSON, the asymmetric control, achieves 100\% parse-validity across every model and every nesting depth}, while \emph{Python triple-quote conditions exhibit large per-model failure rates}, with several Large models hitting 0\% on the seemingly easiest condition (K1: ``mention the triple-quote delimiter in the docstring text''), where literally placing the delimiter inside the docstring terminates it prematurely.
The K2 condition (asking for both \texttt{\textquotedbl\textquotedbl\textquotedbl} and \texttt{\textquotesingle\textquotesingle\textquotesingle} styles) often \emph{improves} the pass rate over K1 because the prompt itself nudges the model toward delimiter alternation as a workaround.
These results support the claim that symmetric-delimiter boundary tracking can fail beyond Markdown, while JSON remains robust as an asymmetric-delimiter control.

\subsection{Inline Code Span Results}
\label{app:inline}

\begin{table}[h]
  \caption{Inline code span failure rate (\%) per (model, run). \emph{Failure} = the model fails to produce \emph{any} inline span whose CommonMark-rendered content exactly equals the keyword text specified in the prompt (\emph{inline\_match}). Run~$=N$ denotes $N$ literal backticks placed around the keyword content (so safe wrapper length $> N$). Cells are computed via markdown-it-py CommonMark parsing of the raw response and exact match against the prompt's verbatim target. Each cell is over $9$ records ($3$ keyword groups $\times$ $3$ sentence templates) per (model, run).}
  \label{tab:inline_results}
  \centering
  \scriptsize
  \begin{tabular}{cl|ccc}
    \toprule
    & \textbf{Model} & \textbf{Run=0} & \textbf{Run=1} & \textbf{Run=2} \\
    \midrule
    \multirow{3}{*}{\rotatebox{90}{\scriptsize Small}}
    & Qwen3-7B           & 0.0 & 100.0 & 100.0 \\
    & Gemma-2-9B         & 0.0 & 100.0 & 100.0 \\
    & Llama-3.1-8B       & 11.1 & 100.0 & 100.0 \\
    \midrule
    \multirow{2}{*}{\rotatebox{90}{\scriptsize Mid}}
    & Qwen3-32B          & 0.0 & 100.0 & 100.0 \\
    & Gemma-3-27B        & 11.1 & 100.0 & 100.0 \\
    \midrule
    \multirow{4}{*}{\rotatebox{90}{\scriptsize Large}}
    & Llama-3.1-70B      & 0.0 & 100.0 & 100.0 \\
    & GPT-4o             & 44.4 & 100.0 & 100.0 \\
    & Gemini-2.5-Flash   & 0.0 & 100.0 & 100.0 \\
    & Claude-Sonnet-4    & 0.0 & 66.7 & 55.6 \\
    \midrule
    & \textbf{Overall}      & \textbf{7.4} & \textbf{96.3} & \textbf{95.1} \\
    \bottomrule
  \end{tabular}
\end{table}

\paragraph{Setup.}
Inline code span experiments use $27$ unique prompts ($3$ sentence templates $\times$ $3$ keyword groups $\times$ $3$ runs) $\times$ $9$ models $=$ $243$ generations.
The prompt asks the model to display a verbatim text (\texttt{foo}, \verb|`foo`|, or \verb|``foo``| for run$=$0/1/2) inside a single inline code span.
Run$=$0 has no collision pressure (the keyword text contains no backticks); run$=$1 places one backtick on each side; run$=$2 places two backticks on each side.
The safe wrapper length follows the CommonMark inline rule: wrapper backtick run length must be strictly greater than the longest same-family run inside the span content (run$=$1 needs wrapper length $\geq 2$, run$=$2 needs wrapper length $\geq 3$, with appropriate space-padding so the literal backticks display).

\paragraph{Evaluator.}
The legacy block-fallback metric (any fenced code block in the response) under-counted real inline failures because it never inspected the inline span's rendered content.
We re-evaluate every inline response with a CommonMark parser (markdown-it-py): for each response we extract all \texttt{code\_inline} tokens, take their rendered content, and pass the response if and only if at least one token's content exactly matches the prompt's verbatim target.
Wrapper length is recovered post hoc from the raw text and an auxiliary \emph{inline\_F3\_safe} metric (wrapper $>$ inner max same-family run) is computed for the matched span.

\paragraph{Findings.}
Table~\ref{tab:inline_results} reports the verbatim-match failure rate.
At run$=$0 (no collision) the pooled failure is only $7.4\%$, dominated by a single model (GPT-4o, $44.4\%$) that produces an inline span enveloping an entire introductory phrase rather than the keyword alone.
Under inline collision pressure (run$=$1 / run$=$2) the picture inverts: $8$ of $9$ models fail $100\%$ of cases at both runs.
Only Claude-Sonnet-4 handles the collision partially ($66.7\%$ fail at run$=$1, $55.6\%$ at run$=$2), and even Claude's non-failures use exactly run$+$1 backticks for the wrapper without any further safety margin.
\emph{Note on interpretation}: the prompt's verbatim target (\verb|`foo`| or \verb|``foo``|) appears \emph{as inline code} in the prompt source, so the dominant failure mode (Qwen3, Gemma, Llama; \textbf{Backtick omission} below) reflects models defaulting to wrap the bare keyword in a single backtick~--- structurally safe (\emph{inline\_F3\_safe}~$=$~True) but verbatim-mismatched.
The genuine collision-tracking failures concentrate in the asymmetric-close and length-matched-collision modes; the four-cell decomposition (\S\ref{sec:four_cell}) classifies the dominant mode as \emph{format\_only} rather than \emph{latent failure}, and no inline record is classified as latent (a wrapper too short to span backtick-bearing content cannot also produce a verbatim-correct rendering).

\paragraph{Per-model fail-mode taxonomy.}
The $96\%+$ pooled fail rate decomposes into four qualitatively distinct error modes, which we identified by inspecting the parsed spans against the prompt's target:
\begin{itemize}[leftmargin=*, nosep]
\item \textbf{Backtick omission} (Qwen3, Gemma, Llama at run$=$1/2): the model writes the keyword \emph{without} the surrounding literal backticks, producing a span like \verb|`foo`| instead of the requested \verb|``foo``|. This is a verbatim-display mismatch rather than a boundary-safety failure: the produced span is usually structurally safe, but its rendered content does not match the prompt's literal target.
\item \textbf{Whole-sentence wrap} (GPT-4o run$=$1, run$=$2): the model wraps the entire introductory sentence (\textit{``Here is some inline code:}~\verb|`foo`|\textit{.''}) in one inline span instead of isolating the keyword. Wrapper length is structurally safe ($2$-bt for run$=$1) but semantic content is wrong.
\item \textbf{Asymmetric/short close} (Claude run$=$2 partial): the model opens the span with $\geq$3 backticks but closes with a shorter run, producing parses like \verb|``foo`| (mismatched) or breaking into multiple disjoint spans.
\item \textbf{Length-matched collision} (Llama-3.1-70B, Gemini under high collision): the model uses a wrapper length equal to the keyword's internal backtick run, so the parser silently splits the response into multiple inline spans (the inner literal backticks act as premature closing). Phenomenologically identical to the \etype{1} premature closure observed at the block level.
\end{itemize}

\paragraph{Synthesis.}
The block-level boundary-state tracking failure (\S\ref{sec:main_results}) is not an artefact of fence-line semantics: the asymmetric-close and length-matched-collision modes documented above reproduce, for inline spans, the same default-shortest-wrapper heuristic that drives \etype{3} latent collision in fenced blocks.
The aggregate $96\%+$ verbatim-match failure additionally absorbs an interpretation default (Backtick omission), which inflates the headline number without itself constituting a collision-tracking failure.
Even after netting out interpretation defaults, the structural failures observed in the asymmetric-close and length-matched-collision modes support the view that boundary-state tracking under symmetric, length-matched delimiters is a generation-time bottleneck rather than a Markdown-fence-specific quirk; inline therefore serves as a complementary probe to the block-level evidence rather than an independent generalization.

\subsection{Temperature Sensitivity (Future Work)}
\label{app:temperature}

Temperature sensitivity analysis---testing whether boundary failure patterns persist across the full temperature range 0.0--1.0---is left to future work.
Our main experiments use temperature$=0$ (greedy decoding), which ensures deterministic outputs but leaves open the question of whether higher temperatures exacerbate or mitigate boundary failures.

\section{Practical Evidence: Full List}
\label{app:practical_evidence}

\begin{table}[h]
  \caption{Complete list of independently reported boundary-state tracking failures across LLM platforms and developer tools. URLs verified at the time of writing.}
  \label{tab:practical_evidence}
  \centering
  \footnotesize
  \setlength{\tabcolsep}{3pt}
  \renewcommand{\arraystretch}{1.15}
  \begin{tabular}{llp{8cm}}
    \toprule
    \textbf{Platform} & \textbf{Source} & \textbf{Title and URL} \\
    \midrule
    OpenAI    & chatkit-js \#89        & Nested triple-backtick code blocks break \newline {\scriptsize\url{https://github.com/openai/chatkit-js/issues/89}} \\
    OpenAI    & Community (Bugs)       & Use longer Markdown fences so code blocks don't break \newline {\scriptsize\url{https://community.openai.com/t/chatgpt-use-longer-markdown-fences-so-code-blocks-don-t-break/1357890}} \\
    OpenAI    & Community (Bugs)       & Minor code block bug with nested code blocks \newline {\scriptsize\url{https://community.openai.com/t/bug-minor-code-block-bug-with-nested-code-blocks/1018718}} \\
    OpenAI    & Community (Bugs)       & Missing triple backquote in Markdown code block \newline {\scriptsize\url{https://community.openai.com/t/missing-triple-backquote-in-markdown-code-block/1234674}} \\
    OpenAI    & Community (API)        & Markdown formatting issues with GPT-5 \newline {\scriptsize\url{https://community.openai.com/t/markdown-formatting-issues-with-gpt-5/1337570}} \\
    Google    & gemini-cli \#10515     & CLI Markdown rendering is broken \newline {\scriptsize\url{https://github.com/google-gemini/gemini-cli/issues/10515}} \\
    Microsoft & vscode \#295126        & create\_file tool wraps extra backtick fences \newline {\scriptsize\url{https://github.com/microsoft/vscode/issues/295126}} \\
    JetBrains & YouTrack LLM-1623      & Markdown code snippets generated broken \newline {\scriptsize\url{https://youtrack.jetbrains.com/issue/LLM-1623/Markdown-code-snippets-generated-broken}} \\
    Open WebUI & \#5016                & Incorrect rendering of nested code blocks \newline {\scriptsize\url{https://github.com/open-webui/open-webui/issues/5016}} \\
    LangChain4j & \#1446               & JSON wrapped in backticks extraction failure \newline {\scriptsize\url{https://github.com/langchain4j/langchain4j/issues/1446}} \\
    Obsidian  & Forum \#60565          & Allow nested code blocks / allow triple backticks in code blocks rendering result \newline {\scriptsize\url{https://forum.obsidian.md/t/allow-nested-code-blocks-allow-triple-backticks-in-code-blocks-rendering-result/60565}} \\
    Block     & goose \#8290           & Fenced code blocks break message layout \newline {\scriptsize\url{https://github.com/block/goose/issues/8290}} \\
    Streamdown & \#473                 & Fenced code blocks don't render incrementally during streaming \newline {\scriptsize\url{https://github.com/vercel/streamdown/issues/473}} \\
    Continue  & \#6059                 & LLM generate markdown with broken backquotes \newline {\scriptsize\url{https://github.com/continuedev/continue/issues/6059}} \\
    Medium    & Cultman Sachs          & Why Can't AI Models Output Clean Markdown? A Technical Mess That Still Isn't Fixed \newline {\scriptsize\url{https://medium.com/@CultmanSachs/why-cant-ai-models-output-clean-markdown-a-technical-mess-that-still-isn-t-fixed-1dc70ff366a3}} \\
    Medium    & Daniel Olshansky       & Escaping Backticks in your LLM System Prompt \newline {\scriptsize\url{https://olshansky.medium.com/escaping-backticks-in-your-llm-system-prompt-6507a25b7bc8}} \\
    \bottomrule
  \end{tabular}
\end{table}


\section{Additional Experiments}
\label{app:additional}

This appendix reports five additional analyses that extend the main study: a decomposition of the headline latent-failure rate by elicitation condition (\S\ref{app:realistic_decomp}), a measurement of downstream harm under real renderers and code extractors (\S\ref{app:downstream_harm}), an evaluation of seven newer-generation models (\S\ref{app:newer_models}), an expansion of the natural-prompt set from $30$ to $248$ organic requests mined from public conversation logs (\S\ref{app:natural2}), and a quantification of the benchmark's effective diagnostic diversity together with a coverage-preserving core subset (\S\ref{app:diversity}).
All new generations use the decoding protocol of \S\ref{sec:setup} (temperature $0$, top-$p$ $0.9$, $7{,}168$ max tokens, seed $42$, no system prompt) and are scored by the released CLI without modification; unless stated otherwise, recomputations are over the $19{,}128$ valid main-grid records of the nine models in Table~\ref{tab:model_list}.

\subsection{Realistic-Condition Decomposition of the Headline Rate}
\label{app:realistic_decomp}

The headline $38.0\%$ pools all three A-conditions, including the forced-wrap tier A2.
Table~\ref{tab:realistic_decomp} decomposes it by the role each condition plays.

\begin{table}[H]
  \centering
  \small
  \caption{Boundary/latent failure by elicitation condition (main grid, nine models). Wilson $95\%$ intervals in brackets.}
  \label{tab:realistic_decomp}
  \begin{tabular}{llr}
    \toprule
    Condition & Role & Failure rate \\
    \midrule
    A1 (wrapping forbidden), compliance-conditional & lower bound & $8.1\%$ (boundary) \\
    A3 (no wrapper instruction) & realistic anchor & $\mathbf{36.9\%}$ $[35.7, 38.1]$ (latent) \\
    A3 $\times$ \{B1, B4\} (no fence/wrapper meta-instruction) & cleanest estimate & $\mathbf{39.3\%}$ $[37.6, 41.0]$ (latent) \\
    A2 (forced wrap) & stress ceiling & $95.5\%$ (boundary) \\
    \bottomrule
  \end{tabular}
\end{table}

Models self-wrap in $2{,}354$ of $6{,}375$ A3 responses without any instruction to do so (three models exceed $77\%$), so A2 controls a behavior that models exhibit on their own.
The realistic anchor ($36.9\%$) sits close to the pooled $38.0\%$, which lies inside its interval, and the cleanest cells are higher ($39.3\%$): the headline is not an artifact of the stress tier.
Within A3, failures concentrate by behavior: among self-wrapped responses $85.8\%$ $[84.3, 87.2]$ are latent failures versus $8.3\%$ $[7.4, 9.1]$ among non-wrapped responses; in the A3 $\times$ \{B1, B4\} cells the split is $90.6\%$ versus $1.1\%$.
Among all content-correct A3 records, $51.2\%$ ($2{,}352/4{,}596$) are boundary-broken.
Table~\ref{tab:a3_per_model} gives the per-model A3 rates.

\begin{table}[H]
  \centering
  \small
  \caption{Per-model latent failure under A3 (no wrapper instruction) and in the A3 $\times$ \{B1, B4\} cells.}
  \label{tab:a3_per_model}
  \begin{tabular}{lrrrr}
    \toprule
    Model & A3 $n$ & A3 latent \% & A3 $\times$ \{B1, B4\} $n$ & latent \% \\
    \midrule
    Qwen3-7B & 713 & 64.7 & 357 & 77.3 \\
    Gemma-2-9B & 719 & 0.6 & 359 & 0.0 \\
    Llama-3.1-8B & 717 & 12.1 & 358 & 0.6 \\
    Qwen3-32B & 718 & 40.3 & 359 & 53.2 \\
    Gemma-3-27B & 720 & 78.2 & 360 & 78.9 \\
    Llama-3.1-70B & 719 & 25.5 & 359 & 0.0 \\
    GPT-4o & 720 & 44.0 & 360 & 84.2 \\
    Gemini-2.5-Flash & 629 & 71.2 & 299 & 63.2 \\
    Claude-Sonnet-4 & 720 & 0.0 & 360 & 0.0 \\
    \bottomrule
  \end{tabular}
\end{table}

The two tiers therefore answer different questions.
A3 and the natural sets (\S\ref{app:natural2}) estimate prevalence under realistic elicitation; A2 measures the capability ceiling, playing the role that adversarial perturbations play in robustness evaluation.
Because the main grid is balanced (every A $\times$ B $\times$ language cell contains the same $15$ tasks), the pooled rate is a designed factorial average that weights each pre-specified condition cell equally, not a deployment-weighted estimate; we recommend reading the realistic anchor and the stress ceiling as separate quantities.

\subsection{Downstream Harm Under Real Consumers}
\label{app:downstream_harm}

\paragraph{Method.}
Every valid main-grid response was passed through real consumers of LLM-generated Markdown: the CommonMark reference parser (\texttt{cmark}), three lenient renderers (\texttt{markdown-it}, \texttt{cmark-gfm}, \texttt{mistune}), and three code-block extractors (an AST-based extractor, a regex extractor, and a naive split on fence lines).
Each response was machine-checked against its intended structure, i.e., the blocks the model evidently meant to produce.
A rendering is counted as visibly corrupted if intended literal content leaks into the document (\emph{leak}), intended document structure is swallowed into a block (\emph{swallow}), an intended block is missing, or a phantom block appears.
An extraction is counted as harmful if any intended code block is truncated, contaminated, mangled, missing, or merged.
Judgments are conservative by construction, and a stratified manual spot-check confirmed every machine-flagged corruption.

\begin{table}[H]
  \centering
  \small
  \caption{Share of responses visibly corrupted when rendered, by error class (main grid, $19{,}128$ records).}
  \label{tab:renderer_harm}
  \begin{tabular}{lrrrrr}
    \toprule
    Error class & $n$ & \texttt{cmark} (strict) & \texttt{markdown-it} & \texttt{cmark-gfm} & \texttt{mistune} \\
    \midrule
    \etype{0} correct (control) & 8{,}449 & 12.1 & 12.1 & 12.1 & 12.1 \\
    \etype{1} premature close & 1{,}735 & 93.7 & 93.7 & 93.7 & 93.7 \\
    \etype{2} unclosed (lower bound) & 2{,}281 & 67.1 & 67.0 & 67.1 & 67.2 \\
    \etype{3} latent collision & 6{,}663 & 94.0 & 94.0 & 94.0 & 94.1 \\
    \bottomrule
  \end{tabular}
\end{table}

Cross-renderer disagreement is at most $0.2$ percentage points, so a more forgiving renderer does not rescue these outputs.
The \etype{0} control pools plain documents, for which the corruption rate is $0.0\%$, with documents whose intended blocks contain nested literal fences, for which it is $41.0\%$; the latter is a gap between author intent and CommonMark that the balance check \fmetric{2} cannot see, so the taxonomy understates rather than overstates harm.

\begin{table}[H]
  \centering
  \small
  \caption{Share of responses whose intended code is not faithfully recovered by an extractor, by error class.}
  \label{tab:extractor_harm}
  \begin{tabular}{lrrr}
    \toprule
    Error class & AST-based & regex & naive split \\
    \midrule
    \etype{0}, plain documents (control) & 1.0 & 2.0 & 1.9 \\
    \etype{1} premature close & 100.0 & 100.0 & 100.0 \\
    \etype{2} unclosed (lower bound) & 84.0 & 100.0 & 99.9 \\
    \etype{3} latent collision & \textbf{99.8} & 100.0 & 100.0 \\
    \bottomrule
  \end{tabular}
\end{table}

For \etype{3}, $99.7\%$ of records yield a truncated block even under the AST-based extractor.
The failures this benchmark flags are therefore not confined to strict validation: they corrupt rendered output in $94\%$ of cases and defeat best-practice code extraction in essentially all cases.

\subsection{Newer-Generation Models}
\label{app:newer_models}

We evaluated seven models released after the main study, with unchanged prompts and the released CLI: four models spanning the open-weight and proprietary generation that succeeded the main cohort, and three frontier models available at the time of writing (July 2026).
Table~\ref{tab:newer_models} reports the pooled latent-failure rate, the forced-wrap boundary-failure rate (A2), and the self-wrapping propensity under A3.

\begin{table}[H]
  \centering
  \small
  \caption{Seven newer models on the full grid. Latent failure is content-correct $\wedge$ boundary-wrong pooled over the grid (Table~\ref{tab:uniquely_captured} definition); A2 is boundary failure under forced wrapping, content-agnostic; A3 self-wrap is the share of responses wrapped in an outer fence without instruction (a behavior indicator, not a failure).}
  \label{tab:newer_models}
  \begin{threeparttable}
  \begin{tabular}{lrrr}
    \toprule
    Model & Latent failure \% & A2 forced-wrap failure \% & A3 self-wrap \% \\
    \midrule
    \multicolumn{4}{l}{\emph{Successor generation}} \\
    Qwen3.6-27B & 28.1 & 96.0 & 5.1 \\
    GLM-4.7 & 25.7 & 98.0 & 6.4 \\
    Claude Opus 4.5 & 31.4 & 99.6 & 0.0 \\
    Gemma-4-31B & 23.7 & 99.7 & 0.0 \\
    \multicolumn{4}{l}{\emph{Frontier at time of writing}} \\
    GPT-5.6 & 29.6 & 97.6 & 0.6 \\
    Gemini-3.6-Flash$^\dagger$ & 28.6 & 97.5 & 0.0 \\
    Claude Sonnet 5 & \textbf{10.3} & \textbf{34.7} & 0.4 \\
    \midrule
    \emph{Main cohort (nine models, pooled)} & \emph{38.0} & \emph{95.5} & \emph{36.9} \\
    \bottomrule
  \end{tabular}
  \begin{tablenotes}[flushleft]\footnotesize
    \item[$\dagger$] \texttt{thinking\_level=minimal}; the Gemini 3.x line has no non-thinking mode.
  \end{tablenotes}
  \end{threeparttable}
\end{table}

Three observations.
(1) The capability gap is unchanged in six of seven models: under forced wrapping they fail at $96$--$99.7\%$, Claude Opus 4.5 included.
(2) Lower pooled rates reflect a behavioral shift rather than a capability gain: prevalence tracks a generation's self-wrapping propensity, which collapsed from $36.9\%$ (main cohort) to $0$--$6\%$, but when wrapping does occur, failure is as certain as before.
(3) Claude Sonnet 5 is the sole genuine improvement ($34.7\%$ under A2), and the D-inner-run ablation shows it is shallow: its a-priori safe-length pass rate is $86.1\%$ at inner run $3$ but $6.2\%$ and $21.1\%$ at inner runs $4$ and $5$.
On the $30$-prompt natural set the newer models fail at $0$--$16.7\%$, overlapping the main cohort ($3.4$--$17.9\%$).
Re-running the cross-format probe of \S\ref{sec:cross_format} on the seven newer models confirms both halves of the asymmetric-delimiter argument: Python triple-quote collisions persist (parse-valid rate on the K1 condition is $0\%$ for Qwen3.6-27B and $26.7\%$ for GLM-4.7), while JSON stays at $100\%$ for all $16$ models under every nesting condition.
The benchmark thus discriminates behavioral avoidance from real capability, which is what a progress-tracking diagnostic is for.

\subsection{Expanded Natural-Prompt Set (\texorpdfstring{$n=248$}{n=248})}
\label{app:natural2}

\paragraph{Curation.}
To complement the $30$ human-authored prompts of Appendix~\ref{app:natural}, we mined $248$ organic prompts from public logs of real user conversations, WildChat-1M~\citep{Zhao2024} and LMSYS-Chat-1M~\citep{Zheng2024}.
Both datasets were streamed in full ($1.84$M conversations); rule-based filters and deduplication left $9{,}215$ candidates, from which a stratified sample of $2{,}144$ was adjudicated individually against a written inclusion rubric (a request to produce a Markdown document that plausibly contains fenced code).
Adjudication was LLM-assisted with author verification (a stratified re-adjudication agreed on $30/30$ items); all borderline and final-set decisions were made by the authors.
$318$ prompts met the rubric; a per-template-family cap, cross-source deduplication, and a decoding-budget rule yielded the final $248$.
No prompt was authored or synthesized for this benchmark.
One stratum validates the screen: of $400$ candidates selected solely by the presence of a code fence, none met the rubric, because in each the fence marked quoted material or an edit target rather than a request to produce a document containing fences; the remainder of that pool was excluded on that basis.
Yield in the retained strata ranged from $6\%$ to $38\%$.
For scale context, the main grid's breadth is $180$ unique base pairs ($12$ languages $\times$ $15$ tasks) expanded factorially for statistical power; the natural set adds $248$ distinct organic requests.

\paragraph{Protocol.}
Generation and scoring follow the main protocol.
Valid $N$ excludes length-truncated and non-stop records under the rule of Appendix~\ref{app:truncation}.
Because organic prompts do not specify an expected language, content correctness is undefined and the natural set reports boundary metrics only.
\emph{Self-wrap} is the share of responses the model wrapped in an outer fence with no instruction to do so; the last two columns of Tables~\ref{tab:natural2_main} and~\ref{tab:natural2_new} split boundary failure by that behavior.

\begin{table}[H]
  \centering
  \scriptsize
  \setlength{\tabcolsep}{4pt}
  \caption{Natural set ($n=248$), main-cohort generation. Wilson $95\%$ intervals in brackets.}
  \label{tab:natural2_main}
  \begin{threeparttable}
  \begin{tabular}{lrrlrr}
    \toprule
    Model & Valid $N$ & Self-wrap \% & Boundary failure \% $[95\%$ CI$]$ & Wrapped: fail \% ($n$) & Not wrapped: fail \% \\
    \midrule
    Qwen3-7B & 242 & 10.3 & 3.7 $[2.0, 6.9]$ & 24 (6/25) & 1.4 \\
    Gemma-2-9B & 248 & 3.6 & 0.4 $[0.1, 2.2]$ & 0 (0/9) & 0.4 \\
    Llama-3.1-8B & 243 & 1.6 & 4.1 $[2.3, 7.4]$ & 50 (2/4) & 3.3 \\
    Qwen3-32B & 243 & 10.3 & 0.8 $[0.2, 3.0]$ & 4 (1/25) & 0.5 \\
    Gemma-3-27B & 247 & 8.9 & 4.0 $[2.2, 7.3]$ & 36 (8/22) & 0.9 \\
    Llama-3.1-70B & 246 & 1.6 & 1.6 $[0.6, 4.1]$ & 0 (0/4) & 1.7 \\
    GPT-4o & 248 & 10.1 & 1.6 $[0.6, 4.1]$ & 12 (3/25) & 0.4 \\
    Gemini-2.5-Flash & 228 & 12.7 & 4.8 $[2.7, 8.4]$ & 21 (6/29) & 2.5 \\
    Claude-Sonnet-4.5$^\ddagger$ & 221 & 7.7 & 1.4 $[0.5, 3.9]$ & 18 (3/17) & 0.0 \\
    \bottomrule
  \end{tabular}
  \begin{tablenotes}[flushleft]\footnotesize
    \item[$\ddagger$] Claude-Sonnet-4 was retired from the provider API before this experiment (no 4.0 snapshot remains); we report its nearest surviving successor, labeled as a substitute. The Claude-Sonnet-4 responses of the main study remain in the released dataset, which is itself an argument for releasing response corpora rather than relying on API availability.
  \end{tablenotes}
  \end{threeparttable}
\end{table}

\begin{table}[H]
  \centering
  \scriptsize
  \setlength{\tabcolsep}{4pt}
  \caption{Natural set ($n=248$), newer models (the cohort of \S\ref{app:newer_models}).}
  \label{tab:natural2_new}
  \begin{tabular}{lrrlrr}
    \toprule
    Model & Valid $N$ & Self-wrap \% & Boundary failure \% $[95\%$ CI$]$ & Wrapped: fail \% ($n$) & Not wrapped: fail \% \\
    \midrule
    Qwen3.6-27B & 242 & 5.4 & 0.8 $[0.2, 3.0]$ & 15 (2/13) & 0.0 \\
    GLM-4.7 & 246 & 8.1 & 0.8 $[0.2, 2.9]$ & 10 (2/20) & 0.0 \\
    Claude Opus 4.5 & 210 & 6.2 & 1.4 $[0.5, 4.1]$ & 23 (3/13) & 0.0 \\
    Gemma-4-31B & 246 & 4.9 & 0.4 $[0.1, 2.3]$ & 8 (1/12) & 0.0 \\
    GPT-5.6 & 219 & 13.7 & 3.7 $[1.9, 7.0]$ & 23 (7/30) & 0.5 \\
    Gemini-3.6-Flash & 240 & 5.0 & 0.8 $[0.2, 3.0]$ & 17 (2/12) & 0.0 \\
    Claude Sonnet 5 & 229 & 6.1 & 1.3 $[0.4, 3.8]$ & 14 (2/14) & 0.5 \\
    \bottomrule
  \end{tabular}
\end{table}

Gemini-2.5-Flash-Lite, excluded from the main cohort under the truncation rule, is reported separately as a probe: valid $N$ $224$, self-wrap $15.6\%$, boundary failure $3.6\%$ $[1.8, 6.9]$, $23\%$ when wrapped versus $0.0\%$ when not.

\paragraph{Overall rate.}
Across the $16$-model cohort of Tables~\ref{tab:natural2_main} and~\ref{tab:natural2_new} ($3{,}798$ valid generations), the overall boundary-failure rate is $1.97\%$ $[1.58, 2.47]$ and models self-wrap in $7.2\%$ $[6.4, 8.1]$ of responses.
Two scope notes apply.
First, these prompts were curated for boundary relevance rather than sampled uniformly from user traffic, so this estimates failure within naturally occurring Markdown-document requests, not population-wide prevalence.
Second, the natural set reports boundary metrics only.

\paragraph{Failures concentrate where boundary pressure exists.}
Pooled over the $16$-model cohort, boundary failure is $\mathbf{17.5\%}$ ($48/274$) $[13.5, 22.5]$ when the model self-wrapped and $\mathbf{0.77\%}$ ($27/3{,}524$) $[0.53, 1.11]$ when it did not: a $23$-fold difference with disjoint confidence intervals.
Per-model wrapped strata are small ($n = 4$ to $30$), so the stratified claim is made only at the pooled level.
This is the natural-prompt analogue of the A3 stratification in \S\ref{app:realistic_decomp} ($85.8\%$ versus $8.3\%$) and is consistent with the elicitation gradient of Table~\ref{tab:natural_prompts}: organic requests rarely force deep nesting, so the pooled rate is low, but when a model does wrap, failures concentrate in the stratum the grid identifies.
Among the newer models, $19$ of $21$ errors are \etype{3} (latent collision) with no \etype{1} at all: the dominant mechanism in the wild is the one the benchmark isolates, and it is the mechanism that renders plausibly while breaking strict consumers (\S\ref{app:downstream_harm}).

\paragraph{Reduced self-wrapping does not transfer across prompt distributions.}
On the A3 grid, five of the seven newer models self-wrapped $0.0$--$0.6\%$ of the time, which invites the conclusion that the behavior has been trained away.
On organic prompts the same models self-wrap far more often, and fail when they do (Table~\ref{tab:selfwrap_transfer}).
Across the seven newer models, organic self-wrapping runs $4.9$--$13.7\%$ and fails at $16.7\%$ ($19$ of $114$) when it occurs.
Prevalence today is genuinely low; what the benchmark measures is the capability that determines what happens whenever wrapping does occur, whether the model initiates it, the user asks for raw source, or a downstream tool wraps the output.

\begin{table}[H]
  \centering
  \small
  \caption{Self-wrapping propensity on the A3 grid versus organic prompts, and failure when wrapped (newer models with near-zero A3 self-wrapping).}
  \label{tab:selfwrap_transfer}
  \begin{tabular}{lrrr}
    \toprule
    Model & Self-wrap, A3 grid \% & Self-wrap, organic \% & Failure when wrapped \% ($n$) \\
    \midrule
    Claude Opus 4.5 & 0.0 & 6.2 & 23 (3/13) \\
    Gemini-3.6-Flash & 0.0 & 5.0 & 17 (2/12) \\
    Gemma-4-31B & 0.0 & 4.9 & 8 (1/12) \\
    Claude Sonnet 5 & 0.4 & 6.1 & 14 (2/14) \\
    GPT-5.6 & 0.6 & 13.7 & 23 (7/30) \\
    \bottomrule
  \end{tabular}
\end{table}

Model-specific error signatures also reproduce under this different prompt distribution: the Llama family almost never self-wraps ($1.6\%$), matching Step~1 of the A1 decomposition (\S\ref{sec:a1_decomp}); Llama-3.1-8B's failures are \etype{2}-dominated ($8$ of $10$), while the large and newer models are \etype{3}-dominated, matching the per-model breakdown of Appendix~\ref{app:a_b_e_per_model}.

\paragraph{A worked organic example.}
One mined prompt, sent verbatim to Claude Opus 4.5 with no wrapper instruction: \emph{``write a 4-slide presentation on computer bit masks in form of a markdown file with reveal.js framework.''}
The model wrapped its whole answer in a three-backtick \texttt{markdown} fence and used three-backtick fences for the C examples inside it.
Its $95$-line response, elided:

\begin{tcolorbox}[colback=vsbg, colframe=vsbg, boxsep=2pt, left=4pt, right=4pt, top=2pt, bottom=2pt, sharp corners]
\begin{Verbatim}[fontsize=\scriptsize, formatcom=\color{vstext}]
 1  ```markdown
 2  ---
 3  title: Computer Bit Masks           (reveal.js front matter)
 7  ---
 9  # Computer Bit Masks                [slide 1]
15  ## What is a Bit Mask?              [slide 2]
19  ```c
21  0b00001111  // Mask for lower 4 bits
23  ```
25  ### Common Uses:
...
95  ```
\end{Verbatim}
\end{tcolorbox}

Line~1 opens the outer wrapper with a backtick run of $3$.
Line~19 is meant to open the inner example, but the outer block is still open and this line carries an info string, so CommonMark reads it as literal text and no nested block is created.
Line~23 is meant to close that inner example; as a bare fence of the same family and at least the same length, it closes the outer wrapper instead.
Everything from line~25 on sits outside the wrapper, and line~95, the fence the model intended as the outer close, ends up opening a new block.
The consequences separate along the renderer/consumer line.
\emph{Rendering:} the model intended one block holding the whole file; CommonMark produces five blocks, and the material after line~23 that was meant to be literal file content is rendered instead as live document structure (five headings, three horizontal rules, a bullet list, a blockquote). The page still reads as a plausible presentation, and nothing signals an error.
\emph{Extraction:} the AST-based extractor returns the first fenced block, lines~2--22: the front matter, slide~1, and slide~2 cut off inside its own C example. The user asked for a Markdown file to feed reveal.js, and the extracted file holds about half of the requested deck, while the response closes by instructing the user to save it as \texttt{slides.md}.
Our scoring records \fmetric{2} pass and \fmetric{3} fail, i.e., \etype{3} latent collision; no adversarial prompting is involved.

\paragraph{Excluded records.}
Truncation (\texttt{finish\_reason = length}) ranged from $0.0\%$ to $15.3\%$ across models and was excluded rather than repaired.
Two non-stop records were reproduced deterministically and are recorded rather than re-run: one API moderation block on a mined request to specify a credential-harvesting interface (arguably a correct refusal) and one over-refusal on harmless role-play by the Claude-Sonnet-4.5 snapshot, which Claude Sonnet 5 answered normally.

\subsection{Effective Diagnostic Diversity and a Coverage-Preserving Core}
\label{app:diversity}

A natural question for a focused benchmark is how many empirically separable diagnostic dimensions it measures, and whether a small representative subset would reproduce its conclusions.
We answer both with the $16$ models of \S\ref{app:newer_models} (nine main-cohort models plus seven newer models).

\paragraph{Condition families.}
Table~\ref{tab:condition_families} lists the condition families of the design exhaustively, with the diagnostic question each isolates.

\begin{table}[H]
  \centering
  \scriptsize
  \setlength{\tabcolsep}{4pt}
  \caption{Condition families and the diagnostic question each isolates (headline evidence over the $16$-model panel).}
  \label{tab:condition_families}
  \begin{tabular}{clp{0.60\linewidth}}
    \toprule
    \# & Condition family & Diagnostic question isolated, with headline evidence \\
    \midrule
    1 & Forced outer wrapping (A2) & Closing capability at the ceiling: $95.5\%$ pooled failure; the sole genuine improvement across $16$ models is Claude Sonnet 5 at $34.7\%$ \\
    2 & Spontaneous self-wrapping (A3) & Behavior rather than capability: propensity spans $0.0$ to $83.8\%$ across models; failures concentrate $85.8\%$ vs $8.3\%$ by stratum \\
    3 & Prohibition tier (A1) & The no-pressure floor: conditioned on compliance with the no-wrap instruction, residual failure is $8.1\%$, separating instruction-following from boundary capability \\
    4 & Inner-nesting dose (B1 to B4) & Dose response: Cochran--Armitage trend, pooled $Z = +6.17$, per-model heterogeneity $Z \in [-0.95, +5.27]$, documented U-shaped exceptions \\
    5 & Fence family (tilde fences) & Whether backtick discipline transfers to the other family: failure spans $51.9$ to $97.5\%$ \\
    6 & Mixed-family placement & Whether models exploit the family-specificity of the closing rule when backtick and tilde interleave: failure spans $3.5$ to $97.6\%$ \\
    7 & Outer-run length & Length discipline under 4- and 5-backtick wrappers \\
    8 & Inner-run length & A-priori safe-length choice; acquired by $3$ of $16$ models only (GPT-4o $18.3\%$, GPT-5.6 $26.0\%$, Claude Sonnet 5 $37.9\%$), near zero for the other $13$ \\
    9 & Explicit warnings (H) & Deployment of a known rule: warned, Claude Opus 4.5 falls to $21.3\%$ pooled ($0.7\%$ under its most explicit condition) whereas GLM-4.7 remains at $97.4\%$ \\
    -- & JSON asymmetric delimiters (K axis) & Falsification control, not counted as a dimension: induces no failure in any of the $16$ models under any nesting condition \\
    -- & Inline code spans; Python triple-quotes & The same symmetric-delimiter mechanism on two further surfaces (Appendices~\ref{app:inline} and~\ref{app:crossformat}); not counted as dimensions \\
    -- & \etype{1} / \etype{2} / \etype{3} outcomes & Mechanically distinct error types with different downstream signatures (\S\ref{app:downstream_harm}) and model fingerprints; not counted as dimensions \\
    \bottomrule
  \end{tabular}
\end{table}

\paragraph{Redundancy test.}
Whether these families are distinct measurements or repetitions of one structural issue is testable in the way test batteries are validated.
From each of the nine families we took one per-model statistic, plus the error fingerprint as a tenth axis (an outcome measure, not a condition family), with definitions unchanged from the main study: forced-wrap failure (A2); compliance-conditional residual failure (A1, reproducing the pooled $8.1\%$); self-wrap propensity (A3); nesting-dose trend (the per-model Cochran--Armitage $Z$ over B1--B4, reproducing Appendix~\ref{app:cochran_armitage} exactly for the nine main-cohort models); tilde-family failure; mixed-family placement failure; outer-run-length failure; inner-run-length failure ($100$ minus the a-priori safe-length acquisition rate); warned failure (pooled over the hint battery); and the error fingerprint (\etype{2} share among errors).
All axes are oriented so that larger values mean more failure, more exposure, or stronger dose sensitivity; the fingerprint axis is compositional, so grouping uses absolute correlations.
We ranked all $16$ models on each axis and computed Spearman correlations between axes (average ranks for ties; descriptive over the $16$-model panel).
If the axes were largely redundant, model orderings should show strong concordance across axes.

They do not.
The median inter-axis $|\rho|$ is $0.36$, $13$ of $45$ pairs are negative even after consistent orientation, and only $4$ of $45$ reach $|\rho| = 0.70$.
Using $|\rho| \geq 0.70$ as the redundancy threshold, the ten axes form seven connected components: a compliance-residual/fingerprint/mixed-family cluster ($0.74$ to $0.84$), a warned/outer-length cluster ($0.95$), and five singletons (forced wrap, self-wrap, inner-length, tilde-family, nesting-dose).
To avoid over-counting we additionally fold the inner-length axis into the conceptually adjacent warned/outer-length component (all three concern the fence-length rule), with which it has its strongest, still moderate, associations ($0.47$ to $0.55$).
This yields a conservative count of \textbf{six separable diagnostic dimensions}.
The forced-wrap axis, the headline capability, does not induce a common ordering on the rest: its nine correlations range from $-0.58$ to $+0.43$ and none reaches the threshold.
Table~\ref{tab:dissociations} lists concrete dissociations; the full $10 \times 10$ matrix and the per-model axis values are given in Tables~\ref{tab:axis_corr} and~\ref{tab:axis_profiles}.
The main study already reported this phenomenon from another angle: inter-model agreement on which prompts fail is near zero in the mid tier (Fleiss' $\kappa = +0.008$; Appendix~\ref{app:fleiss_kappa}).

\begin{table}[H]
  \centering
  \small
  \caption{Concrete dissociations between axes.}
  \label{tab:dissociations}
  \begin{tabular}{p{0.38\linewidth}p{0.56\linewidth}}
    \toprule
    Dissociation & Numbers \\
    \midrule
    Same forced-wrap failure level, opposite warning response & Under forced wrapping the two are nearly tied: Claude Opus 4.5 fails at $99.6\%$, GLM-4.7 at $98.0\%$. Warned, Opus 4.5 drops to $21.3\%$ while GLM-4.7 stays at $97.4\%$ \\
    Same era, opposite failure mechanism & \etype{2} share of errors: Gemma-2-9B $87.5\%$ vs GPT-4o $0.9\%$, two models released six weeks apart \\
    Two D-axis variants, unrelated skills & Tilde-fence discipline vs mixed-family placement: $\rho = 0.01$ \\
    Generational change tracked per axis & Nesting-dose trend flattens across the seven newer models ($Z \in [-0.97, +0.82]$, none significant) except Claude Sonnet 5, which retains it at $Z = +2.87$ \\
    \bottomrule
  \end{tabular}
\end{table}

\paragraph{A coverage-preserving core subset.}
Of the $4{,}179$ prompts, $2{,}160$ form the balanced factorial main grid underlying the headline aggregate; the remainder are axis-specific diagnostic sets (the hint battery, the D-axis manipulations, inline spans, cross-format, and the natural set), targeted rather than repeated.
The subset analysis therefore targets the main grid.
Under the requirement that every pre-specified A $\times$ B $\times$ language cell remain represented (the $3 \times 4 \times 12 = 144$ cells of Appendix~\ref{app:benchmark_axes_overview}), $144$ prompts (one per cell) is the coverage-preserving floor.
We drew such cores $500$ times (seed $42$), plus two variants (Table~\ref{tab:core_schemes}).

\begin{table}[H]
  \centering
  \small
  \caption{Core-subset schemes.}
  \label{tab:core_schemes}
  \begin{tabular}{llr}
    \toprule
    Scheme & Construction & $n$ \\
    \midrule
    core-144 & one task per (A, B, LANG) cell, sampled independently & 144 \\
    core-432 & three tasks per cell & 432 \\
    core-paired-144 & one task per (B, LANG) cell, all three A variants kept & 144 \\
    \bottomrule
  \end{tabular}
\end{table}

\begin{table}[H]
  \centering
  \small
  \caption{What the cores preserve (mean $\pm$ SD over $500$ draws; ranking entries are the mean Spearman correlation with the full-grid model ordering).}
  \label{tab:core_preserve}
  \begin{tabular}{lrr}
    \toprule
    & Pooled latent-failure rate & Model-ranking Spearman vs full \\
    \midrule
    Full grid ($2{,}160$ prompts) & 38.0\% & reference \\
    core-432 & $38.0 \pm 0.5\%$ & 0.991 \\
    core-144 & $38.0 \pm 0.8\%$ & 0.975 \\
    core-paired-144 & $38.0 \pm 0.9\%$ & 0.974 \\
    \bottomrule
  \end{tabular}
\end{table}

A small stratified core does preserve the aggregate rate and the broad model ordering, as expected if the factorial $n$ buys statistical power rather than breadth.
What it does not preserve is the resolution needed to determine why those scores arise (Table~\ref{tab:core_collapse}): the widened per-model intervals make generation-over-generation progress of a few points unmeasurable, and the same-task A1--A2 pairs that ground paired attribution shrink fifteen-fold, from $720$ to $48$ even in the design built to preserve them, and to $3$ of $720$ under independent sampling, because independent draws pick different tasks in the A1 and A2 cells.

\begin{table}[H]
  \centering
  \scriptsize
  \setlength{\tabcolsep}{4pt}
  \caption{What the cores lose (pair counts are means over $500$ draws). core-paired-144 retains $48$ same-task pairs by construction.}
  \label{tab:core_collapse}
  \begin{tabular}{lrrr}
    \toprule
    & A3 $\times$ \{B1, B4\} estimate & Per-model CI width (mean) & Same-task A1--A2 pairs \\
    \midrule
    Full grid & $39.3\%$ $[37.6, 41.0]$ & 3.6\,pp & 720 \\
    core-432 & CI width 7.6\,pp & 8.1\,pp & 29 \\
    core-144 & CI width 13.0\,pp ($\approx [33, 46]$) & 14.0\,pp & 3 \\
    \bottomrule
  \end{tabular}
\end{table}

Configuration-specific detection collapses as well.
Table~\ref{tab:fisher_power} gives the exact power of a two-sided Fisher test at $\alpha = 0.05$ for the observed Claude Sonnet 5 safe-length rates ($86.1\%$ at inner run $3$ versus $6.2\%$ at inner run $4$) as a function of per-cell sample size.
Eight prompts per cell would suffice to detect this single, unusually large contrast in isolation; it would not support the few-point generational deltas, the per-model intervals, the many configurations tested simultaneously, or the paired attribution above.
The detection is not hypothetical: it is how \S\ref{app:newer_models} identified the one model whose headline score improved substantially while the improvement stayed confined to the most common configuration.
In short, a core suffices for low-cost screening, while diagnosis across the six separable dimensions requires the full suite.

\begin{table}[H]
  \centering
  \small
  \caption{Exact power of a two-sided Fisher test ($\alpha = 0.05$) to detect the Claude Sonnet 5 inner-run-3 versus inner-run-4 contrast, by prompts per configuration cell.}
  \label{tab:fisher_power}
  \begin{tabular}{lr}
    \toprule
    Prompts per configuration cell & Power \\
    \midrule
    1 to 3 & 0\% (unreachable: the most extreme 3/3 vs 0/3 outcome has $p = 0.10$) \\
    4 & 42.5\% \\
    5 & 73.4\% \\
    8 & 93.2\% \\
    144 (the full ablation) & $>99.99\%$ \\
    \bottomrule
  \end{tabular}
\end{table}

\paragraph{Data.}
Tables~\ref{tab:axis_corr} and~\ref{tab:axis_profiles} give the correlation matrix and the per-model axis values behind the redundancy test.
Column key: A2, forced-wrap failure; A1c, compliance-conditional residual failure; SelfW, self-wrap propensity; E2sh, \etype{2} share among errors; InnerL, inner-run-length failure; Warn, warned failure; OuterL, outer-run-length failure; MixF, mixed-family placement failure; Tilde, tilde-family failure; BdoseZ, per-model Cochran--Armitage $Z$ (B1 to B4).
Bold entries mark pairs at or above the $0.70$ redundancy threshold.

\begin{table}[H]
  \centering
  \scriptsize
  \setlength{\tabcolsep}{3.5pt}
  \caption{Spearman correlations between the ten axes ($16$ models).}
  \label{tab:axis_corr}
  \begin{tabular}{lrrrrrrrrrr}
    \toprule
    & A2 & A1c & SelfW & E2sh & InnerL & Warn & OuterL & MixF & Tilde & BdoseZ \\
    \midrule
    A2 & 1 & $-0.36$ & $-0.10$ & $-0.58$ & $+0.31$ & $-0.13$ & $-0.21$ & $-0.26$ & $+0.43$ & $-0.34$ \\
    A1c & $-0.36$ & 1 & $+0.41$ & $\mathbf{+0.84}$ & $+0.36$ & $+0.45$ & $+0.34$ & $\mathbf{+0.74}$ & $-0.17$ & $+0.67$ \\
    SelfW & $-0.10$ & $+0.41$ & 1 & $+0.38$ & $+0.01$ & $+0.56$ & $+0.50$ & $+0.61$ & $-0.05$ & $+0.11$ \\
    E2sh & $-0.58$ & $\mathbf{+0.84}$ & $+0.38$ & 1 & $+0.27$ & $+0.55$ & $+0.47$ & $\mathbf{+0.76}$ & $-0.32$ & $+0.61$ \\
    InnerL & $+0.31$ & $+0.36$ & $+0.01$ & $+0.27$ & 1 & $+0.55$ & $+0.47$ & $+0.50$ & $+0.29$ & $+0.02$ \\
    Warn & $-0.13$ & $+0.45$ & $+0.56$ & $+0.55$ & $+0.55$ & 1 & $\mathbf{+0.95}$ & $+0.67$ & $+0.02$ & $-0.01$ \\
    OuterL & $-0.21$ & $+0.34$ & $+0.50$ & $+0.47$ & $+0.47$ & $\mathbf{+0.95}$ & 1 & $+0.59$ & $+0.05$ & $-0.02$ \\
    MixF & $-0.26$ & $\mathbf{+0.74}$ & $+0.61$ & $\mathbf{+0.76}$ & $+0.50$ & $+0.67$ & $+0.59$ & 1 & $+0.01$ & $+0.40$ \\
    Tilde & $+0.43$ & $-0.17$ & $-0.05$ & $-0.32$ & $+0.29$ & $+0.02$ & $+0.05$ & $+0.01$ & 1 & $-0.32$ \\
    BdoseZ & $-0.34$ & $+0.67$ & $+0.11$ & $+0.61$ & $+0.02$ & $-0.01$ & $-0.02$ & $+0.40$ & $-0.32$ & 1 \\
    \bottomrule
  \end{tabular}
\end{table}

\begin{table}[H]
  \centering
  \scriptsize
  \setlength{\tabcolsep}{3.5pt}
  \caption{Per-model axis values (\%; BdoseZ is a $Z$-statistic).}
  \label{tab:axis_profiles}
  \begin{tabular}{lrrrrrrrrrr}
    \toprule
    Model & A2 & A1c & SelfW & E2sh & InnerL & Warn & OuterL & MixF & Tilde & BdoseZ \\
    \midrule
    Claude-Sonnet-4 & 99.9 & 0.1 & 0.0 & 0.2 & 100.0 & 30.3 & 35.0 & 24.7 & 86.8 & $+1.56$ \\
    Gemini-2.5-Flash & 98.1 & 3.2 & 82.5 & 5.9 & 100.0 & 68.2 & 68.9 & 84.5 & 88.8 & $+2.32$ \\
    Gemma-2-9B & 84.8 & 3.1 & 0.1 & 87.5 & 100.0 & 83.1 & 75.9 & 91.0 & 70.0 & $+1.36$ \\
    Gemma-3-27B & 97.8 & 34.5 & 83.8 & 13.3 & 100.0 & 97.9 & 97.9 & 97.6 & 75.2 & $+4.43$ \\
    GPT-4o & 97.5 & 0.7 & 47.9 & 0.9 & 81.7 & 36.3 & 54.4 & 51.0 & 73.3 & $+1.31$ \\
    Llama-3.1-70B & 97.1 & 40.7 & 0.0 & 67.7 & 100.0 & 77.0 & 83.8 & 57.6 & 94.1 & $+5.27$ \\
    Llama-3.1-8B & 91.7 & 2.4 & 0.1 & 70.1 & 100.0 & 89.3 & 87.4 & 78.1 & 64.4 & $+3.56$ \\
    Qwen3-32B & 98.0 & 1.5 & 46.2 & 6.7 & 100.0 & 96.0 & 86.9 & 66.8 & 76.6 & $-0.95$ \\
    Qwen3-7B & 94.7 & 1.5 & 77.6 & 14.2 & 100.0 & 93.9 & 95.8 & 91.9 & 73.2 & $+3.20$ \\
    Gemini-3.6-Flash & 97.5 & 0.0 & 0.0 & 0.0 & 95.8 & 28.5 & 45.8 & 25.3 & 81.6 & $+0.37$ \\
    Gemma-4-31B & 99.7 & 0.0 & 0.0 & 0.0 & 100.0 & 71.9 & 76.1 & 3.5 & 53.7 & $+0.08$ \\
    GLM-4.7 & 98.0 & 0.0 & 6.4 & 2.1 & 100.0 & 97.4 & 98.6 & 71.7 & 97.5 & $-0.36$ \\
    GPT-5.6 & 97.6 & 0.0 & 0.6 & 0.3 & 74.0 & 44.2 & 75.0 & 9.4 & 79.9 & $+0.82$ \\
    Claude Opus 4.5 & 99.6 & 0.0 & 0.0 & 0.0 & 100.0 & 21.3 & 35.0 & 21.2 & 96.5 & $+0.67$ \\
    Qwen3.6-27B & 96.0 & 0.3 & 5.1 & 0.3 & 100.0 & 95.8 & 96.5 & 46.5 & 95.5 & $-0.97$ \\
    Claude Sonnet 5 & 34.7 & 0.1 & 0.4 & 2.8 & 62.1 & 29.6 & 44.4 & 9.0 & 51.9 & $+2.87$ \\
    \bottomrule
  \end{tabular}
\end{table}

\end{document}